\documentstyle[psfig]{elsart}

\def\pc{{\rm\thinspace pc}}
\def\yr{{\rm\thinspace yr}}
\def\kmps{\hbox{$\km\s^{-1}\,$}}
\def\Mpc{{\rm\thinspace Mpc}}
\def\Msun{\hbox{$\rm\thinspace M_{\odot}$}}
\def\keV{{\rm\thinspace keV}}

\def\km{{\rm\thinspace km}}
\def\s{{\rm\thinspace s}}
\def\g{{\rm\thinspace g}}
\def\as{\hbox{$a^*$}}
\def\keV{{\rm\thinspace keV}}
\def\eV{{\rm\thinspace eV}}
\def\K{{\rm\thinspace K}}
\def\erg{{\rm\thinspace erg}}
\def\ergps{\hbox{$\erg\s^{-1}\,$}}
\def\arcsec{{\rm\thinspace arcsec}}

\begin{document}
\begin{frontmatter}
\title{Fluorescent iron lines as a probe of astrophysical black hole systems}
\author[ad1]{Christopher S. Reynolds\thanksref{corr}},
\author[ad2]{Michael A. Nowak}
\thanks[corr]{Corresponding author. E-mail: chris@astro.umd.edu}
\address[ad1]{Department of Astronomy, University of Maryland, 
College Park, MD~20742, USA}
\address[ad2]{MIT, Center for Space Research, NE80-6077, 77 
Massachusetts Ave., Cambridge, MA~02139, USA}

\begin{abstract}
  With most physicists and astrophysicists in agreement that black
  holes do indeed exist, the focus of astrophysical black hole
  research has shifted to the detailed properties of these systems.
  Nature has provided us with an extremely useful probe of the region
  very close to an accreting black hole --- X-ray irradiation of
  relatively cold material in the vicinity of the black hole can
  imprint characteristic features into the X-ray spectra of black hole
  systems, most notably the K$\alpha$ fluorescent line of iron.
  Detailed X-ray spectroscopy of these features can be used to study
  Doppler and gravitational redshifts, thereby providing key
  information on the location and kinematics of the cold material.
  This is a powerful tool that allows us to probe within a few
  gravitational radii, or less, of the event horizon.
  
  Here, we present a comprehensive review of relativistic iron line
  studies for both accreting stellar mass black holes (i.e., Galactic
  Black Hole Candidate systems; GBHCs), and accreting supermassive
  black holes (i.e., active galactic nuclei; AGN).  We begin with a
  pedagogical introduction to astrophysical black holes, GBHCs, AGN,
  and accretion disks (including a brief discussion of recent work on
  the magnetohydrodynamical properties of accretion disks).  We then
  discuss studies of relativistic iron lines in the AGN context, and
  show how differences between classes of AGN can be diagnosed using
  X-ray spectroscopy.  Furthermore, through a detailed discussion of
  one particular object (MCG--6-30-15), we illustrate how the exotic
  physics of black hole spin, such as the Penrose and Blandford-Znajek
  processes, are now open to observational study.  We proceed to
  discuss GBHCs, which turn out to possess rather more complicated
  X-ray spectra, making robust conclusions more difficult to draw.
  However, even in these cases, modern X-ray observatories are now
  providing convincing evidence for relativistic effects.  We conclude
  by discussing the science that can be addressed by future X-ray
  observatories.
\end{abstract}


\end{frontmatter}

\section{Introduction : The astrophysics of relativistic compact objects}
\label{Intro}

Gravitational collapse of normal matter can produce some of the most
exotic objects in the universe --- neutron stars and black holes.
Proving that these objects exist in Nature occupied theoretical and
observational astrophysicists for much of the 20th century.  Most of
the detailed debate centered around understanding the possible final
states of massive stars.  On his now famous sea voyage from India to
England in 1930, Subrahmanyan Chandrasekhar considered the structure
of white dwarf stars --- compact stellar remnants in which
gravitational forces are balanced by electron degeneracy pressure.  He
realized that, if the white dwarf was sufficiently massive, the
degenerate electrons will become relativistic thereby rendering the
star susceptible to further gravitational collapse
\cite{chandrasekhar:31a,chandrasekhar:31b}.  Although hotly debated by
Arthur Eddington, Chandrasekhar correctly deduced that a white dwarf
would undergo gravitational collapse if its mass exceeded $M_{\rm
ch}\approx 1.4\Msun$ (where $1\Msun=2.0\times 10^{33}\g$ is the mass
of the Sun), a limit now known as the Chandrasekhar limit\footnote{The
Russian physicist Lev Landau independently calculated the upper mass
limit of a white dwarf star, and his results were published in 1932
\cite{landau:32a}.}.

Once gravity overwhelms electron degeneracy pressure, it is thought
that neutron degeneracy pressure is the last, best hope for averting
total gravitational collapse.  Objects in which gravitational forces
are balanced by neutron degeneracy pressure are called {\it neutron
stars}.  Although there was initial hope that nuclear forces would
always be sufficient to resist gravity, the upper limit to the mass of
a neutron star is now believed to be in the range $M\approx
1.8-2.2\Msun$\cite{akmal:98a,srinivasan:02a}.  Uncertainties arising
from the equation of state at super-nuclear densities continue to
plague our determination of this critical mass, but an absolute upper
limit of $M\approx 4\Msun$ arises from very general considerations,
i.e., the validity of General Relativity and the principle of
causality\cite{rhoades:74a}.  Above this mass, it is thought that
complete gravitational collapse cannot be avoided.  In particular,
Hawking's singularity theorems\cite{hawking:73a} show that the
formation of a spacetime singularity is unavoidable (irrespective of
the mass/energy distribution) once the object is contained within the
light trapping surface.  The result is a {\it black hole}, i.e., a
region of spacetime bounded by an event horizon and, at its heart,
possessing a spacetime singularity.

While the above considerations now have a firm theoretical base,
observational astrophysics was, and continues to be, critically
important in guiding our understanding of such extreme objects.  In
the case of both neutron stars and black holes, the very existence of
these objects was only widely accepted when compelling observational
evidence was forthcoming.  For neutron stars, the pivotal observation
was the discovery of pulsars by Jocelyn Bell and Anthony Hewish via
radio observations taken from Cambridge.  Black holes gained wide
acceptance after it was demonstrated that the X-ray emitting compact
object in the binary star system Cygnus~X-1 did, in fact, possess a
mass in excess of the maximum possible neutron star
mass\cite{murdin:71a,hutchings:78a,balog:81a,gies:86a,herrero:95a}.
This made it the first of the so-called Galactic Black Hole Candidates
(GBHCs), a class that has now grown to include some two dozen objects.

We now know of another class of black holes --- the supermassive black
holes, with masses in the range of $10^5-10^{9.5}\Msun$, that reside
at the dynamical centers of most, if not all, galaxies\footnote{There
  have been recent suggestions of a possible population of
  ``intermediate mass'' black holes, with masses in the range of
  $10^3$--$10^5$\Msun.  (See the discussion in \cite{miller:02d}, for
  example.)  The existence of such objects is still speculative, and
  detailed studies of their spectra, of the kind to be discussed in
  this review, do not yet exist.  We shall therefore not consider
  these objects further in this review.}.  Today, by far the strongest
case for a supermassive black hole can be made for our own Galaxy.
Modern high-resolution, infra-red imaging reveals that the stars in
the central-most regions of our Galaxy are orbiting an unseen mass of
$2.6\times 10^6\Msun$\cite{eckart:97a,ghez:98a,schoedel:02a}.
Furthermore, studies of the orbital dynamics (which now include
measured accelerations as well as velocities;
\cite{ghez:00a,eckart:02a}) constrain the central mass to be extremely
compact.  According to conventional physics, the only long-lived
object with these properties is a supermassive black hole.
Alternatives, such as a compact cluster of neutron stars, would suffer
a dynamical collapse on short time scales \cite{genzel:97a}.

\begin{figure}[t]
\label{fig:tanaka}
\centerline{
\hspace{2cm}
\psfig{figure=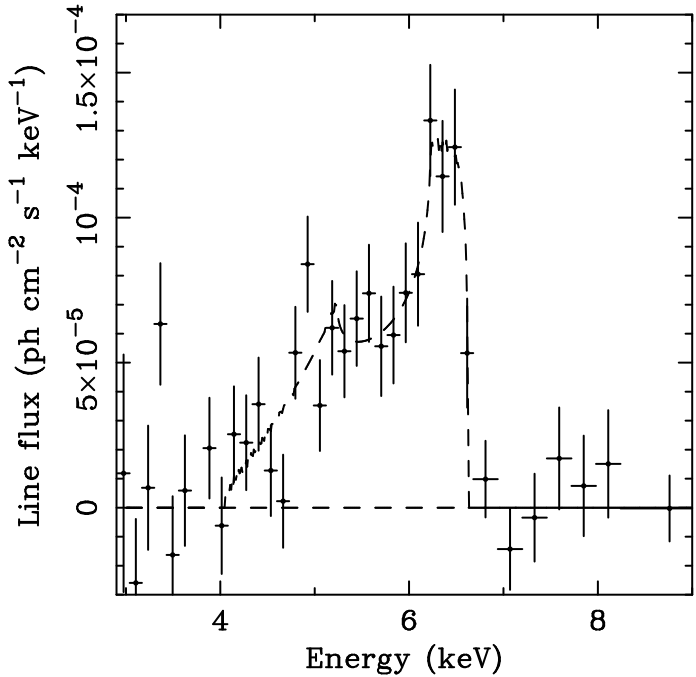,angle=270,width=1.0\textwidth}
}
\caption{Continuum subtracted iron line from the
  long July-1994 {\it ASCA} observation of the Seyfert-1 galaxy
  MCG--6-30-15 \cite{tanaka:95b}. The dashed line shows a model
  consisting of iron line emission from a relativistic accretion disk
  around a non-rotating (Schwarzschild) black hole, with a disk
  inclination of $i=30^\circ$, and an emissivity profile of $r^{-3}$
  extending down to the radius of marginal stability ($6\,{\rm GM/c^2}$).}
\end{figure}

Having established beyond reasonable doubt that black holes exist, it
is obviously interesting to perform detailed observational studies of
them.  The regions in the immediate vicinity of a black hole bear
witness to complex interactions between matter moving at relativistic
velocities, electromagnetic fields, and the black hole spacetime
itself.  Given that the apparent angular scales of even the biggest
black hole event horizons are $\sim 10^{-6}\arcsec$, direct imaging
studies of these regions will not be possible for many
years\footnote{However, on time scales of 10--20 years, there are plans
for both radio wavelength and X-ray interferometers which will be able
to image nearby supermassive black holes on spatial scales comparable
with that of the event horizon (see \S\ref{sec:maxim}).}.  In the
meantime, we must study these regions using more indirect methods,
chief among which are spectroscopic methods.

As we will detail in this review, Nature has provided us with a
well-understood and extremely useful spectral diagnostic of matter in
the near vicinity of astrophysical black holes.  In essence,
relatively cold matter in the near vicinity of an astrophysical black
hole will inevitably find itself irradiated by a spectrum of hard
X-rays \cite{guilbert:88a,lightman:88a}.  The result can be a spectrum
of fluorescent emission lines, the most prominent being the K$\alpha$
line of iron at an energy of $6.40-6.97\keV$ (depending upon the
ionization state of the iron) \cite{basko:78a,george:91a,matt:91a}.
Ever since the launch of the {\it Advanced Satellite for Cosmology and
  Astrophysics} (ASCA) in February 1993, X-ray astrophysicists have
had the capability to identify this emission line and measure its
spectral profile.  Figure~\ref{fig:tanaka} shows the iron line in the
X-ray emissions originating near the supermassive black hole in the
galaxy MCG--6-30-15 \cite{tanaka:95b}.  Bearing in mind that the line
is intrinsically narrow with a rest-frame energy of 6.4\,keV, it can
be seen that the line has been dramatically broadened and skewed to
low-energies.  It is now widely accepted that the line originates from
material that is just a few gravitational radii from the black hole,
and possesses a profile that is shaped by (relativistic) Doppler
shifts and gravitational redshift effects.  Investigating these
spectral features in X-ray luminous black hole systems has given us
the clearest window to date on the physics that occurs in the
immediate vicinity of astrophysical black holes.

The intent of this review is to describe our current understanding of
black hole astrophysics, with an emphasis on what has been learnt by
utilizing these X-ray spectral signatures.  We begin by discussing the
basic theoretical framework within which we understand the
astrophysical environment around black holes.  A central and important
part of this discussion is an introduction to the modern theory of
accretion disks.  Hand-in-hand with the theoretical discussion, we
will introduce the necessary phenomenology associated with stellar
mass and supermassive black holes.  We then describe the array of
past, current, and future X-ray observatories which have bearing on
relativistic studies of black holes before discussing how iron line
spectroscopy has dramatically improved our current understanding of
black hole astrophysics.  We conclude by presenting the prospects for
future research in this field.

\section{Black Holes: Theory and Astrophysical Setting}

\subsection{Review of some basic theoretical issues}
\label{basic_theory}

Throughout this review, we shall assume that standard General
Relativity is valid for all regions of interest.  Under this
conservative assumption, it is a remarkable fact that the relatively
simple Kerr metric \cite{kerr:63a} precisely describes the spacetime
outside of almost any astrophysical black hole\footnote{A dramatic
exception to this statement occurs when two black holes of comparable
mass merge.  Large and complicated deviations from the Kerr metric can
be readily obtained during such an event.  However, such events are
extremely rare and settle down to the Kerr metric on only a few light
crossing times of the event horizon.}.  The Kerr metric is described
by only two parameters which can be chosen to be the mass $M$ and
angular momentum $J$ of the black hole.  The no-hair theorem of
General Relativity tells us that black holes can also possess
electrical charge.  However, in any astrophysical setting, a black
hole with appreciable charge will rapidly discharge via vacuum
polarization. In Boyer-Lindquist coordinates (the natural
generalization of spherical polar coordinates), the line element of
the Kerr metric is given as
\begin{eqnarray}
  ds^2&=&-\left(1-\frac{2Mr}{\Sigma}\right)dt^2 -
  \frac{4aMr\sin^2\theta}{\Sigma}dt\,d\phi +
  \frac{\Sigma}{\Delta}dr^2\\\nonumber
&+&\Sigma\,d\theta^2+\left(r^2+a^2+
\frac{2a^2Mr\sin^2\theta}{\Sigma}\right)\sin^2\theta\,d\phi^2,
\end{eqnarray}
where $a=J/M$, $\Delta=r^2-2Mr+a^2$, $\Sigma = r^2+a^2\cos^2\theta$
and we have chosen units such as to set $G=c=1$.  The event horizon of
the black hole is given by the outer root of $\Delta=0$, i.e.
$r_+=(M+\sqrt{M^2-a^2})$.  The special case of $a=0$ corresponds to
non-rotating black holes, reducing the metric to the Schwarzschild
(1916) metric \cite{schwarzschild:16a}
\begin{equation}
  ds^2=-\left(1-\frac{2M}{r}\right)dt^2 +
  \left(1-\frac{2M}{r}\right)^{-1}dr^2+r^2\,d\theta^2+
  r^2\sin^2\theta\,d\phi^2,
\end{equation}
with the event horizon at $r=2M$.

A feature of the black hole spacetime that is crucially important for
accretion disk models is the existence of a radius of marginal
stability, $r_{\rm ms}$.  This is the radius within which circular
test particle orbits are no longer stable.  We shall refer to the
region $r<r_{\rm ms}$ as the {\it plunging region} since, in most
astrophysical settings, it will be occupied by material that is in the
process of plunging into the black hole.  For a given black hole mass,
this radius is a function of the black hole's angular momentum and,
for orbits in the equatorial ($\theta=\pi/2$) plane, is given by
\begin{eqnarray}
r_{\rm ms}&=&M\left(3+Z_2\mp \left[(3-Z_1)(3+Z_1+2Z_2) \right]^{1/2} \right),
\end{eqnarray}
where the $\mp$ sign is for test particles in prograde and retrograde
orbits, respectively, relative to the spin axis of the black hole and we have defined,
\begin{eqnarray}
Z_1&=&1+\left(1-\frac{a^2}{M^2}\right)^{1/3}\left[\left(1+\frac{a}{M}\right)^{1/3}+\left(1-\frac{a}{M}\right)^{1/3}\right],\\
Z_2&=&\left(3\frac{a^2}{M^2}+Z_1^2\right)^{1/2}.
\end{eqnarray}

Effects of the black hole spin that arise as a result of the
$dt\,d\phi$ term in the line element correspond to the dragging of
inertial frames.  An extreme manifestation of frame-dragging is the
existence of a region (the {\it ergosphere}; given by
$r<M+\sqrt{M^2-a^2\cos^2\theta}$) in which all time-like particles
must rotate in the same sense as the black hole as seen by an observer
at infinity.  The frame-dragging becomes more extreme as one
approaches the event horizon until, at the horizon, all time-like
particles orbit with an angular velocity of $\Omega_{\rm H}=a/2Mr_+$.
This is referred to as the angular velocity of the event horizon.
Another peculiar feature of the Kerr metric is the existence of
negative energy orbits within the ergosphere.  If a particle is placed
on such an orbit via some interaction within the ergosphere, it will
fall into the black hole and {\it diminish} the mass and angular
momentum of the black hole.  This gives rise to the Penrose-process
for extracting the rotational energy of a spinning black hole
\cite{penrose:71a}.  A consideration of black hole thermodynamics
readily shows that, in principle, an energy of up to 29\% of the mass
energy of the hole, i.e. 0.29\,M, can be extracted from a
maximally-rotating black hole, thereby reducing its angular momentum
to zero (e.g., see the treatment in \cite{shapiro:83a}).

In principle, a Kerr black hole can possess a spin parameter
arbitrarily close to $a/M=1$ without violating the cosmic censorship
hypothesis.  However, as argued by Kip Thorne \cite{thorne:74a}, a
rapidly spinning black hole that is actively accreting matter will
preferentially capture photons with negative angular momentum, thereby
limiting the spin parameter to $a/M\approx 0.998$ (depending somewhat on
the angular distribution of the photons emitted by the accreting
matter).  For this reason, most of the astrophysical literature uses
the term ``extremal'' or ``near-extremal'' Kerr black hole to refer to
a black hole with $a/M=0.998$.  We shall follow this convention.

\subsection{Stellar mass black hole systems and X-ray 
binaries}
\label{gbhc_sys}

Now that we have introduced the basic theory of black holes, we
present some of the phenomenology of real-life astrophysical black
holes.  In fact, the scientific communities studying stellar mass and
supermassive black holes have remained rather distinct from each
other, leading to the development of rather different languages and
phenomenological descriptions.  We begin by discussing the stellar
mass black holes that we believe arise from the collapse of a massive
star.

Although most observed stellar mass black hole candidates reside in
binary star systems, it is expected that the vast majority of
such black holes occur as isolated systems.  They remain undetected
due to their lack of any significant radiation.  Current estimates
indicate that there are between 0.1--1 billion such systems within our
Galaxy, a very small fraction of which, $< 0.001\%$, might be
detectable in a deep X-ray survey of the Galactic plane
\cite{agol:01a}.  Even a so-called isolated black hole will accrete at
some level from the surrounding gaseous interstellar medium (ISM) ---
this mode of accretion is called Bondi-Hoyle accretion after Herman
Bondi and Fred Hoyle who calculated the details of this process
\cite{bondi:44a}.  Low-level X-ray emission is expected due to
Bondi-Hoyle accretion, although there are large uncertainties
regarding both the mass flux and radiative efficiency of the accretion
flow \cite{agol:01a}.  Currently, the best prospects of detecting such
isolated black holes is via their gravitational effects, specifically
via gravitational micro-lensing --- the amplification of the observed
light from a background star due the gravitational focusing (i.e.,
bending) of that light in the potential of a foreground compact
object.  The amplification is expected to be time-symmetric,
achromatic (i.e., independent of wavelength), and of a duration from
weeks to months for expected stellar distances, velocities, and
compact object masses. A number of surveys for such micro-lensing
events are currently being undertaken, and there are claims of candidate
detections of stellar mass black holes \cite{bennett:01a,mao:01a}.

\subsubsection{The identification of Galactic Black Hole Candidates (GBHCs)}

By contrast to these isolated black hole systems, the first GBHC
discovered was in a binary star system.  A rocket borne X-ray
detector, launched in 1964, discovered the persistently bright X-ray
source Cygnus X-1 \cite{bowyer:65a}.  Its exact nature was not clear,
however, until its optical identification with the O-type star
HDE~226868 \cite{webster:72a,bolton:72a,hjellming:73a}.  Based upon
measurements of the optical companion's orbital parameters, mass
estimates for the compact object have ranged from $8-16\Msun$, with
recent estimates placing the mass at approximately $10\Msun$
\cite{hutchings:78a,balog:81a,gies:86a,herrero:95a}.  The chief
uncertainties in the compact object mass determination are due to some
uncertainty in the mass of the normal companion star, and due to the
unknown inclination of the system's orbital plane, with estimates of
the latter being scattered from 26--67$^\circ$
\cite{bochkarev:86a,ninkov:87a,ninkov:87b,dolan:92a}.  All estimates,
however, place the mass of the compact object above the theoretical
maximum mass for a neutron star.  The tremendous X-ray luminosities of
GBHCs (the X-ray luminosity of Cyg X-1 is approximately $10^5$ times
larger than the total luminosity of the Sun) are thought to be due to
accretion from the companion star onto the compact object.  As we will
discuss in \S3, accretion can be a very efficient process for
converting a mass flux into a persistent luminosity.  Prior to the
recent micro-lensing studies, \emph{all} known GBHCs were discovered
as X-ray sources in binary systems.

The identification of Cyg X-1 as a GBHC was followed by a series of
discoveries of other persistent or quasi-persistent X-ray sources that
were later identified as GBHCs.  These discoveries included the
systems LMC X-1
\cite{giacconi:74a,cowley:78a,cowley:95a,hutchings:87a}, LMC X-3
\cite{giacconi:74a,cowley:83a}, GX~339$-$4
\cite{markert:73a,samimi:79a}, and 4U~1957+11 \cite[the ``U'' standing
for \textsl{Uhuru}, an early 70's X-ray satellite that scanned much of
the sky]{giacconi:74a,margon:78a}.  The first two objects reside
approximately 50\,kpc away in the nearby satellite galaxy, the Large
Magellanic Cloud.  Similar to Cyg~X-1, the black hole classifications
of LMC X-1 and LMC X-3 rest upon dynamical mass estimates of 6 and
9\Msun, respectively \cite{cowley:95a,cowley:83a}.  The GBHC
classifications of GX~339$-$4 and 4U~1957+11, on the other hand, come
from the fact that their X-ray spectral and variability properties are
very similar to other GBHCs that do have dynamical mass estimates.
Specifically, GX~339$-$4 is very similar to Cyg~X-1 as well as other
systems \cite{zdziarski:98a,nowak:02a}, while 4U~1957+11 (whose black
hole candidacy is still vigorously being debated) is very similar to
LMC~X-3 \cite{nowak:99d,wijnands:02c}.  These analogies among the
properties of X-ray sources has led to the concept of `X-ray states'
with distinct spectral and temporal properties.

\subsubsection{The X-ray states of GBHCs}

Broadly speaking, the states of stellar mass black hole candidates are
described in terms of their `spectral hardness' and their short time
scale ($\approx 10^{-3}$--$10^3$\,s) variability properties.
Typically, the spectra at energies greater than $10$\,keV are
roughly described by a power law (which may or may not have a
detectable high energy `break'). The slope of this power-law is
often described via the photon index, $\Gamma$, where the photon
number flux per unit energy (${\rm photons~cm^{-2}~s^{-1}~keV^{-1}}$) is
\begin{equation}
F_{\rm N}(E)\propto E^{-\Gamma},
\end{equation}
where $E$ is the photon energy.  The luminosities of `hard states' are
dominated by emission from energies above $10$\,keV, and their spectra
have $\Gamma < 2$ \cite{miyamoto:92a}.  The luminosities of `soft
states' are dominated by emission from energies less than $3$\,keV.
Their spectra are usually described by a quasi-thermal component with
characteristic temperature between $\approx 0.5$--3\,keV and,
occasionally in addition to the low energy X-ray emission, high-energy
power-law spectra with $\Gamma > 2$ \cite{miyamoto:93a}.  At low
luminosities, stellar mass black hole systems tend to exhibit strongly
variable X-ray lightcurves, with root mean square [rms] fluctuations
of $\approx 40\%$, and `hard' X-ray spectra.  At high luminosities,
these systems tend to show more weakly variable (rms $\le 20\%$) X-ray
lightcurves and soft X-ray spectra.  There are a number of reviews
describing in more detail the properties of these states, with the
hard and soft states often being further divided into various
sub-states \cite[and references
therein]{tanaka:95b,nowak:95a,done:01b}.

\begin{figure}
\centerline{
\psfig{figure=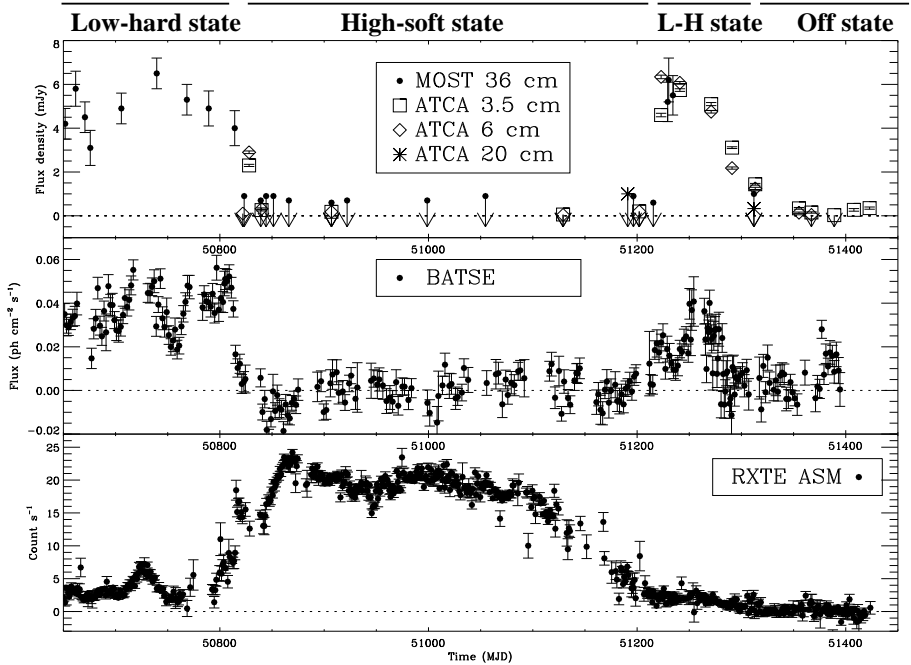,width=1.0\textwidth}
}
\caption{X-ray and radio observations of the Galactic Black Hole
Candidate GX~339$-$4 \cite{corbel:00a}, as the source transits between
a ``hard/radio-loud state'' and a ``soft/radio-quiet state''.  Top
panel refers to the measured radio flux (MOST=Molongo Observatory
Synthesis Telescope; ATCA=Australia Telescope Compact Array).  Middle
panel refers to the flux in hard (30--300\,keV) X-rays as measured by
the Burst and Transient Source Experiment (BATSE) on-board the {\it
Compton Gamma Ray Observatory}.  Bottom panel refers to the flux in
soft X-rays (1--12\,keV) as measured by the All Sky Monitor (ASM)
onboard the {\it Rossi X-ray Timing Experiment}.  The horizontal axis
is the time (Modified Julian Date) measured in units of days.}
\label{fig:gx339_states}
\end{figure}

Cyg~X-1, GX339$-$4, and LMC~X-3 each show transitions between the less
variable soft states, and highly variable hard states
\cite{nowak:02a,coe:76a,zhang:97b,cui:96a,ilovaisky:86a,grebenev:91a,miyamoto:91a,wilms:01a}.
Whereas Cyg~X-1 exhibits relatively little luminosity evolution during
its state transitions \cite{zhang:97b,cui:96a}, GX~339$-$4 exhibits
large luminosity fluctuations \cite{nowak:02a,kong:02a}, with the
source at times being nearly undetectable in the X-ray sky
\cite{ilovaisky:86a,kong:00a}.  Figure~\ref{fig:gx339_states}
illustrates these points with radio and X-ray data taken during a
typical state transition of GX~339$-$4.  From this perspective,
GX~339$-$4 is sometimes classified as a transient X-ray source.  In
contrast to sources such as Cyg~X-1, in fact the vast majority of
identified stellar mass black hole systems are transient X-ray
sources, with many of these transients falling into the class of
`X-ray novae'.

\subsubsection{X-ray transient sources}

As an example of an X-ray nova, we shall describe the system
GS~1124$-$683, a.k.a.  Nova Muscae.  Its behavior was characteristic
of many X-ray transients observed prior to the launch of the
\textsl{Rossi X-ray Timing Explorer (RXTE)}.  In the span of less than
10 days, the 1--10\,keV source flux rose by more than a factor of 100
\cite{kitamoto:92a}. At its maximum, the X-ray spectrum exhibited a
strong, soft quasi-thermal component and a power law tail with $\Gamma
\approx 2.5$ \cite{miyamoto:93a,miyamoto:94a}.  The flux then
exponentially decayed with a time constant of $\approx 30$\,days for
the soft X-ray flux and $\approx 13$\,days for the hard flux.  The
X-ray variability was also seen to decrease as the power-law component
flux decreased \cite{miyamoto:93a,miyamoto:94a}. Approximately 5
months after maximum flux, the source switched into a `hard state' with
$\Gamma \approx 1.7$ and then continued to decay.

After the X-ray source decayed away, optical observations revealed
modulations on an orbital period of 10.4 hours.  Combining this with
velocity measurements of the companion star, the lower limit to the
compact object mass was determined to be $3.1\Msun$ and its most
likely value was determined to be $6\Msun$ \cite{remillard:92a}.
Prior to those observations, several other quiescent X-ray novae had
revealed dynamical evidence for massive compact objects.  GS 2023+33,
a.k.a.  V404~Cyg was found to have a compact object mass of $10\Msun$,
and A0620$-$00 was found to have a compact object mass of $7.3\Msun$
(lower limit $3.2\Msun$) \cite{wagner:92a,mcclintock:86a}. There have
been approximately two dozen such transients, roughly half of which
have measured masses.  To date, our greatest source of information
about the mass distribution of stellar mass black holes has come from
optical observations of transient systems in quiescence
\cite{charles:01a}.  The X-ray nova outburst acts like a beacon
calling attention to the system, while the subsequent decay into
quiescence allows the orbital parameters of the binary system to be
measured via observations of the typically very faint companion star.

\begin{figure}
\centerline{
\hspace{0.5cm}
\psfig{figure=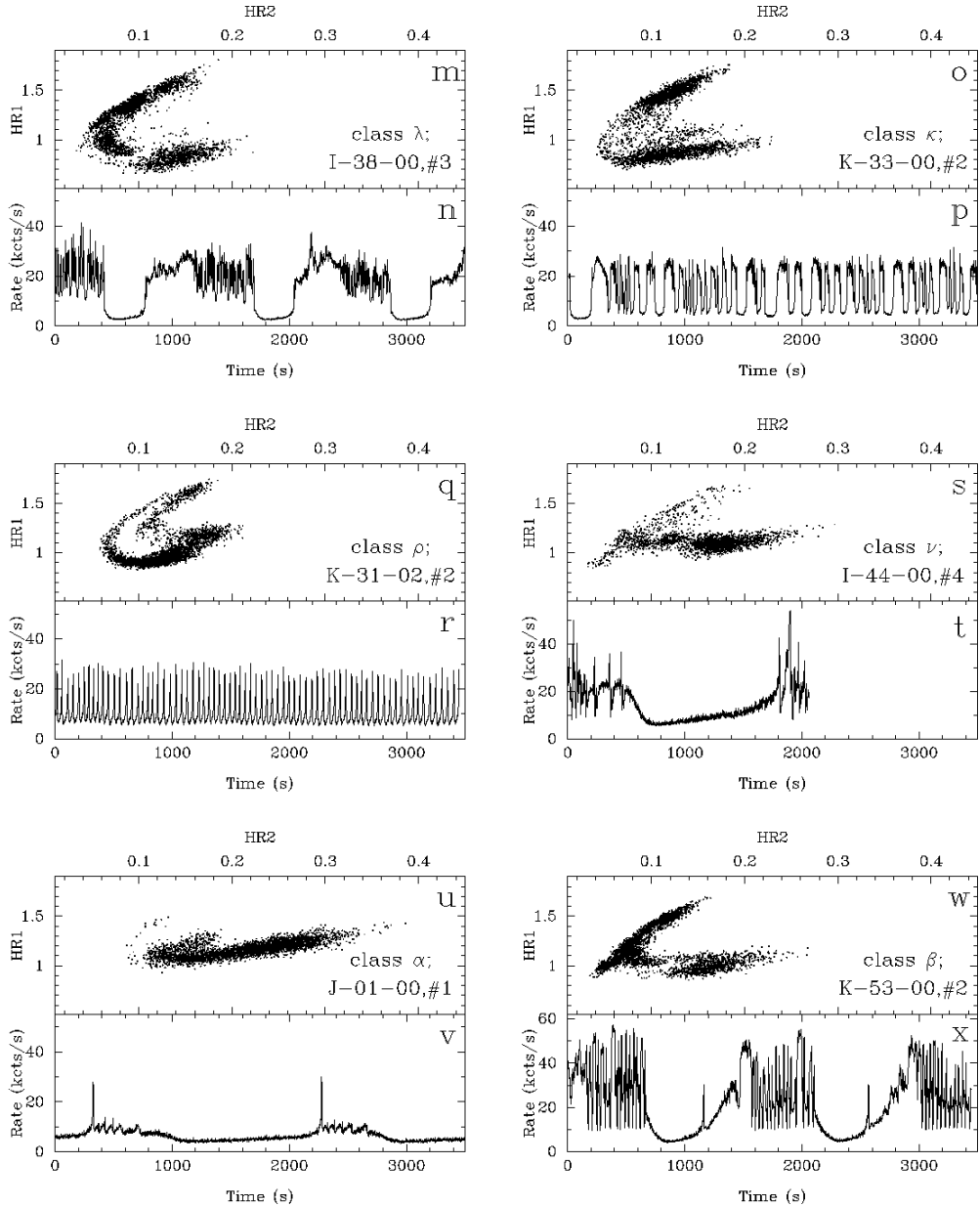,width=1.1\textwidth}
}
\caption{Sampling of lightcurves (1\,s bins) from various {\it RXTE}
observations of the Galactic ``micro-quasar'' GRS 1915+105
\cite{belloni:00a}.  Bottom panels show 2--60\,keV count rates in
units of thousand of counts per second.  (Counting noise is negligible
for these observations.)  Top panels show ``hardness ratios'' over the
course of these observations. HR$_1$ is defined as the ratio of the
5--13\,keV count rate to the 2--5\,keV count rate, while HR$_2$ is
defined as the ratio of the 13--60\,keV count rate to the 2--5\,keV
count rate.}
\label{fig:grs1915}
\end{figure}

X-ray transients that deviate strongly from the `fast-rise/exponential
decay' behavior of Novae Muscae also have been discovered.  A wide
variety of lightcurve morphologies have been observed by \textsl{RXTE}
\cite{belloni:00a}. Among the most interesting examples are
GRO~J1655$-$40 and GRS~1915+105, both of which were found to exhibit
relativistic jets of plasma that were observed in the radio band with
inferred velocities in excess of 0.9$c$
\cite{hjellming:95a,mirabel:94a}.  Both of these sources have shown
dramatic variability (see Fig.~\ref{fig:grs1915}), and the latter
source remains in an active X-ray state to the present day.  It has
since been realized that the formation of radio jets or outflows is
often associated with accretion in stellar mass black hole systems
\cite{fender:00b}.  Typically, the radio flux is intermittent and
`optically thin' (i.e., $F_\nu \propto \nu^{-\alpha}$, with $\alpha =
\Gamma - 1 > 0$, where $F_\nu$ is the radio energy flux per unit
frequency, ${\rm ergs~cm^{-2}~s^{-1}~Hz^{-1}}$) at the highest fluxes
within the soft state, becomes quenched at lower luminosities in the
soft state, and then re-occurs in the hard state.  The radio spectrum
in the X-ray hard state is usually persistent and `flat' or `inverted'
(i.e. $F_\nu\propto \nu^{-\alpha}$ with $\alpha \le 0$), and the radio
flux is positively correlated with the X-ray flux.
\cite{corbel:00a,fender:00b,hanni:98a,fender:99b,stirling:01a,fender:01a}.

There is currently a great deal of controversy regarding the
underlying mechanisms behind these state transitions, the formation of
the radio outflow, as well as the basic geometry of the accretion flow
itself.  Observations of relativistically distorted X-ray emission
lines, the prime focus of this review, are being used as probes of the
accretion flow properties near the event horizons of stellar mass
black holes in binary systems precisely for the purpose of gaining
more information about these issues.

\subsection{Supermassive black hole systems}
\label{smbh_sys}

We now turn to the basic phenomenology of supermassive black holes
(SMBHs).  Although it is artificial to place strict bounds on any
definition, the term SMBH usually refers to black holes with a mass
greater than $10^5\Msun$.  These objects are always found to reside at
the dynamical center of their host galaxy.  This is readily understood
through the action of {\it dynamical friction} \cite{binney:87a}.
Suppose the SMBH was not located at the center of the host galaxy and,
instead, was orbiting within the galaxy's gravitational potential.  As
the SMBH moves through and gravitationally interacts with the
background of stars (which have much lower mass) and dark matter, it
gravitationally induces a wake in its trail in which the stellar and
dark matter densities are enhanced.  The resulting gravitational force
between the wake and the SMBH itself acts as a drag force on the SMBH.
The effective drag force scales as $M^2$ and hence, for a sufficiently
massive black hole, is able to extract energy from the orbit of the
SMBH on a fairly short time scale and make it come to rest at the
bottom of the gravitational potential well, i.e., at the dynamical
center of the galaxy.

Unlike the case of typical stellar mass black holes, it is far from clear
how supermassive black holes formed.  While it is clear that accretion
has been responsible for significant growth of supermassive black
holes \cite{soltan:82a}, the origin of the initial massive ``seed''
black holes is highly controversial.  The interested reader is pointed
to the seminal review by Martin Rees \cite{rees:84a}.

\subsubsection{The Galactic Center}

While the proximity of our Galactic Center (only 8\,kpc away) is
obviously a big advantage for any study of the SMBH residing there, it
is a difficult region to study for other reasons.  The substantial
quantities of dust in the plane of the Galactic disk (and hence
between us and the Galactic Center) extinguish the optical light from
the Galactic Center by a factor of $10^{10}$, rendering it basically
invisible.  Progress must be made by utilizing wavelengths that can
more readily penetrate the Galactic dust --- in particular, radio,
infra-red, hard X-rays and $\gamma$-rays.

In recent years, dramatic progress has been made by high spatial
resolution studies of the near infra-red emission from bright stars in
the central-most regions of the Galaxy.  By achieving
diffraction-limited near infra-red (K-band) images of the Galactic
Center (with resolutions of 60~milli-arcsecs) over the time span of
several years, one can directly observe the motions of rapidly-moving
stars within the central stellar cluster that inhabits the Galactic
Center.  One finds that the velocities of stars within the central
regions of this cluster drops as the square-root of the distance from
the Galactic Center, implying a gravitationally dominant mass of
$2.6\times 10^6\Msun$ confined to the central $10^{-6}\pc^3$
\cite{eckart:97a,ghez:98a}.  A supermassive black hole is the only
long-lived object of this mass and compactness --- a compact cluster
of stellar-mass objects would undergo dynamical collapse on a time
scale of $10^7\yr$ or less \cite{genzel:97a}.  More recently, these
proper motion studies have also detected acceleration of several stars
in the Galactic Center region\cite{ghez:00a,eckart:02a}.  The
supermassive black hole hypothesis survives the powerful consistency
check allowed by a vector analysis of these accelerations.  Any
remaining doubt seems to have been removed by recently reported
observations that show a star passing within 17 light hours of the
putative SMBH \cite{schoedel:02a}.  The star remains on a Keplerian
orbit, even at the peri-center of the orbit where it achieves a
velocity exceeding $5000\kmps$.  This raises the implied mass-density
to more than $10^{17}\Msun\,{\rm pc}^{-3}$ --- if this were a cluster
of compact stellar remnants, it would evaporate or collapse on a
timescale of less than $10^5\yr$.  

\subsubsection{Other kinematic studies of SMBHs, and the SMBH census}

Kinematic studies of stars around the SMBH in the center of our own
Galaxy have only been possible in recent years with the development of
diffraction limited near infra-red imaging on large ground-based
telescopes. Before that, studies of the central regions of other
galaxies (where the obscuration problem is much less severe) were
crucial for establishing the existence of SMBHs.  In particular, there
were two studies published in the mid-1990s that were pivotal in
providing compelling evidence for SMBHs --- the Hubble Space Telescope
(HST) observations of M87 and the radio observations of the H$_2$O
MASERs in NGC~4258.

The center of M87 was an important target for HST since a central SMBH
had long been suspected based upon previous ground-based optical
observations of the galaxy's central regions
\cite{young:78a,sargent:78a}, as well as due to the presence of a
prominent synchrotron emitting jet of plasma that flows away from the
galaxy's core at relativistic speeds.  HST imaged a disk of ionized
gas, with a radius of $\sim 50\pc$ centered on the galactic core
\cite{ford:94a}.  The high resolution of HST allowed the spectrum of
this ionized gas to be measured as a function of position across the
gas disk, thereby allowing the kinematics of the disk to be determined
\cite{harms:94a}.  It was found that the velocity profile of the
central $20\pc$ of the gas disk possessed a Keplerian profile (i.e.,
$v\propto r^{-1/2}$) as expected if the gas was orbiting in the
gravitational potential of a point-like mass
\cite{harms:94a,macchetto:97a}.  From the measured velocities of this
disk ($\sim 1000\kmps$), the central mass was determined to be
$3\times 10^9\Msun$.  The only known and long-lived object to possess
such a large mass in a small region of space, and be as under-luminous
as observed, is a SMBH.

Similar conclusions were drawn for the center of the galaxy NGC~4258,
albeit using very different observational techniques
\cite{greenhill:95a,myoshi:95a}.  The center of this galaxy contains
molecular gas which is subjected to heating by a central X-ray source.
Collisional pumping of water within the molecular gas leads to a
population inversion thereby driving a naturally occurring maser.  Both
the spatial position and line-of-sight velocity of the masing blobs
can be measured very accurately using radio observations (in
particular, the Very Long Baseline Array; VLBA).  It is found that the
masers lie in a very thin disk (oriented almost edge-on to us) that is
orbiting a central object of mass $3.6\times 10^7\Msun$ with an almost
perfect Keplerian velocity profile (with the velocity of the inner
masing region being $\sim 1000\kmps$).  The accuracy with which the
velocity profile follows a Keplerian law tightly constrains the
spatial extent of the central mass, again rendering any explanation
other than a SMBH very problematic.  The direct detection of
centripetal accelerations within the masing regions provides a
powerful consistency check \cite{greenhill:95b}.

While the Galactic Center, M87 and NGC~4258 are important cases in the
argument for the existence of SMBHs, it is difficult to generalize
these observational techniques to all galaxies.  A more generally
applicable technique is to examine the ensemble velocity distribution
of stars in the central region of a galaxy via observations of stellar
absorption lines in galactic spectra.  By carefully comparing detailed
galactic models (that include the distribution of stars across the
possible phase space of orbits within a given gravitational potential)
with high quality imaging and spectral data, one can constrain the
mass of any central black hole.  Using these techniques, several
authors have performed relatively large surveys of nearby galaxies in
order to examine the demographics of SMBHs.  There are two exciting
results from these studies.  Firstly, it appears that every galaxy
that possesses a well defined bulge contains a SMBH.  Secondly, there
is a good correlation between the mass of the central black hole $M$
and the mass of the galactic bulge ($M_{\rm bulge}$), with $M/M_{\rm
bulge}\sim 0.003$ \cite{magorrian:98a,gebhardt:00b}.  In fact, it has
been shown that the more fundamental (and better) correlation is
between the mass of the central SMBH and the velocity dispersion
$\sigma$ (or ``temperature'', considering an analogy between stellar
kinematics in a galaxy and particle kinematics in a gas) of the
stellar population; $M\propto \sigma^{b}$, with $b=4-5$
\cite{gebhardt:00a,ferrarese:00a}.  The underlying cause for this
correlation is still the subject of intense work and much debate,
since these stars are far enough from the center of the galaxy to have
negligible direct influence from the gravitational field of the SMBH.
It is widely regarded that this correlation argues for a connection
between the formation of the SMBH and the galaxy itself.

\subsubsection{Active Galactic Nuclei and Quasars}

Active Galactic Nuclei (AGN) were known and studied before the concept
of SMBHs became firmly established.  Observationally, an AGN is
defined as a galactic nucleus which displays energetic phenomena such
as large electromagnetic luminosities and/or powerful jets.  The first
known AGN was the ``radio quasar'' 3C~273.  This object was first
identified by radio observations, but radio imaging alone was
insufficient to localize its position on the sky well enough to allow
follow-up investigation with optical telescopes.  A major breakthrough
was the use of lunar occultation techniques (i.e., precision
measurement of the time at which the radio emissions from 3C~273 were
blocked as the Moon passed in front of it) by Cyril Hazard and
collaborators \cite{hazard:63a} which localized 3C~273 to within
1\,arcsec.  This allowed an identification and subsequent spectroscopy
of the corresponding optical object.  The optical spectrum was
initially confusing, displaying emission lines at wavelengths that did
not correspond to any known atomic transition, until it was realized
by M.Schmidt that the spectrum was redshifted by a seemingly enormous
factor \cite{schmidt:63a}
\begin{equation}
z\equiv\frac{(\lambda_{\rm obs}-\lambda_{\rm emit})}{\lambda_{\rm
emit}}=0.158.
\end{equation}
Interpreting such a redshift as cosmological (i.e., due to the
expansion of the Universe), suggests the object to be extremely
distant and, therefore, extremely luminous.  These enormous
luminosities were the clue that led some researchers, most notably
Salpeter, Zeldovich and Lynden-Bell, to suggest the existence of
accreting SMBHs \cite{salpeter:64a,zeldovich:64a,lynden-bell:69a}.
Accretion of matter into a relativistically deep gravitational
potential well is one of the few processes in nature efficient enough
to produce such luminosities without having to process unreasonably
large amounts of mass.  The supermassive nature of the black holes
follows from the requirement for the AGN to remain gravitationally
bound (and hence long-lived) despite the enormous outward pressure
exerted by the electromagnetic radiation.

\begin{figure}
\centerline{
\psfig{figure=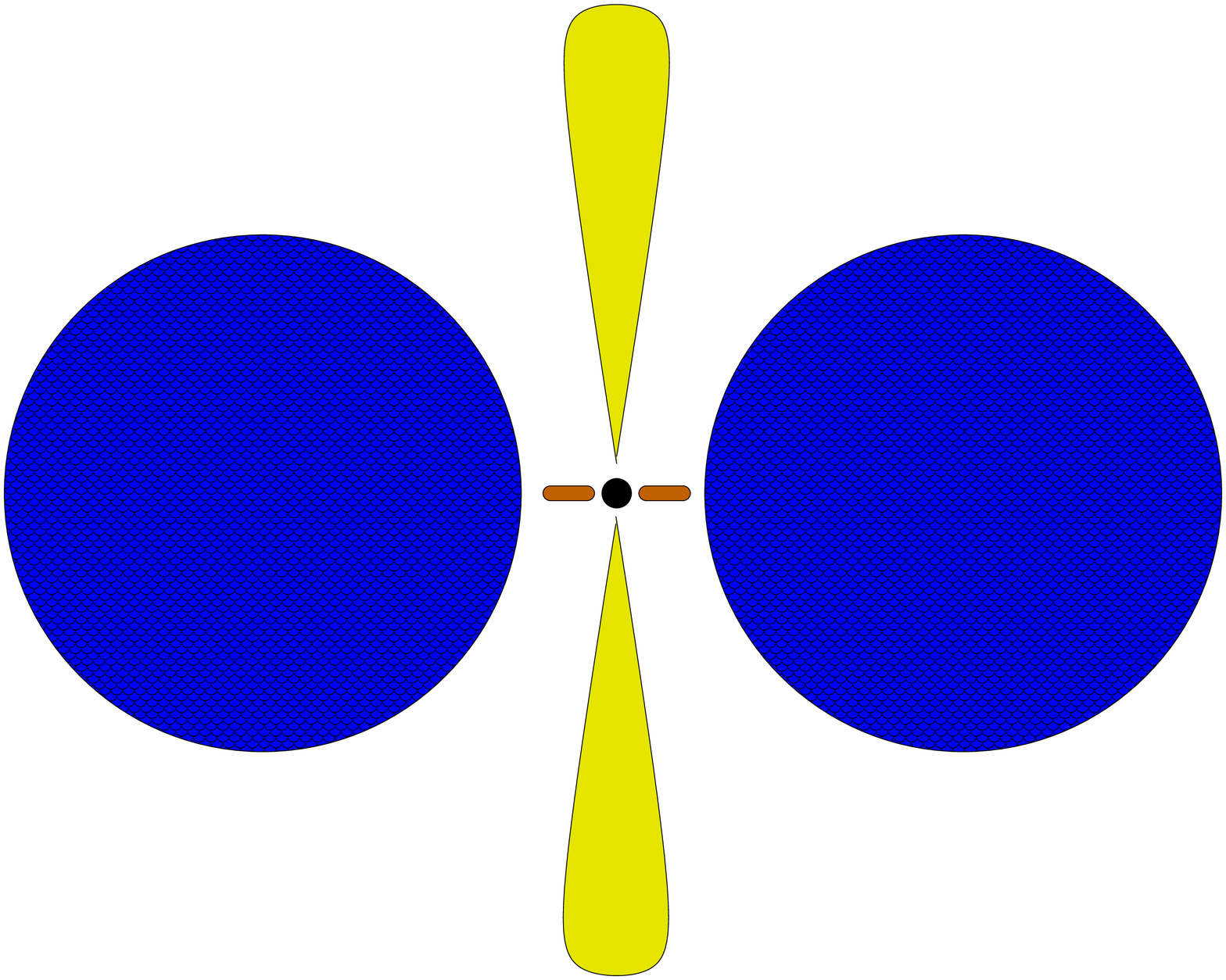,width=0.8\textwidth}
}
\centerline{
\psfig{figure=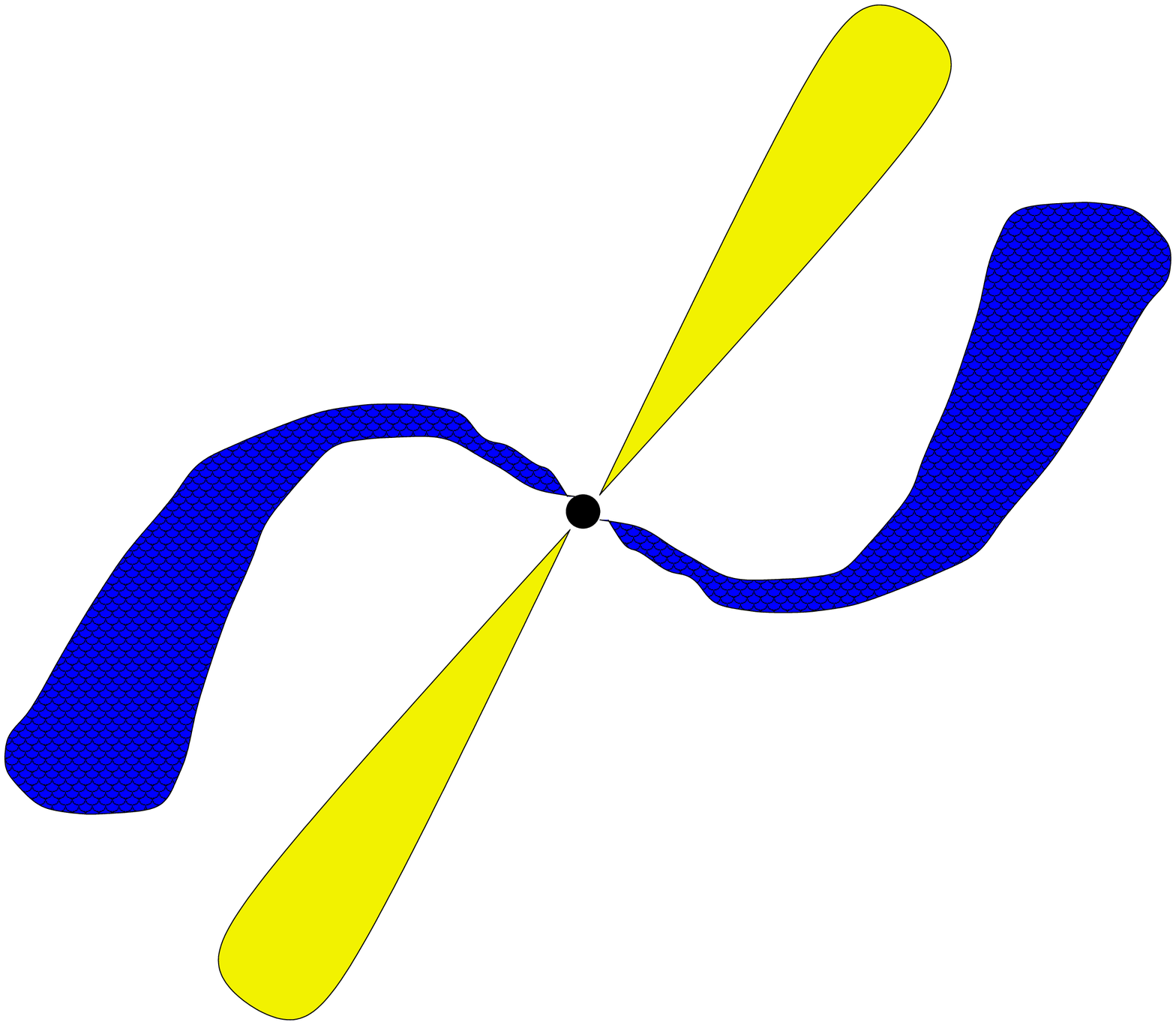,width=0.8\textwidth}
}
\caption{Two possible geometries relevant for active galactic nuclei (AGN).  
  The dark shading represents dusty molecular gas that is in either a
  toroidal geometry (top) or warped disk geometry (bottom).  The
  lighter shading represents the jets that are known to exist in
  radio-loud AGN and may well exist (albeit in a weaker form) in
  radio-quiet AGN as well.}
\label{fig:agn_geom}
\end{figure}

After four decades of study, there is a large and complex
phenomenology associated with AGN.  It has proven useful to classify
AGN according to their luminosity, electromagnetic spectrum, and
spatial (radio) morphology\footnote{We will only describe the most
coarse aspects of the classification scheme here --- the interested
reader is directed to Chapter 1 of the book by Krolik
\cite{krolik:99a} for further details.}.  A basic dichotomy seems to
exist between radio-loud AGN and radio-quiet AGN.  Radio-loud AGN (of
which 3C~273 and M87 are examples) possess back-to-back twin jets of
plasma that are produced in the vicinity of the SMBH and propagate
outwards at relativistic velocities (with Lorentz factors of 5--10 or
more) for, in some cases, distances of $10^5-10^6\pc$.  These jets
tend to be copious sources of radio and X-ray emission due to,
respectively, synchrotron and inverse Compton emission by high-energy
electrons within the magnetized plasma.  The primary physical
mechanisms initiating, accelerating and collimating these jets are
still far from clear, although it seems likely that they are launched
from the inner accretion disk or ergosphere of the SMBH and are
accelerated/collimated by hydromagnetic processes.  In the rare cases
where one of the jets is directed straight at us, special relativistic
beaming strongly enhances the observed jet emission, often diluting
and thus making it difficult to observe any other emissions from the
AGN.  Such objects are known as blasars (of which the well-studied
BL-Lac objects are a sub-class) and can be sources of extremely
energetic emissions (with photon energies up to $\sim 10$\,TeV having
been detected from the BL-Lac objects Mrk501 and Mrk~421
\cite{quinn:96a,catanese:97a}).

Radio-quiet objects, on the other hand, seem not to possess these
powerful jets.  Radio-quiet AGN with moderate electromagnetic
luminosities ($L\sim 10^{43}-10^{45}\ergps$), commonly referred to as
Seyfert galaxies, are a particularly important and well studied
sub-class.  This is due to the fact that they are reasonably common
(constituting 1--10\% of all major galaxies), leading to some
relatively nearby and easily studied examples.  Seyfert galaxies
themselves have been classified into two broad categories.  Those
objects in which we can directly view the energetic regions
immediately around the SMBH are called type-1 Seyfert galaxies (often
referred to as Seyfert-1 galaxies).  On the other hand, if the SMBH
region is obscured by large amounts of dust and gas (situated near to
the AGN and/or within the host galaxy), it is called a type-2 Seyfert
galaxy.  Spectropolarimetry of some nearby Seyfert galaxies (in
particular NGC~1068; \cite{antonucci:85a}) provides evidence that the
same AGN may appear as a type-1 or type-2 depending upon the
orientation at which we view the system.  Two possible geometries that
are often discussed are the dusty torus geometry or the warped disk
geometry (see Fig.~\ref{fig:agn_geom}).  In these geometries, an
observer viewing the AGN along a line of sight that intercepts the
dusty torus or the warped disk will be obscured, leading to a type-2
classification.  An observer who views the SMBH region unobscured will
assign a type-1 classification.  Since our discussion
focuses on observational constraints on the region very close to the
black hole, we shall focus on type-1 AGN.

In the absence of strongly beamed jet emissions, the overall
electromagnetic spectrum of type-1 radio-quiet and radio-loud AGN are
qualitatively similar.  There is often a distinct bump in the spectrum
at optical/UV wavelengths, generically referred to as the ``big blue
bump'', that is thought to correspond to thermal radiation from the
optically-thick portions of the accretion flow.  Furthermore, a
power-law spectral component extends from the big blue bump up to hard
X-ray energies.  As we will discuss in \S\ref{sec:coronae}, we believe
that this component arises from inverse Compton scattering of
optical/UV photons in a very hot corona associated with the accretion
flow. A high-energy exponential cut-off at $kT\sim 100-200\keV$,
corresponding to the temperature of the coronal plasma, has been
observed directly in some Seyfert-1 galaxies.  

\section{Accretion disks and disk atmospheres}

\subsection{Introduction to accretion and accretion disks}
\label{disks_intro}

As we have discussed in \S\ref{gbhc_sys} and \S\ref{smbh_sys}, many of
the observable effects from black hole systems arise due to the
accretion of gas.  At its most basic level, gravitational potential
energy can be released from the accreting matter and produce
electromagnetic radiation or power bi-polar outflows or jets.  More
exotic is the possibility that accreting matter may produce and
support magnetic fields which interact with the rotating black hole
thereby facilitating the extraction of its spin energy via, for
example, the process elucidated by Roger Blandford and Roman Znajek
\cite{blandford:77a}.

In any realistic scenario, matter will be accreting from a substantial
distance and will possess significant angular momentum.  In this case,
we expect the accreting gas to form a rotationally-flattened structure
orbiting the black hole, an {\it accretion disk}.  A particular
element of gas in the accretion disk will then lose angular momentum
through a process which is loosely referred to as {\it viscosity},
thereby allowing it to flow slowly inwards towards the black hole and
eventually accrete.  While the nature of this angular momentum
transport was mysterious for some time, we now believe it to be due to
magneto-hydrodynamical (MHD) turbulence driven by a well-studied
instability, the magnetorotational instability (MRI).  In this
section, we shall present the ``standard'' black hole accretion disk
model and discuss these MHD phenomena.  We then proceed to discuss
disk atmospheres and the X-ray spectral signatures that result from
these surface layers.  Readers who are already familiar with accretion
disk physics and X-ray reflection may wish to skip to
\S\ref{sec:lines_from_disks}.  An excellent review of disk
hydrodynamics and MHD, with emphasis on turbulent transport within
astrophysical disks, is given by Steven Balbus and John Hawley
\cite{balbus:98a}.

Before launching into a detailed discussion of accretion disks, we
must introduce the concept of the {\it Eddington limit}.  Consider
spherically-symmetric accretion onto an object of mass $M$, producing
a radiative luminosity of $L$.  If the accreting material is totally ionized,
its opacity will be dominated by electron-scattering.  If $L$ exceeds
a critical luminosity (the Eddington luminosity) given by
\begin{equation}
L_{\rm Edd}=\frac{4\pi c G M \mu_e}{\sigma_{\rm T}},
\end{equation}
where $\mu_e$ is the mass per electron and $\sigma_{\rm T}$ is the
electron scattering cross section, then the outward pressure of
radiation will exceed the inward force of gravity making prolonged
accretion an impossibility.  While the Eddington luminosity strictly
applies to spherically-symmetric systems, it also probably sets a
practical limit for the luminosity of disk-like accretion flows
(although there are recent suggestions that MHD processes may allow
the limit to be somewhat violated \cite{begelman:01a}).

\subsection{The structure of accretion disks}

We begin by discussing the structure and dynamics of the so-called
standard model of radiatively-efficient accretion disks.  We shall
focus on such disks since they turn out to be the most relevant
for iron K$\alpha$ studies of black hole systems.  Our discussion will
be brief and aims at providing a review of the salient points required
for the observationally-oriented discussion later.  The interested
reader is referred to the articles cited near the beginning of each
segment of this discussion.

\subsubsection{Non-relativistic radiative accretion disks \& the $\alpha$-prescription}

Initially, we shall discuss non-relativistic models of accretion
disks.  Here, we follow the presentations and arguments presented in
\cite{shakura:73a}, \cite{pringle:81a}, and \cite{frank:92a}.
Throughout this discussion, we assume cylindrical polar coordinates,
$(r,\phi,z)$.

We consider a rotating mass of gas around a point mass $M$, which
represents the black hole.  We assume that the gas can
radiate-efficiently and, hence, that it maintains a temperature well
below the virial temperature of the system.  Furthermore, we assume
that an angular momentum transport mechanism operates and can be
characterized by a kinematic viscosity $\nu$, but we suppose that the
time scale for redistributing angular momentum is long compared with
either the orbital or radiative time scales.  Under these assumptions,
the gas should form a geometrically-thin disk (with vertical scale
height $h$) in which fluid elements are orbiting in almost circular
paths with angular velocity $\Omega(r)$.  A small radial velocity
$v_r(r,t)$ corresponds to the actual accretion flow.

If we denote the surface mass density of the disk as $\Sigma(r,t)$,
the conservation of mass and angular momentum can be written as
\begin{equation}
r\frac{\partial \Sigma}{\partial t}+ \frac{\partial}{\partial r}(r\Sigma v_r)=0,
\end{equation}
and
\begin{equation}
r\frac{\partial}{\partial t}(\Sigma r^2\Omega)+\frac{\partial}{\partial r}(r^3\Sigma v_r\Omega)= \frac{1}{2\pi}\frac{\partial {\cal G}}{\partial r},
\end{equation}
where
\begin{equation}
{\cal G}(r,t)=2\pi r^3\nu\Sigma\frac{\partial\Omega}{\partial r}
\end{equation}
is the torque exerted on the disk interior to radius $r$ by the flow
outside of this radius.  These equations can be combined, using the
fact that in Newtonian gravity $\Omega\propto r^{-3/2}$, to obtain
\begin{equation}
\frac{\partial \Sigma}{\partial t}=\frac{3}{r}\frac{\partial}{\partial r}\left[r^{1/2}\frac{\partial}{\partial r}(\nu\Sigma r^{1/2}) \right].
\end{equation}
Noting that $\nu$ may be a function of the local variables in the
disk, this has the form of a non-linear diffusion equation governing
the behavior of $\Sigma(r,t)$ given some initial state.  Given a
solution for $\Sigma(r,t)$, the radial velocity is 
\begin{equation}
\label{radial_vel_eqn}
v_r=-\frac{3}{\Sigma r^{1/2}}\frac{\partial}{\partial r}(\nu \Sigma r^{1/2}).
\end{equation}

We need to know the behavior of $\nu$ in order to close these
equations and permit a full determination of the radial structure of
the accretion disk.  All of the complexities currently ignored (such
as the $z$-structure of the accretion disk, and the detailed
microscopic physics) enter into the problem via $\nu$.  However, in
the case of time-independent accretion disks (i.e. $\partial/\partial
t = 0$), it is possible to determine some of the most interesting
observables independent of our ignorance of $\nu$.  In particular, it
is readily shown (e.g. Chapter 5 of \cite{frank:92a}) that the viscous
dissipation per unit disk face area is
\begin{equation}\label{eq:disk_diss}
D(r)=\frac{3GM\dot{M}}{8\pi r^3}\left[  1-\left(\frac{r_{\rm in}}{r}\right)^{1/2}\right],
\end{equation}
where $\dot{M}=2\pi r\Sigma(-v_r)$ is the rate at which mass is
flowing inwards through the disk, and $r_{\rm in}$ is some inner
boundary to the disk where the torque ${\cal G}$ goes to zero.  One
peculiarity of viscous accretion flows becomes apparent upon an
examination of this last expression --- for $r\gg r_{\rm in}$, the
rate that energy is lost from the disk is three times the local rate
of change of binding energy.  
%That is to say that a ring of the disk
%located at $r\gg r_{\rm in}$ dissipates twice as much energy that is
%``viscously'' transported into it by the interior disk than is
%released locally.  
That is to say that for a ring of the disk located at $r\gg r_{\rm
in}$, two thirds of the dissipation is attributable to energy that is
`viscously'' transported from the interior disk, while the remaining
one third is attributable to the local rate of change of binding
energy.  Of course, to compensate for this effect, the innermost
regions of the disk radiate less than the local rate of change of
binding energy.  Integrating over the whole disk, the total luminosity
of the disk is $L_{\rm disk}=GM\dot{M}/2r_{\rm in}$.  An equal amount
of energy remains in the form of kinetic energy of the accreting
material as it flows through $r=r_{\rm in}$.  If there is a
(slowly-rotating) star at the center of the disk (as opposed to a
black hole), this remaining energy can be radiated in the disk-star
boundary layer.

If we suppose that the viscously dissipated energy is radiated as a
black body spectrum, energy conservation [$\sigma_{\rm SB}
T(r)^4=D(r)$; where $\sigma_{\rm SB}$ is the Stephan-Boltzman
constant] gives the temperature of the disk surface as a function of
radius,
\begin{equation}\label{eq:disk_temp}
T(r)=\left(\frac{3GM\dot{M}}{8\pi\sigma_{\rm SB} r^3}\left[ 1-\left(\frac{r_{\rm in}}{r}\right)^{1/2}\right]\right)^{1/4}.
\end{equation}
For a mass accretion rate that scales with mass (i.e. for a fixed
ratio between the source luminosity and the Eddington limit) and a scaled
radius $r/M$, the temperature of the disk scales weakly with the black
hole mass $T\propto M^{-1/4}$.  The accretion disks around
supermassive black holes generally possess a characteristic disk
temperature of $T\sim 10^5-10^6\K$, making them optical and
ultraviolet sources.  On the other hand, accretion disks around
stellar mass black holes are appreciably hotter with $T\sim 10^7\K$,
making them sources of thermal soft X-rays.

For a thin disk, there are no appreciable motions or accelerations in
the $z$-direction.  Hence the $z$-structure (i.e., the vertical
structure) of the disk is determined by a hydrostatic equilibrium
condition,
\begin{equation}
\frac{1}{\rho}\frac{\partial p}{\partial z}=\frac{\partial}{\partial z}\left[ \frac{GM}{(r^2+z^2)^{1/2}}\right],
\end{equation}
where $p$ is the pressure.  Noting that the sound speed is $c_{\rm
  s}\sim \sqrt{p/\rho}$, the hydrostatic equilibrium expression can be
used to see that the vertical scale height $h$ is,
\begin{equation}
h\sim \frac{c_{\rm s}r}{v_{\rm K}},
\end{equation}
where $v_{\rm K}\equiv (GM/r)^{1/2}$ is the local Keplerian velocity.
Thus, an accretion disk is geometrically thin ($h/r\ll 1$) if the
Keplerian velocity is highly supersonic.

In order to study the detailed physical structure of accretion disks,
or any aspect of their time-variability (including the stability of
accretion disks), we require a knowledge of the viscosity.  In an
extremely successful application of dimensional analysis, Shakura \&
Sunyaev \cite{shakura:73a} parameterized the viscosity as $\nu=\alpha
c_{\rm s}h$, where $c_{\rm s}$ is the sound speed in the disk, $h$ is
the vertical scale height and $\alpha$ is a dimensionless parameter
which, on general hydrodynamical grounds, is argued to be less than
unity.  Furthermore, they made the supposition that $\alpha$ is a
constant for a given accretion disk.  The resulting family of disk
models are referred to as $\alpha$-models (or Shakura-Sunyaev models).
Given the $\alpha$-prescription, we can solve for the detailed radial
disk structure (the results of this exercise are given in
\cite{shakura:73a} and nicely reviewed by \cite{frank:92a}).  In
addition, the assumptions underpinning the basic equations of the
radiatively-efficient thin accretion disk can now be checked
post-priori.  From eqn.~\ref{radial_vel_eqn} it can be seen that the
radial velocity is given by
\begin{equation}
v_r\sim \frac{\nu}{R}\sim \frac{\alpha c_{\rm s} h}{r}\ll c_{\rm s}.
\end{equation}
Thus, the radial inflow is very subsonic.  Using the radial force
equation, this also implies that the orbital velocity is very close to
Keplerian.  Furthermore, we can relate the various fundamental
time scales governing the disk.  Let the {\it dynamical time scale} be
defined as $t_{\rm dyn}=\Omega^{-1}$.  It can be shown that the
time scale on which the dissipated energy is radiated from the disk,
the {\it thermal time scale}, is given by $t_{\rm th}\sim t_{\rm
dyn}/\alpha>t_{\rm dyn}$.  Finally the time scale on which angular
momentum is redistributed, the {\it viscous time scale}, is given by
$t_{\rm visc}\sim r^2/\nu\sim (r/h)^2t_{\rm th}\gg t_{\rm th}>t_{\rm
dyn}$.  The ordering of these time scales, together with the fact that
the disk material orbits the black hole in almost Keplerian orbits,
justifies the assumptions listed at the beginning of this section.

While the $\alpha$-prescription is extremely useful, allowing us to
construct simple analytic models of accretion disks, it is important
to realize the severe limitations of this approach.  As will be
discussed in \S~\ref{sec:mhd_disks}, we now believe that MHD
turbulence is responsible for the angular momentum transport that
drives accretion.  While the angular momentum transport resulting from
MHD turbulence can be approximately parameterized in terms of an
``effective $\alpha$'', there are circumstances where it is incorrect
to treat the accretion flow as an unmagnetized gas with an
anomalously high kinematic viscosity (\cite{balbus:98a,balbus:02a}.
Even when used within its domain of validity, the actual value of
$\alpha$ applicable to real systems is very unclear with estimates
covering full range from essentially zero to unity (see discussion in
\cite{frank:92a}).

\subsubsection{Relativistic disks}
\label{sec:rel_disks}

Of course, the Shakura-Sunyaev disk models presented above are
non-relativistic and, hence, can only be viewed as a crude
approximation to real black hole accretion disks.  The formal
relativistic generalization of the steady-state Shakura-Sunyaev disk
model to a relativistic accretion disk in a Kerr metric was made in
1974 by Kip Thorne, Don Page and Igor Novikov
\cite{novikov:74a,page:74a,riffert:95a}, whereas time-dependent
relativistic disks were first treated by Douglas Eardley and Alan
Lightman \cite{eardley:75b}.  We shall refer to these relativistic,
geometrically-thin and radiatively-efficient disks as the ``standard
model'' for black hole accretion disks.

Several new features to the model arise once the fully relativistic
situation is considered.  Most importantly, there is a natural choice
of inner boundary.  In the standard model of a geometrically-thin
relativistic accretion disk, the torque is assumed to vanish at the
radius of marginal stability, $r=r_{\rm ms}$; we shall refer to this
standard assumption as the zero-torque boundary condition (ZTBC).  It
is envisaged that, within $r=r_{\rm ms}$, the material plunges into
the black hole conserving its energy and angular momentum.  When the
ZTBC is employed, the viscous dissipation per unit proper disk face
area per unit proper time is \cite{page:74a}
\begin{equation}
D_{\rm PT}(r;\as)=\frac{\dot{M}}{4\pi r}{\cal F},
\label{eq:pt}
\end{equation}
where we have defined the function ${\cal F}$
\begin{eqnarray}
{\cal F}=&\frac{3}{2M}&\frac{1}{x^2(x^3-3x+2\as)}\biggl[
x-x_0-\frac{3}{2}\as\ln\left(\frac{x}{x_0}\right)\\\nonumber
&-&\frac{3(x_1-\as)^2}{x_1(x_1-x_2)(x_1-x_3)}\ln\left(\frac{x-x_1}{x_0-x_1}\right)\\\nonumber
&-&\frac{3(x_2-\as)^2}{x_2(x_2-x_1)(x_2-x_3)}\ln\left(\frac{x-x_2}{x_0-x_2}\right)\\\nonumber
&-&\frac{3(x_3-\as)^2}{x_3(x_3-x_1)(x_3-x_2)}\ln\left(\frac{x-x_3}{x_0-x_3}\right)
\biggl],
\end{eqnarray}
with,
\begin{eqnarray}
x&=&\sqrt{r/M}\\
x_0&=&\sqrt{r_{\rm ms}/M}\\
x_1&=&2\cos(\frac{1}{3}\cos^{-1}\as-\pi/3)\\
x_2&=&2\cos(\frac{1}{3}\cos^{-1}\as+\pi/3)\\
x_3&=&-2\cos(\frac{1}{3}\cos^{-1}\as).
\end{eqnarray}
Here, we have defined the dimensionless spin parameter to be
$\as=a/M$.  This is the relativistic generalization of
eqn.~\ref{eq:disk_diss}.  As we shall discuss in
\S\ref{sec:innerdisk_mhd}, there are likely situations in which the
ZTBC does not apply and that the region in which angular momentum and
energy transport occurs penetrates significantly within $r=r_{\rm
ms}$.  Nevertheless, the ``standard model'' is still a useful
reference point against which more sophisticated disk models can be
judged.

The properties of the inner edge of the disk are important in
determining the efficiency of the accretion disk.  We define the
efficiency of the disk, $\eta$, by $L=\eta \dot{M}c^2$, where
$\dot{M}$ is the mass accretion rate onto the black hole and $L$ is
the total power output of the accretion disk (summing over all forms
including radiation and the kinetic energy of any outflows).  The
assumptions of the standard model (and, in particular, the ZTBC),
fixes the efficiency $\eta$ to be the change in specific binding
energy as a fluid element moves from infinity down to $r=r_{\rm ms}$.
For a non-rotating black hole, the radius of marginal stability is
relatively distant ($r_{\rm ms}=6M$) and the radiative efficiency is
only $\eta\approx 0.06$.  On the other hand, a standard disk around a
Kerr black hole with $\as\equiv a/M=1$ has an efficiency of
$\eta\approx 0.42$.  For a realistic rapidly spinning black hole, with
$\as\approx 0.998$, the efficiency is $\eta\approx 0.30$.

\subsubsection{The magneto-rotational instability and MHD turbulence}
\label{sec:mhd_disks}

Until the early 1990s, there was no clear candidate for the actual
angular momentum transport mechanism in black hole (or, indeed,
neutron star, white dwarf or protostellar) accretion disks --- normal
atomic/molecular viscosity turns out to be orders of magnitude too
small to drive the accretion that was presumed to cause the prodigious
radiative powers observed from such sources.  Detailed theoretical
studies of accretion disks were plagued by our lack of understanding
of the basic angular momentum transport mechanism.  However, in 1991,
Steve Balbus and John Hawley \cite{balbus:91a} showed that the
presence of a (arbitrarily) weak magnetic field in the accretion disk
material will render it susceptible to a powerful local MHD shear
instability.  Assuming that the disk is in the MHD limit, they showed
that the only criterion for linear instability is that the angular
velocity of the disk decreases outwards, $d\Omega^2/dr<0$; this
condition is easily satisfied for any accretion disk around a compact
object.  It is now widely accepted that this instability, known as the
magneto-rotational instability (MRI), lies at the heart of angular
momentum transport in the inner regions of accretion disks around
compact objects (for a detailed review of accretion disk MHD, see
\cite{balbus:98a}).

A full exploration of the non-linear evolution of the MRI requires
numerical MHD simulations.  The initial era of simulations
\cite{hawley:91a,hawley:95a,stone:96a} examined the dynamics in a
local patch of the disk, with an extent much smaller than the radius
of the disk.  These simulations confirmed the linear MRI analysis, and
allows the evolution to be followed into the non-linear regime.  It is
found that the MRI enters the non-linear regime (some aspects of which
have been studied analytically by Goodman \& Xu \cite{goodman:94a})
and MHD turbulence results after a small number of dynamical
time scales.  The process becomes a self-sustaining dynamo, and the
turbulence does not decay away with time.  Of course, Cowling's
antidynamo theorem dictates that non-axisymmetric simulations are
required in order to produce self-sustained dynamo action.  In recent
years, simulators have been able to leap beyond these local
simulations and tackle complete accretion disks (so-called ``global''
simulations, \cite{armitage:98b,hawley:00a,hawley:01a}).  As in the
local simulations, the global models also produce sustained MHD
turbulence.

The importance to accretion disks comes from the fact that the MRI has
the properties of an exchange instability in which high angular
momentum material is exchanged with low angular momentum material,
thereby facilitating angular momentum transport.  From the
simulations, one can extract effective values of the Shakura-Sunyaev
$\alpha$-parameter.  For local simulations, one obtains
$\alpha=10^{-3}-10^{-2}$ if there is no net vertical magnetic flux
threading the shearing box, and $\alpha=10^{-2}-10^{-1}$ when there is
a significant (but still dynamically weak) net vertical flux
\cite{hawley:95a}.  The global simulations (which, to date, have only
been performed for the case of zero net vertical flux) show
$\alpha=10^{-2}-10^{-1}$.  There is no inconsistency between these
global and local results; large-scale field structures generated
within the global models can effectively impose a net magnetic flux on
the local level, thereby enhancing the local angular momentum
transport.

\subsubsection{The role of magnetic fields in the innermost disk}
\label{sec:innerdisk_mhd} 

While stating the assumption of zero-torque at the radius of marginal
stability, Page \& Thorne \cite{page:74a} noted that significant
magnetic fields might allow torques to be transmitted some distance
within the radius of marginal stability.  This idea has been
rejuvenated by recent work on MRI driven turbulence.

\begin{figure}
\centerline{
\psfig{figure=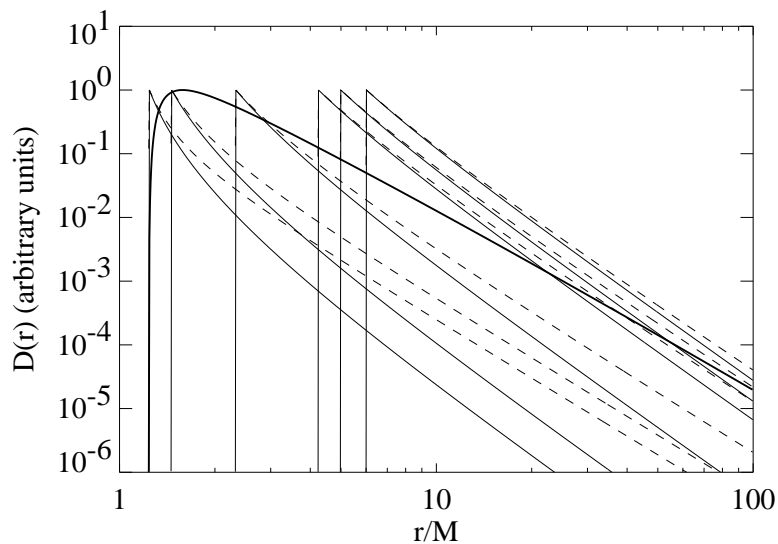,width=0.8\textwidth}
}
\caption{Model dissipation profiles for torqued time-independent accretion 
  disks (from \cite{agol:00b}).  The thick line shows the dissipation
  profile for a standard non-torqued disk (i.e., one in which the ZTBC
  applies) around a near-extremal Kerr black hole with spin parameter
  $a=0.998$.  The thin solid lines show the torque-induced dissipation
  component $D_{\rm tor}$ for spin parameters of (from left to right)
  $a=0.998,0.99,0.9,0.5,0.3$, and $0$.  Note how the torque-induced
  dissipation profile becomes more centrally concentrated as the spin
  parameter is increased.  The dashed lines show the torque-induced
  component of the emitted flux, including the effects of returning
  radiation.}
\label{fig:pt_agol}
\end{figure}

Significant insight can be gained from analytic considerations,
assuming that the inner regions of the accretion disk are
time-independent and axisymmetric.  In separate treatments by Julian
Krolik and Charles Gammie \cite{agol:00b,krolik:99b,gammie:99a}, it
was shown that magnetic fields might be effective at transporting
angular momentum from material in the plunging region into the main
body of the accretion disk.  Accompanying this angular momentum
transport, we would expect either local dissipation/radiation of the
liberated binding energy, or energy transport into the
radiatively-efficient portions of the disk.  Either way, the principal
consequence is to enhance the efficiency $\eta$ of the disk above the
value predicted by the Page \& Thorne models.  The simplest analytic
approach (followed by Agol \& Krolik \cite{agol:00b}) is to model this
effect as a finite torque applied at the radius of marginal stability,
i.e, an explicit violation of the ZTBC.  In addition to modeling the
magnetic connection between the accretion disk and the plunging
region, this torque can also model any magnetic connection between the
accretion disk and the rotating black hole itself \cite{li:02a}. This
torque performs work on the accretion disk, thereby adding to the power
ultimately dissipated in the disk and increasing the disk efficiency.

This salient aspects of this model are illustrated in
Fig.~\ref{fig:pt_agol} which displays the dissipation rate in the disk
per unit disk face area as a function of Boyer-Lindquist radius,
$D(r)$.  The thick line in this plot shows the dissipation law for a
standard Page \& Thorne accretion disk around a black hole with spin
parameter $a=0.998$ [$D_{\rm PT}(r;a=0.998M)$ given by
eqn.~\ref{eq:pt}].  Also shown are the additional dissipation profiles
resulting from non-zero torques applied at the radius of marginal
stability $D_{\rm tor}(r;a)$.  It can be shown that \cite{agol:00b}
\begin{equation}
D_{\rm tor}(r;\as)=\frac{3\dot{M}r_{\rm ms}^{3/2}C_{\rm ms}^{1/2}\Delta \eta}{8\pi r^{7/2}C(r)}
\label{eq:agol}
\end{equation}
where $C(r)=1-3M/r+2\as M^{3/2}/r^{3/2}$ and $\Delta\eta$ is the
additional disk efficiency induced by the torque.  Strictly, this is
true if the additional torque does not directly affect the dynamics of
the disk.  In this case, the total dissipation is just the sum
$D(r;\as)=D_{\rm PT}(r;\as)+D_{\rm tor}(r;\as)$.  The torque-induced
component to the dissipation is much more centrally concentrated than the
standard disk dissipation profile.  In fact, it is so centrally
concentrated that a significant fraction of the radiative flux emitted
from the disk surface will return to the disk surface due to strong
light bending effects.  This ``returning radiation'' will be absorbed
or scattered from the disk surface and, hence, adds to the local
energy budget relevant for determining the observed amount of
radiative flux from a given radius.  Including this effect, the total
emitted flux per unit area of the disk can be written as
\begin{equation}
D_{\rm tot}(r;\as)=D_{\rm PT}(r;\as)+D_{\rm tor}(r;\as)+D_{\rm ret}(r;\as),
\end{equation}
where $D_{\rm ret}(r;\as)$ is the term accounting for returning
radiation.  From the point of view of an observer riding on the disk
at a large radius, the returning radiation produces a ``mirage'' of
the central disk raised to some height above the disk plane.  Thus, at
large radii, irradiation due to returning radiation scales as $r^{-3}$
while the dissipated flux dominates at small radii.  Thus, as shown in
\cite{agol:00b}, the term $D_{\rm ret}$ is approximately given by
\begin{equation}
D_{\rm ret}(r;\as)\approx\frac{3M\dot{M}\Delta\eta R_\infty(\as)}{8\pi r^3},
\end{equation}
where $R_\infty(\as)$ is well described by the fitting formula given
by Agol \& Krolik \cite{agol:00b}.  The function $D_{\rm tot}(r;\as)$,
which includes the effects of returning radiation, is also shown in
Fig.~\ref{fig:pt_agol}.   

The analytic models of Gammie, Agol and Krolik
\cite{gammie:99a,agol:00b} suggest that these torqued accretion disks
can become super-efficient ($\eta>1$).  In this case, some fraction of
the radiative power must be extracted from the spin of the black hole.
This is a particular realization of the Penrose process, with the
accreting material in the innermost parts of the plunging region being
placed on negative energy orbits by magnetic connections with the
accretion flow at larger radii.

Simulations are required to move beyond the restrictive geometry and
physics of the analytic models.  To date, almost all of the relevant
MHD simulations have been inherently non-relativistic and have mocked
up the effects of General Relativity via the use of the
Paczy\'nski-Wiita \cite{pacynsky:80a} pseudo-Newtonian potential ,
\begin{equation}
\Phi=\frac{M}{r-2M},
\end{equation}
where, to recall, we have set $G=c=1$.  Newtonian computations using
this modified point-source potential serve as remarkably effective
toy-models for examining dynamics in the vicinity of a Schwarzschild
black hole.  Simulations of MHD accretion disks within this potential
have been presented in a number of
papers\cite{hawley:00a,hawley:01a,armitage:01a,reynolds:01c,hawley:02a}.
These works support the hypothesis that energy and angular momentum
can be transported out of the plunging region, although the
universality of these processes is still the subject of debate
\cite{armitage:01a,reynolds:01c}.   

We note in brief that, at the time of writing, fully relativistic MHD
codes capable of studying the evolution of the MRI and MHD turbulence
around rapidly rotating black holes are just starting to become
available \cite{devilliers:02b}.  Simulations performed with such codes
will be extremely important in assessing the true nature of the
innermost regions of black hole accretion disks.

\subsection{Accretion disk coronae}
\label{sec:coronae}

An accretion disk described exactly by the standard model produces a
relatively soft, quasi-thermal spectrum (dominated by optical/UV
radiation in AGN, and soft X-rays in GBHCs).  However, accreting black
hole systems (both stellar and supermassive) often exhibit power-law
components to their spectra which extend to hard X-ray energies with
exponential rollovers at high energy ($E\sim 300\keV$).  A promising
mechanism for producing such a spectrum is unsaturated inverse Compton
scattering \cite{thorne:75a,sunyaev:79a}.  Extensive descriptions of inverse
Comptonization and its applications to astrophysical sources can be
found in the literature \cite{rybicki:79a,pozd:83a}. Here we provide a
brief summary of these descriptions.

Compton scattering is the scattering of a photon by an electron. In
the rest frame of the electron, the differential scattering cross
section for a photon of energy $E\ll m_{\rm e}c^2$ is given by
\begin{equation}
\frac{{\rm d} \sigma_T}{{\rm d}\Omega} = \frac{3}{8 \pi} \sigma_T 
     \left ( \frac{1 + \cos^2 \theta}{2} \right ) ~~,
\end{equation}
where 
\begin{equation}
\sigma_T = \frac{8 \pi}{3} \left ( \frac{e^2}{m_e c^2} \right )^2 = 
  6.67 \times 10^{-25} ~{\rm cm^{2}}
\end{equation}
is the Thomson cross section and $\theta$ is the angle between the
photon's initial and final trajectories.  In the initial rest frame of
the electron, the photon imparts kinetic energy to the recoiling
electron, and hence the photon loses energy,
\begin{equation}
\Delta E = - \frac{E^2}{m_e c^2} ( 1 - \cos \theta )~~.
\end{equation}
In the laboratory frame, however, the
electron can impart energy to the photon, up to $\Delta E = (\gamma -
1) m_e c^2$, where $\gamma$ is the electron's relativistic Lorentz
factor.

It can be shown \cite{rybicki:79a,pozd:83a} that for a non-relativistic thermal
distribution of electrons with temperature $T_e$, the average photon
energy change for a single scattering is given by
\begin{equation}
\langle \Delta E \rangle = (4 k T_e - E) \frac{E}{m_e c^2} ~~.
\end{equation}
If $E \gg 4 kT_e$, the electron's kinetic energy is negligible and we
recover the angle average of the previous equation for photon energy
loss.  If, however, $E \ll 4 k T_e$, the photon gains energy, with the
fractional energy gain being approximately proportional to $\gamma^2$.
This proportionality can be understood as one factor of $\gamma$
coming from the boosting of the photon into the initial electron rest
frame, and another factor of $\gamma$ coming from a boosting of the
scattered photon back into the lab frame \cite{rybicki:79a,pozd:83a}.  This
gain of photon energy is referred to as inverse Compton
scattering.

So long as $E \ll 4 k T_e$, the photon continues to gain energy after
each scattering event with $\Delta E/E \approx 4 k T_e / m_e c^2$.  Thus,
after $N$ scattering events, the average final photon energy, $E_f$,
compared to the initial photon energy, $E_i$, is given by
\begin{equation}
E_f \approx E_i \exp \left ( N \frac {4 k T_e}{m_e c^2} \right ) ~~.
\end{equation}
For a medium with an optical depth of $\tau_{\rm es}$, the average
number of scatters is roughly max($\tau_{\rm es}$, $\tau_{\rm es}^2$).
This leads one to define the Compton $y$ parameter, 
\begin{equation}
y \equiv {\rm max}(\tau_{\rm es}, \tau_{\rm es}^2) (4 k T_e/m_e c^2), 
\end{equation}
such that $E_f \approx E_i \exp(y)$ \cite{rybicki:79a}. For $y > 1$,
the average photon energy increases by an `amplification factor' $A(y)
\approx \exp (y)$.  For $y \gg 1$, the average photon energy reaches
the thermal energy of the electrons.  This latter case is referred to
as saturated Compton scattering, while the former case is referred to
as `unsaturated inverse Comptonization'.  Unsaturated inverse
Comptonization is of most interest in the study of black hole systems.

As discussed above, the `standard' optically-thick, geometrically-thin
disk model produces a maximum temperature that scales with the compact
object mass as $M^{-1/4}$, so one naturally expects lower disk
temperatures in AGN, as opposed to stellar mass black hole systems.
However, one can instead define a `virial temperature' which refers to
the average accretion energy \emph{per particle}.  Specifically, the
gravitational energy released per particle of mass $m$ scales as $GM m
/R$, and therefore scales as $m c^2/[R/(GM/c^2)]$.  Thus, in terms of
geometrical radii, the virial temperature is independent of compact
object mass and furthermore can reach a substantial fraction of the
rest mass energy of the accreted particles.  Electron virial
temperatures of tens to hundreds of keV can be readily achieved in the
innermost regions of compact object systems.  (The proton virial
temperature is a factor of $\approx 2000$ greater; therefore, it is
conceivable that electrons can achieve even higher temperatures
depending upon the temperature of the protons and the coupling between
the electrons and protons.)

It is such a distribution of low-to-moderate optical depth, high
temperature electrons that is suspected of creating the observed power
law spectra in black hole systems.  Although multiple photon
scatterings in such a corona become exponentially unlikely, multiple
scatterings lead to exponential photon energy gain.  The two effects
balance, to some degree, and produce a power law spectrum.  Simple
estimates \cite{rybicki:79a} show that for a given Compton $y$
parameter, the photon index of the resulting power law is
approximately given by
\begin{equation}
\Gamma =  -\frac{1}{2} + \sqrt{ \frac{9}{4} + \frac{4}{y} } ~~,
\end{equation}
i.e., $\Gamma \approx 2$ for $y = 1$\footnote{These particular
expressions come from a solution of the Kompaneets equation, which
under certain approximations describes the diffusion in phase space of
photons undergoing Comptonization.  Excellent discussions of the
Kompaneets equation can be found in the review by Pozdnyakov et al.
\cite{pozd:83a}, the book by Rybicki and Lightman \cite{rybicki:79a},
and Chapter 24 of the book by Peebles \cite{peebles:93a}.}.  Of
course, the exact function $\Gamma(y)$ is dependent upon geometry and
other assumptions. A power law form only holds for photon energies
somewhat less than the electron thermal energy.  As photons approach
the electron thermal energy, they no longer gain energy from
scattering, and a sharp rollover is expected in the spectrum.  Thus,
in inverse Comptonization models, the observed high energy spectral
cutoff yields information about the temperature of the underlying
electron distribution.

\begin{figure}
\centerline{
\psfig{figure=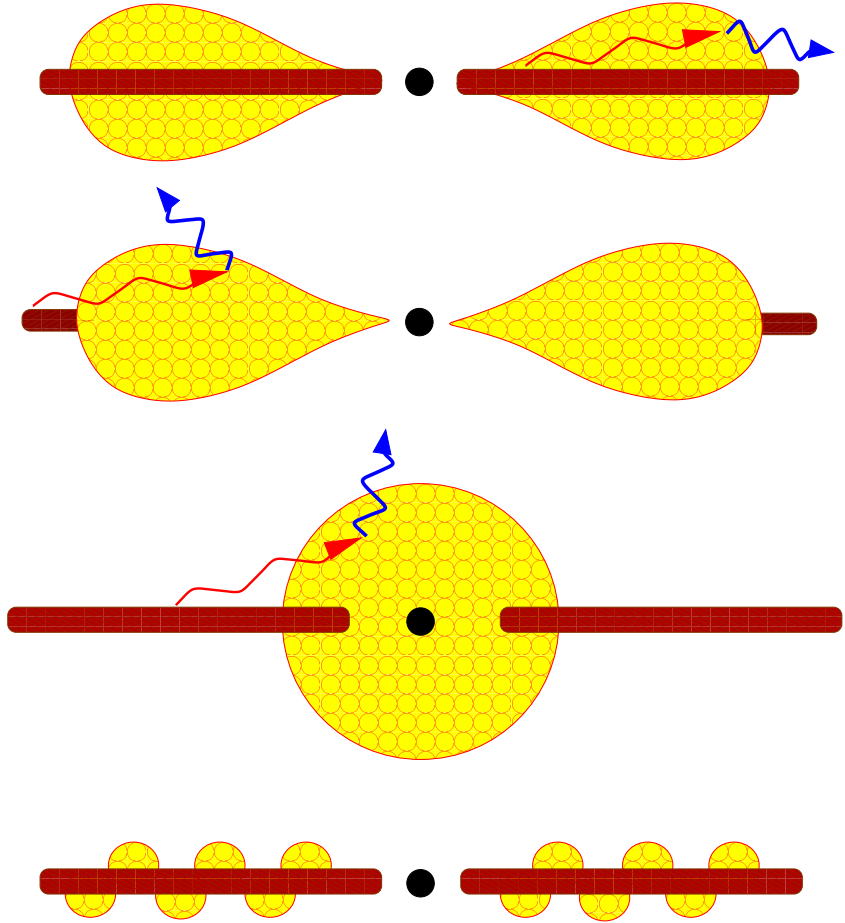,width=0.8\textwidth}
}
\caption{Suggested geometries for an accretion disk and Comptonizing
corona for predominantly spectrally hard states.  The top figure is
referred to as a ``slab'' or ``sandwich'' geometry; however, it tends
to predict spectra softer than observed.  The remaining three show
``photon starved geometries'' wherein the corona is less effectively
cooled by soft photons from the disk.  The middle two geometries are
often referred to as ``sphere+disk geometries'', while the bottom
geometry is often referred to as a ``patchy corona'' or ``pill box''
model \protect\cite{stern:95a}.}
\label{fig:coronal_geom}
\end{figure}

There are two principle uncertainties in applying coronal models to
observed black hole systems, and both are substantial.  First, the
mechanism for heating the electrons to near virial temperatures is
currently unknown.  Current hypotheses invoke magnetic processes,
perhaps akin to solar flares on the Sun or heating of the solar
corona.  Contrary to solar models, however, black hole coronae may be
energetically dominant.  Second, the geometry of the corona is also
completely unknown.  Thus, models have taken to hypothesizing specific
geometries (see Fig.~\ref{fig:coronal_geom}), and parameterizing the
energy input into the corona.

An early suggestion \cite{shapiro:76a} for the geometry of coronal
systems was that the accretion disk consisted of two zones--- a cool,
outer disk, and a hot inner, disk.  It was envisioned that energy
release was actively occurring within the disk itself, but that a
``two-temperature'' disk was being formed within the inner regions.
Ions would reach temperatures of $\approx 10^{11-12}$\,K, while
electrons would achieve temperatures of only $\approx 10^9$\,K.  Hard
X-ray emission would be due to Comptonization of soft photons from the
outer disk by the hot electrons in the inner disk.  It was later shown
\cite{pringle:76a,piran:78a} that this specific disk model was
thermally unstable; however, the basic geometry, which we will refer
to as the `sphere+disk' geometry, has been proposed in a variety of
forms since that time.  The attraction of such a geometry is that the
hot, inner corona sees only a fraction of the soft flux from the
outer, cool disk, and is therefore not strongly Compton cooled.  This
allows the corona to remain very hot and to produce hard spectra
\cite{dove:97a}.

The above disk models postulate a radial separation between the
geometrically thin, optically thick disk and the geometrically thick,
small optical depth corona.  Other sets of coronal models exist,
however, wherein both `disk' and `corona' exist at the same radii, but
the accretion energy is postulated to be dissipated predominantly
within the corona
\cite{ostriker:76a,liang:77a,bisno:77a,liang:79a}. It was thought
early on that a promising source for the necessary coronal energy
would be dissipation of magnetic flux tubes as they buoyantly rose
above the disk into the corona \cite{galeev:79a}.  The recent studies
of MHD instabilities in accretion flows have revived interest in such
mechanisms \cite{balbus:91a,balbus:98a,miller:00a}.

The simplest such model is one where the corona sandwiches the
accretion disk \cite{haardt:91a,haardt:93a,field:93a,haardt:96a}.  For
a uniform and plane parallel corona and disk, the basic energetics of
this geometry can be obtained from conservation principles.  Let us
decompose the radiative flux into that from the corona $F_c$ and that
from the optically-thick part of the disk $F_d$.  The coronal flux is
comprised of two parts, an outward component, $F_c^+$, and an inward
component, $F_c^-$, which will be intercepted and reprocessed by the
accretion disk.  The total flux from the disk, $F_d^t$, is comprised
of any intrinsic dissipation within the disk, $F_d^i$, plus the flux
intercepted from the corona.  Thus $F_d^t = F_d^i + F_c^-$.  The total
flux from the corona, $F_c^+ + F_c^-$, is comprised of any intrinsic
dissipation within the corona, $F_c^i$, plus the total outward flux of
the disk.  By definition, this is set equal to the outward flux from
the disk multiplied by the Compton amplification factor, $A_c$.
Specifically,
\begin{equation}
  F_c^+ + F_c^- = A_c F_d^t ~~.
\end{equation}
After multiple scatterings within the corona, the radiation field
becomes nearly isotropic, therefore $F_c^+ \approx F_c^- + \Theta
F_d^t$, where $0 \le \Theta \le 1$ is a measure of the anisotropy of
the coronal flux.  As the optical depth of the corona, $\tau_{\rm es}
\rightarrow 0$, $\Theta \rightarrow 0$.  If one now assumes that a
fraction, $f$, of the total (local) accretion energy is dissipated in
the corona, while $1-f$ is dissipated within the disk, one obtains
$F_d^i/F_c^i = (1-f)/f$, and
\begin{equation}
  A_c = \frac{1 + \Theta f / 2}{1 - f/2} ~~.
\end{equation}

The above highlights two important considerations of this slab
geometry.  First, even under the most extreme conditions of $\Theta
\approx 1$ and $f \approx 1$ (i.e., all accretion energy is dissipated
within the corona), the Compton amplification factor does not exceed
3.  This is equivalent to saying that the Compton $y$ parameter does
not exceed $\approx 1$. Second, 100\% of the soft, seed photons from
the disk pass through the corona.  Thus, the corona is relatively
easily Compton cooled and it is difficult for it to obtain very high
temperatures \cite{stern:95a,dove:97a}.  Even if the optical depth of such a
corona is made small, pair production within the corona will ensue
until the rate of heating is balanced by Compton cooling.  With a pure
slab geometry, extremely hard spectra (i.e., with very high energy
cutoffs) are impossible to achieve \cite{stern:95a,dove:97b}.

There have been several suggestions put forward that allow the above
basic geometry to achieve higher coronal temperatures, and thus harder
spectra.  If the corona is `patchy', such that small coronal regions
sit atop the disk (sometimes referred to as a `pill box' geometry),
then a large fraction of the coronal flux can be intercepted by the
disk, with a relatively small fraction of the (reprocessed) disk flux
being intercepted by the corona.  Thus, the coronal patches are not
very effectively Compton cooled \cite{stern:95a}.  Another method for
varying the amount of soft flux reprocessed by the corona is to invoke
relativistic motion within the corona.  Downward motion beams the
radiation towards the disk and increases the local disk reprocessing,
while motion away from the disk leads toward beaming away from the
disk, thereby decreasing reprocessing
\cite{reynolds:97e,beloborodov:99a}.  Finally, as we will further
discuss below, if the disk surface is highly ionized, then rather than
being reprocessed by the disk into soft radiation, the hard radiation
will merely reflect back through the corona, without substantially
cooling it \cite{ross:99a,nayakshin:01a,done:01c,done:01d}.

\subsection{Radiatively-inefficient accretion disks}
\label{sec:adafs}

Here we briefly discuss an important class of accretion disks models
that have received much attention in recent years --- the
radiatively-inefficient disks.  Such disks, in which the
radiative-efficiency $\eta$ becomes small, can occur when the
accretion rate is either very low or very high.  The physics is rather
different in these two cases.  For low accretion rates, the accreting
plasma becomes so tenuous that the electrons and ions may lose thermal
contact with each other (assuming that they are principally coupled
via Coulomb collisions).  The ions are very poor radiators; thus if
most of the accretion energy is channeled to the ions rather than the
electrons, it will not be radiated and instead remain as thermal
energy advected along within the accretion flow
\cite{ichimaru:77a,rees:82a}. In the so-called Advection Dominated
Accretion Flows (ADAFs), this energy is advected right through the
event horizon \cite{narayan:94a,narayan:95a,abramowicz:95a}.  Such
models were in fact first proposed to explain the hard state spectrum
of Cyg~X-1 \cite{ichimaru:77a}, but were later applied to AGN
\cite{rees:82a}, and they have been used to model the Galactic Center
\cite{narayan:95aa,narayan:98a,quataert:99a}.

It was later realized that the situation described by these low
luminosity ADAF models was dynamically unlikely --- the viscous
transport of energy within this flow could readily unbind material
further out, possibly leading to a powerful wind \cite{blandford:99a}
or strong convection \cite{quataert:00a}.  These suggestions remain
controversial \cite{balbus:02a} and are the subject of active research
\cite{stone:99a,igumenshchev:00a,abramowicz:02a}, but in any case,
such a flow is likely to be extremely hot (electron temperatures of
$\sim 10^9\K$) and optically-thin.  

Although these models possess the sphere+disk geometry (the inner
region is advection dominated while the outer region represents a
standard disk), ADAF models postulate that a large fraction of the
seed photons for Comptonization come from synchrotron radiation
internal to the radiatively inefficient flow \cite{narayan:96e}.  In
addition, their low-levels of emission are concentrated towards small
radii \cite{narayan:94a}, while their outer radii have been postulated
to extend anywhere from 10-$10^4\,GM/c^2$ \cite{esin:97c}.  In the
context of the disk atmosphere modeling described below, most (but
not all) variants of the ADAF model predict weak spectral lines from
the innermost regions of the accretion flow.

For very high accretion rates, comparable to that needed to produce
the Eddington luminosity, the accretion inflow time scale can become
less than the time it takes for radiation to diffuse out of the disk,
and hence the photons can become trapped in the accretion flow
\cite{begelman:79a}.  The inability of these disks to radiate the
gravitational potential energy, together with the viscous transport of
energy and strong radiation pressures present, will almost certainly
lead to strong outflows.  The appearance of such disks is highly
uncertain --- it is unclear whether an X-ray emitting corona forms,
and whether the atmosphere of such a disk is in a state capable of
producing X-ray reflection spectral signatures.  At the very least, it
is highly unlikely that strong line features will emanate from the
innermost, highly relativistic regions of the accretion flow.

\subsection{Disk atmospheres and X-ray reprocessing features}

As opposed to the ADAFs described above, standard,
radiatively-efficient accretion disks are optically-thick (i.e.,
opaque) across all frequencies of interest --- thus all of the
radiation directly attributed to the accretion disk must be
reprocessed through its outer layers.  To be slightly more precise, we
shall refer to the outer few Thomson depths\footnote{The Thomson depth
is the distance over which the electron scattering optical depth is
unity.} of the accretion disk surface as the ``atmosphere''.  It is
obviously crucial to study the disk atmosphere if we are to use
observations of accretion disks to study the environment of black
holes.  Due to the scope of this review, we focus on the reprocessing
and ``reflection'' of X-rays by the disk atmosphere.

\subsubsection{Simple X-ray reflection models}

Initially, we explore a model for the disk atmosphere that is clearly
oversimplified but helps elucidate the physics of X-ray reflection.
We suppose that the surface of the accretion disk can be modeled as a
semi-infinite slab of uniform density gas, irradiated from above by a
continuum X-ray spectrum produced in the disk corona via thermal
Comptonization.  Furthermore, we assume that the hydrogen and helium
are fully ionized, but all other elements (collectively referred to as
``metals'' by astronomers) are neutral.  This is a crude approximation
to the situation found in a ``cold'' AGN accretion disk.  Now we
consider the possible fates of an incident X-ray photon.  Firstly, the
photon can be Compton scattered by either the free-electrons
associated with the ionized hydrogen and helium, or the outer
electrons of the other elements.  Secondly, the photon can be
photoelectrically absorbed by one of the neutral atoms.  For this to
happen, the photon must possess an energy above the threshold energy
for the particular photoelectric transition.  The transitions with the
largest cross-sections are those associated with the photo-ejection of
a K-shell (i.e. $n=1$ shell) electron.  Following K-shell
photoionization, the resulting ion commonly de-excites in one of two
ways, both of which start with an L-shell ($n=2$) electron dropping
into the K-shell.  In the first case, the excess energy is radiated as
a $K\alpha$ line photon\footnote{In fact, the K$\alpha$ emission line
  is a doublet (comprising of the K$\alpha_1$ and K$\alpha_2$ lines).
  Since the splitting between the K$\alpha_1$ and K$\alpha_2$ lines is
  only a few eV, it is not relevant for the current discussion of
  relativistically broadened emission lines.  Thus, we will ignore the
  fine structure of the K$\alpha$ line for the remainder of this
  review.} (i.e., fluorescence).  In the second case, the extra energy
is carried away via the ejection of a second L-shell electron (i.e.,
autoionization or the Auger effect).  The {\it fluorescent yield} of a
species gives the probability that the excited ion will de-excite via
fluorescence rather than autoionization.

\begin{figure}
\centerline{
\hspace*{2cm}
\psfig{figure=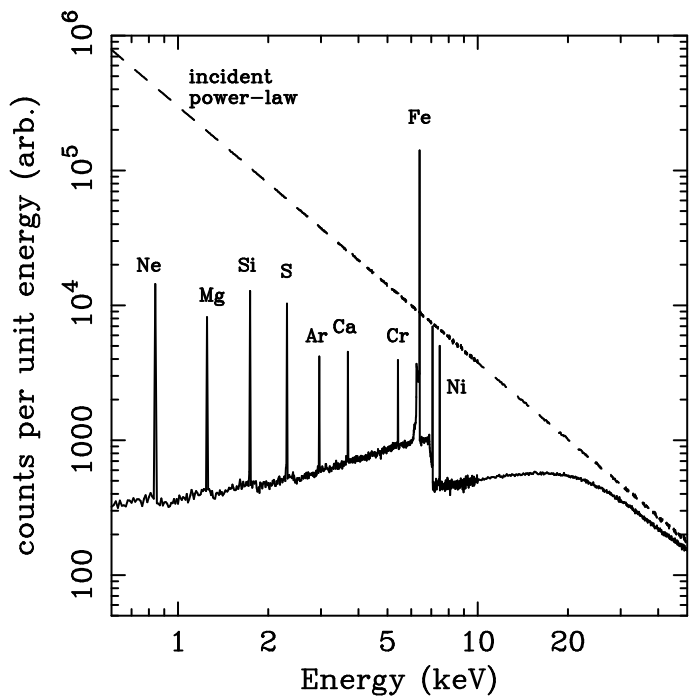,width=1.0\textwidth,angle=270}
}
\caption{Results of a simple Monte Carlo simulation demonstrating the
``reflection'' of an incident power-law X-ray spectrum (shown as a
dashed line) by a cold and semi-infinite slab of gas with cosmic
abundances.  In the accretion disk setting, one would observe the sum
of the direct power-law continuum and the reflection spectrum --- the
principal observables are then the cold iron K$\alpha$ fluorescent
line at 6.40\,keV and a ``Compton reflection hump'' peaking at $\sim
30$\,keV.  Figure from reference \cite{reynolds:96b}.}
\label{fig:simple_refl}
\end{figure}

Figure~\ref{fig:simple_refl} shows the results of a Monte-Carlo
simulation modeling these processes when a power-law X-ray continuum
with photon index $\Gamma=2$ is incident on a gaseous slab
\cite{george:91a,reynolds:96b}, assuming the slab has cosmic
abundances \cite{anders:82a}.  At soft X-ray energies, the albedo of
the slab is very small due to the photoabsorption by the metals in the
slab.  However, at hard X-ray energies, this photoabsorption becomes
unimportant (i.e., the photoelectric cross section falls to small
values) and most of the X-rays incident on the slab are Compton
scattered back out of the slab.  Associated with the photoionization
of metals in the slab, there is a spectrum of fluorescent emission
lines.  Due to the combination of high fluorescent yield and large
cosmic abundance, the most prominent such fluorescence is the
K$\alpha$ line of iron at 6.40\,keV.  It is interesting to note in
Fig.~\ref{fig:simple_refl} the weak ``shoulder'' on the low-energy
side of the iron-K$\alpha$ line.  This feature corresponds to line
photons that have Compton scattered, and hence lost energy due to
electron recoil, before escaping the disk.  It is often referred to as
the ``Compton shoulder''.

It is customary to measure the relative strengths of astrophysical
emission lines via the use of {\it equivalent widths}.  The equivalent
width of an emission line is the energy (or wavelength) range over
which the continuum radiation contains a flux equal to that contained
in the emission line.  For the case presented in
Fig.~\ref{fig:simple_refl} the equivalent width of the iron line when
viewed in the combined direct$+$reflected spectrum is approximately
180\,eV.  The other lines are much weaker, with equivalent widths at
least an order of magnitude less \cite{matt:97a}.

\subsubsection{Physical X-ray reflection models}
\label{sec:reflection}

The physical situation with a real disk atmosphere is much more
complex.  We can consider the atmosphere as a layer of gas that is
heated from below by the main body of the accretion disk (which, in
turn, is heated by the ``viscous'' dissipation of gravitational
potential energy) and irradiated from above by high-energy radiation
from the corona.  To determine the structure of the disk atmosphere,
we must model the radiative transfer, and solve for the thermal and
ionization balance of each fluid element.  Furthermore, the atmosphere
almost certainly responds dynamically to the heating/irradiation,
thereby requiring us to solve the equations of hydrodynamics (or, more
generally MHD) at the same time.  Finally, we must acknowledge the
fact that radiation pressure may well be important in many disks of
interest.

\begin{figure}
\centerline{
\psfig{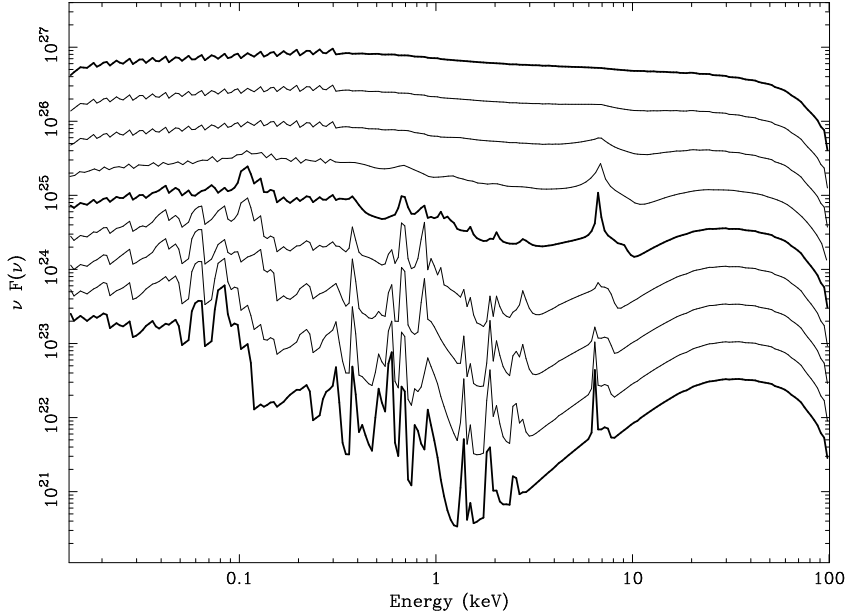}
}
\caption{Theoretical ionized reflection spectra for different ionization
  parameters, as computed using the constant density models of David
  Ballantyne and collaborators \cite{ballantyne:01a}.  From bottom to
  top, the curves show the reflected X-ray spectrum for ionization
  parameters of $\log\xi=1.0, 1.5, 2.0, 2.5, 3.0, 3.5, 4.0, 4.5$ and
  $5.0$.  The detailed behavior is discussed in the main text.}
\label{fig:ionized_disk}
\end{figure}

This is clearly a very complex problem, and progress to date has been
made only by considering rather special cases.  The first step towards
realistic disk reflection models maintained the assumption of a
stationary, time-independent, uniform density structure, and assumed
that the atmosphere was in thermal and ionization equilibrium.  By
solving the radiative transfer problem in this fixed background, the
thermal and ionization structure as a function of depth in the
atmosphere can be obtained
\cite{ross:93a,matt:93a,zycki:94a,magdziarz:95a}.  Hence, the
resulting X-ray reflection spectrum can be determined.  A useful
quantity in the discussion of photoionized atmospheres is the
ionization parameter,
\begin{equation}
\xi(r)=\frac{4\pi F_{\rm x}(r)}{n(r)},
\end{equation} 
where $F_{\rm x}(r)$ is the X-ray flux received per unit area of the
disk at a radius $r$, and $n(r)$ is the comoving electron number
density: it measures the ratio of the photoionization rate (which is
proportional to $n$) to the recombination rate (proportional to
$n^2$).  The constant density ionization models suggest that the value
of $\xi$ delineates four regimes of behavior (see
Fig.~\ref{fig:ionized_disk}).
\begin{enumerate}
\item $\xi <100$ ergs cm s$^{-1}$: This is the ``neutral reflection''
regime since the X-ray reflection resembles that from cold gas
containing neutral metals (as in Fig~\ref{fig:simple_refl}).  There is
a cold iron line at 6.4 keV, and the Compton backscattered continuum
only weakly contributes to the observed spectrum at this energy.
There is a weak iron K=shell edge at $7.1\keV$.
\item 100 ergs cm s$^{-1}<\xi <500$ ergs cm s$^{-1}$: In this
``intermediate ionization'' regime, the iron is in the form of
Fe\,{\sc xvii}--Fe\,{\sc xxiii} and there is a vacancy is the L-shell
($n=2$) of the ion.  These ions can resonantly absorb the
corresponding K$\alpha$ line photons.  Following each absorption,
there can be another fluorescent emission (with the consequence that
the photon has effectively been scattered) or de-excitation via the
Auger effect. A given iron K$\alpha$ photon can be trapped in the slab
via the resonant scattering until it is terminated by a Auger event.
Thus, only a few line photons can escape the disk leading to a very
weak iron line.  The reduced opacity below the iron edge due to
ionization of the lower-$Z$ elements leads to a moderate iron
absorption edge.
\item 500 ergs cm s$^{-1}<\xi <5000$ ergs cm s$^{-1}$: In this ``high
ionization'' regime, the ions are too highly ionized to permit the
Auger effect (which requires at least 2 L-shell electrons prior to
photoionization).  While the line photons are still subject to
resonant scattering, the lack of a destruction mechanism ensures that
they can escape the disk.  The result is K$\alpha$ iron emission of
Fe\,{\sc xxv} and Fe\,{\sc xxvi} at 6.67\,keV and 6.97\,keV
respectively. Due to the high levels of ionization and subsequent lack
of photoelectric opacity characterizing the low-$Z$ elements, the
Compton backscattered continuum is a significant contributor to the
observed emission at $6\keV$. Thus, there is a large iron absorption
edge in the observed spectrum.
\item $\xi >5000$ ergs cm s$^{-1}$: In this ``fully ionized'' regime,
the disk is too highly ionized to produce any atomic signatures.
There is no iron emission line or edge.
\end{enumerate}
More recent variants of these fixed density models have improved the
treatment of Compton scattering in the upper atmosphere
\cite{ballantyne:01a}, and have assumed a Gaussian density profile
$\rho(z) = \rho(0) \exp[-(z/z_0)^2]$ \cite{ross:99a}.  The qualitative
behavior is, however, unchanged.

Yet more realistic models require us to relax the assumption of a
fixed density structure.  The next simplest, yet physical, assumption is
to suppose that the disk atmosphere is in hydrostatic (or, more
generally, hydromagnetic) equilibrium.  The density, temperature and
pressure must then adjust so as to achieve hydrostatic equilibrium.
When such hydrostatic, irradiated disk-atmosphere models are
constructed, a qualitatively new phenomenon occurs --- the thermal
ionization instability (TII) \cite{nayakshin:00a}.  

The origin of the TII is as follows.  The equations defining local
thermal and ionization equilibrium for a specified density admit a
unique solution.  On the other hand, if the {\it pressure} is
specified, there are certain values of pressure for which there exist
more than one equilibrium thermal and ionization state (with the
intermediate temperature equilibria often proving to be thermally
unstable).  This is the thermal ionization instability.  The
primary consequence of the TII is that, over certain ranges of
pressure, a multiphase medium can form with comparatively cool regions
co-existing in pressure equilibrium with very hot regions.  

Now consider moving to greater and greater depths in a hydrostatic
disk atmosphere.  As the pressure increases monotonically, one will
encounter layers of the atmosphere in which the material is subject to
the TII and, hence, multiple temperature and ionization structures are
allowed. The exact solution actually realized by a particular fluid
element depends upon the history of that element, but there are
general arguments suggesting that it will usually exhibit the highest
temperature solution\cite{nayakshin:00a,nayakshin:01a}.  As one
transits from regions (i.e., pressures) where the TII operates to
regions where it does not, there will be abrupt changes in the
temperature, density and ionization state of the material.  In
practice, a highly-ionized and well-defined hot ``skin'' will develop,
overlying the colder parts of the disk that are more relevant for
producing interesting X-ray reflection features.  If this ionized skin
has a low optical depth, it will be largely irrelevant to the observed
reflection spectrum.  However, if the electron scattering optical
depth of the skin approaches unity, it will cause a smearing of any
sharp features in the reflection spectrum (due to Compton recoil) as
well as an apparent dilution of the observed reflection features
(since it will Compton scatter X-ray continuum photons back towards
the observer without any appreciable spectral change).  For strongly
irradiated disks, the hot skin will be optically-thick and act as a
``Compton mirror'', simply reflecting back the irradiating spectrum
(with slight modifications caused by electron recoil which only become
important for very hard X-rays; see for example \cite{poutanen:96a}).
If the irradiating spectrum is sufficiently hard ($\Gamma<2$), the
Compton mirror can form before the disk is capable of producing
observable H- or He- like K$\alpha$ iron line emission.

When faced with real data, how is one to make progress given the
complexities involved in modeling the reflection spectrum?  A major
simplification, pointed out by David Ballantyne and collaborators
\cite{ballantyne:01a}, is that self-consistent reflection spectra can
be approximately mapped to more manageable constant-density models,
albeit with different inferred reflection fractions (due to dilution
by an ionized skin) and ionization parameters.  Thus, for all
practical purposes, X-ray observers to date have used the constant
density models when attempting to fit real data.

\section{Emission lines from accretion disks}
\label{sec:lines_from_disks}

Having discussed accretion disk theory, the properties of accretion
disk atmospheres, and X-ray reflection, we move to the main thrust of
our discussion --- the use of emission lines from the disk surface as
a probe of the inner regions of black hole accretion disks.  In this
section, we treat the basic physics underlying the energy
(frequency) profiles of disk emission lines.  We closely follow the
discussion presented in \cite{wilms:98c}.

\subsection{Formalisms for computing line profiles}

Since the spectrum is emitted from an accretion disk close to the
black hole, an observer at infinity will see the reflected X-ray
spectrum of Fig.~\ref{fig:simple_refl} distorted by relativistic
effects, namely the Doppler effect, relativistic aberration, and
gravitational light-bending/red-shift.  

The specific flux $F_{\nu_0}$ at frequency $\nu_0$ as seen by an
observer at infinity is defined as the (weighted) sum of the observed
specific intensities $I_{\nu_0}$ from all parts of the accretion-disk,
\begin{equation}\label{eq:spec}
F_{\nu_0} = \int_{\Omega} I_{\nu_0} \cos\theta\,d\Omega,
\end{equation}
where $\Omega$ is the solid angle subtended by the accretion disk as
seen from the observer and $\theta$ is the angle between the direction
to the disk and the direction of the observed photon. Since the black
hole is assumed to be very far away from the observer, we can safely
set $\cos\theta=1$. Thus, we have to compute the specific intensity
$I_{\nu_0}$ at infinity from the spectrum emitted on the surface of
the accretion disk, $I_{\nu_e}$. In an axisymmetric accretion disk,
$I_{\nu_e}$ is a function of the radial distance of the point of
emission from the black hole, $r_{\rm e}$, and of the inclination
angle, $i_{\rm e}$, of the emitted photon, measured with respect to
the normal of the accretion disk.

The emitted and observed frequencies differ due to Doppler boosting
and gravitational red-shift.  We define the frequency shift factor
$g$, relating the observed frequency $\nu_0$ and the emitted frequency
$\nu_e$,
\begin{equation}\label{eq:g}
g = \frac{\nu_0}{\nu_e} = \frac{1}{1+z},
\end{equation}
where $z$ is the red-shift of the photon. According to Liouville's
theorem, the phase-space density of photons, proportional to
$I(\nu)/\nu^3$, is constant along the photon path. It is therefore
possible to express eqn.~(\ref{eq:spec}) in terms of the emitted
specific flux on the accretion disk:
\begin{equation}\label{eq:fnuo}
F_{\nu_0}=\int_{\Omega} \frac{I_{\nu_0}}{\nu_0^3} \nu_0^3\,d\Omega
        =\int_{\Omega} \frac{I_{\nu_e}}{\nu_e^3} \nu_0^3\,d\Omega
        =\int_{\Omega} g^3 I_{\nu_e}(r_{\rm e},i_{\rm e})\,d\Omega.
\end{equation}
In other words, the computation of the emerging spectrum breaks down
to the computation of $g$.  In the weak field limit, when $r\gg 3$,
and in the Schwarzschild metric, $g$ and therefore the line profile
emitted by the accretion disk, can be evaluated analytically.
Profiles computed this way have been presented, e.g., by Andy Fabian
and collaborators \cite{fabian:89a} for the Schwarzschild case, and by
Chen and Halpern \cite{chen:89b} in the weak field limit. In general,
however, the computation has to be done in the strong-field Kerr
metric.

The brute-force approach to the computation of $g$ in the Kerr metric
is the direct integration of the trajectory of the photon in the Kerr
metric \cite{bromley:97a,karas:92a}. This approach allows the
computation of exact line profiles even in the case of very
complicated geometries, such as geometrically-thick or warped
accretion disks \cite{hartnoll:00a} or disks with non-axisymmetric
(e.g., spiral wave) structures \cite{hartnoll:02a,karas:01a}.  The
second way to compute the observed flux was first used by Cunningham
in 1975 \cite{cunningham:75a} who noted that the observed flux from
can be expressed as
\begin{equation}
F_{\nu_0} = \int T(i_{\rm e},r_{\rm e},g) I_{\nu_e}(r_{\rm e},i_{\rm e})\,dg\,r_{\rm e}\,dr_{\rm e},
\end{equation}
where the integration is carried out over all possible values of $g$
and over the whole surface of the accretion disk. This form is well
suited for fast numerical evaluation.  All relativistic effects are
contained in the \emph{transfer-function} $T$ which can be evaluated
once for a given black hole spin parameter. We refer to
\cite{cunningham:75a,laor:91a,speith:95a}, and the references in these
works for the technical details.

\subsection{The emerging line profile}

\begin{figure}
\centerline{
\psfig{figure=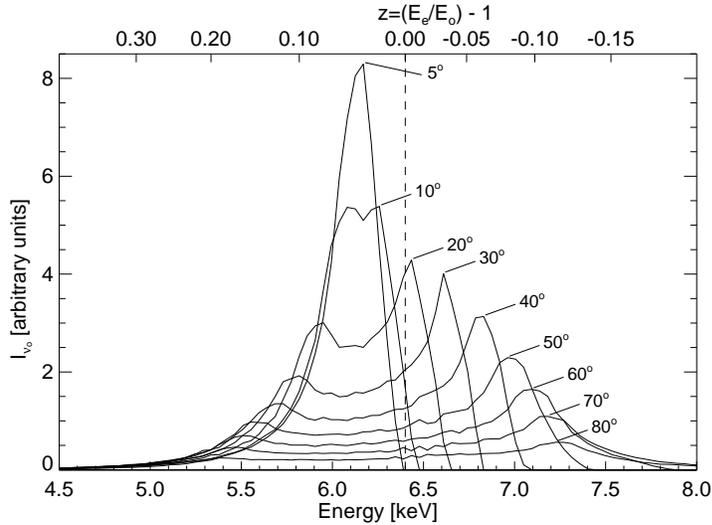,width=0.7\textwidth}
}
\caption{Iron line profiles from relativistic accretion disk models
  as a function of disk inclination (as measured by the angle between
  the normal to the disk plane and the observers line of sight).  The
  black hole is assumed to be rapidly-rotating ($a=0.998$), and the
  disk is assumed to possess a line emissivity index of $\beta=0.5$
  down to the radius of marginal stability $r\approx 1.23\,GM/c^2$.}
\label{fig:relline_incl}
\end{figure}

\begin{figure}
  \centerline{ \psfig{figure=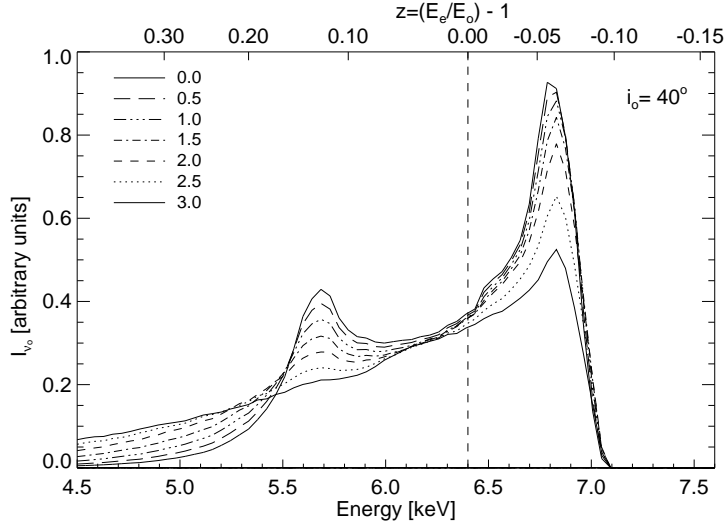,width=0.7\textwidth} }
\caption{Iron line profiles from relativistic accretion disk models
  as a function of emissivity index, $\beta$.  The black hole's spin
  parameter is assumed to be $a=0.5$ and, again, it is assumed that
  the disk extends to the radius of marginal stability.}
\label{fig:relline_beta}
\end{figure}

\begin{figure}
\centerline{
\psfig{figure=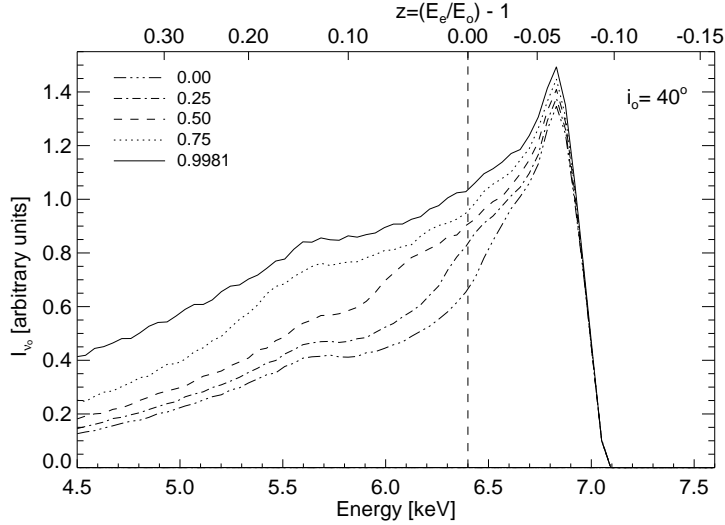,width=0.7\textwidth}
}
\caption{Iron line profiles from relativistic accretion disk models
  as a function of black hole spin parameter.  The disk is assumed to
  extend to the radius of marginal stability with an emissivity index
  of $\beta=3$.}
\label{fig:relline_spin}
\end{figure}

In Figs.~\ref{fig:relline_incl} to \ref{fig:relline_spin} we
illustrate the relativistic effects on the emerging line profile.  All
line profiles have been computed with the code of Roland Speith
\cite{speith:95a}.  In our computations we assumed a geometrically
thin, but optically thick Keplerian accretion disk to be the source of
the line radiation.  The local emissivity of the line on the disk was
parameterized as
\begin{equation}\label{eq:emiss}
  I_{\nu_e}(r_{\rm e},i_{\rm e})\propto f(i_{\rm e})\,r_{\rm e}^{-\beta},
\end{equation}
where $f(i_{\rm e})$ is a function that can be chosen to model various
limb-darkening scenarios.  This parameterization is sufficient for
most practical work \cite{bao:94a}.  For optically thick material,
$f={\rm constant}$.  

In all line profiles shown here, the emitting part of the disk is
assumed to extend from the radius of marginal stability to an outer
radius of $r_{\rm out}=50GM/c^2$.  For $\beta>2$, the location of this
outer radius is unimportant to the line profile since the line
emission is dominated by the inner regions.  However, for $\beta<2$,
this choice of outer radius is of some importance since the bulk of
the line emission comes from these outer regions.  In the latter case,
the separation between the red and blue peaks of the iron profile
diminishes as the outer radius is increased (due to the diminishing
line of sight velocity difference between the approaching and receding
sides parts of the disk).

Common to all line profiles is a characteristic double-horned shape
(Fig.~\ref{fig:relline_incl}).  To see the origin of this, consider
the emission line profile that arises from a narrow annulus of the
disk $r\rightarrow r+dr$.  By assumption, the emitting matter in this
annulus has a single orbital speed but, clearly, the observed Doppler
shift from different parts of the annulus will differ due to the
changing angle between the observer's line-of-sight and the velocity
vector of the matter (which is undergoing near-circular motion).  Line
emitting material on the portion of the orbit moving away from the
observer will contribute lower-frequency emission than material on the
approaching portion of the orbit.  Considering this simple case, it
can be seen that the line flux in a given frequency range $d\nu$ will
be at a maximum when the Doppler shifts are at an extremum, i.e., for
the minimum and maximum Doppler shifts from the annulus.  These
``red'' and ``blue'' peaks\footnote{It is customary in astronomy to
call lower energies ``red'' and higher energies ``blue'', even when
talking about lines in energy bands other than the optical} in the
line profile of the ring-element correspond to extrema in the
``pathlength'' through material at a given line-of-sight velocity ---
these extrema are found on the receding and approaching parts of the
annulus.  Once these horns are formed in the line profile, special
relativistic aberration (i.e., the beaming of emission in the direction
of motion) strongly enhances the blue horn over the red.  This effect
is clearly visible in Fig.~\ref{fig:relline_incl}.

In addition to the line-of-sight Doppler boosting, the line emission
is also red-shifted due to the gravitational potential and the
transverse Doppler effect (i.e., special relativistic time
dilation). The influence of all of these effects on the line profile
depends on the observer's inclination angle $i_0$: For a disk seen
almost face on (i.e.  $i_0$ close to $0^\circ$), the
gravitational/time-dilation redshifts dominate.  Even with a face-on
disk, integrating the observed emission across all radii including the
gravitational/time-dilation effects can still produce a rather broad
emission line, unlike the case of a face-on Newtonian disk which would
display a very narrow line.  With progressively larger inclinations
$i_0$, line-of-sight Doppler effects become dominant and the line
becomes broader.

The broadest parts of the profile are due to material emitted very
close to the black hole, as is evident from
Fig.~\ref{fig:relline_beta}, where line profiles for different
emissivity coefficients $\beta$ are shown. For large $\beta$, most of
the line emission takes place close to the last stable orbit, so that
these profiles are the broadest. Note that for values of $\beta<2$ the
red wing of the profile gets weaker until it is almost undetectable.

If one assumes that the line emitting disk extends down to, and is
then truncated, at the radius of marginal stability, the strong
dependence of $r_{\rm ms}$ on the black hole spin translates into a
strong dependence of line profile on black hole spin.  In particular,
the red-wing of the iron line can extend to much lower energies in the
case of a rapidly-rotating black hole as compared with a non-rotating
black hole.  This is illustrated in Fig.~\ref{fig:relline_spin}.
However, as we discuss in more detail in \S\ref{sec:mcg6}, it is
unclear that the effects of the radius of marginal stability can be
modeled so trivially, leading to many subtleties to using these line
profiles to probe black hole spin.

\section{The X-ray astronomer's arsenal}

\subsection{Past and current}

Of course, the Earth's atmosphere is completely opaque to X-rays from
space --- thus, X-ray astronomy must be conducted from the upper
atmosphere or, preferably, space.  Early X-ray astronomy utilized
detectors on sub-orbital rockets and high-flying balloons. Over the
past 30 years, however, X-ray astronomy has been conducted primarily
via earth-orbiting satellites, launched by a wide range of nations and
international collaborations.  Countries participating in these
missions have included the United States of America, the United
Kingdom, Japan, the European Union, Germany, the Soviet Union and
Russia, Italy, the Netherlands, and India.

Below we discuss some of those satellites that have made the greatest
contribution to the study of accretion disk features in compact object
systems.  The list is not meant to be exhaustive or exclusive, but it
does provide a fair sampling of the observational capabilities at the
disposal of X-ray astronomers in the past, present, and near future.
We also discuss some of the future
missions being planned by US National Aeronautics and Space
Administration (NASA) and the European Space Agency (ESA).

\subsection{Past and current X-ray observatories}

{\small
\begin{table}
\begin{center}
\begin{tabular}{llccc}\hline\hline
Observatory & Instrument & Area & Band pass & Resolution \\
(lifetime) & & (cm$^2$) & (keV) & ($E/\Delta E$ at 6\,keV) \\\hline\hline
EXOSAT (ESA) & GS  & 100 & 2--20 & 10 \\
May 1983--Apr. 1986 & ME & 1600 & 1--50 & 10 \\\hline
Ginga (Japan) & LAC & 4000 & 1.5--37 & 10 \\
Feb. 1987--Nov. 1991\\\hline
ASCA (Japan$+$NASA) & GIS & $2\times 50$ @ 1\,keV & 0.8--11 & 10 \\
Feb. 1993--Mar. 2001 & SIS & $2\times 100$ @ 6\,keV & 0.5--10 & 50 \\\hline 
RXTE (NASA) & PCA & 6500 & 2--60 & 10 \\
Dec. 1995--present & HEXTE & $2\times 800$ & 15--250 & - \\\hline
BeppoSAX (IT+NL) & LECS & 22 @ 0.28\,keV & 0.1--10 & 8\\
Apr. 1996--Apr.2002 & MECS & 150 @ 6\,keV & 1.3--10 & 8\\
& PDS & 600 @ 80\,keV & 15--300 & --\\\hline
Chandra (NASA) & ACIS & 340 @ 1\,keV & 0.2--10 & 50 \\
Jul. 1999--present & HETG & 59 @ 1\,keV & 0.4--10 & 200 \\\hline
XMM-Newton (ESA) & EPIC-MOS & $2\times 920$ @ 1\,keV & 0.2--12 & 50 \\
Dec. 1999--present & EPIC-PN & 1220 @ 1\,keV & 0.2--12 & 50 \\\hline
\end{tabular}
\caption{Observatories and instruments that have been important for 
  studies of X-ray reflection from black hole sources.  Note that some
  of these observatories possess other instruments/detectors that we
  have not listed since they are not of direct relevance to iron line
  studies.  Instrument abbreviations: ACIS=AXAF Charged Coupled
  Imaging Spectrometer; EPIC=European Photon Imaging Camera;
  GIS=Gas Imaging Spectrometer; GS=Gas Scintillation Proportional
  Counter; HETG=High Energy Transmission Grating; HEXTE=High Energy
  X-ray Timing Experiment; LAC=Large Area Proportional Counter;
  LECS=Low Energy Concentrator Spectrometer; ME=Medium Energy
  Proportional Counter; MECS=Medium Energy Concentrator Spectrometer;
  PCA=Proportional Counter Array; PDS=Phoswich Detection System;
  SIS=Solid-State Imaging Spectrometer.}
\label{tab:observatories}
\end{center}
\end{table}
}

Table~\ref{tab:observatories} lists the capabilities of the major
X-ray observatories relevant to the present discussion.  Before the
launch of {\it ASCA}, the most important observatories for iron line
studies were the European's {\it EXOSAT} and Japan's {\it Ginga}
satellites.  As we will discuss in the following sections, these
observatories provided the first clear evidence for iron fluorescence
and X-ray reflection in accreting black hole sources
\cite{fabian:89a,barr:85a,nandra:89a,nandra:91a}.  However, we had to
await the major increase in spectral resolution brought about by the
launch of {\it ASCA} before the relativistic effects could truly be
explored.

The fourth Japanese X-ray satellite, initially called Astro-D, was
renamed to {\it ASCA}\footnote{From Asuka, pronounced ASCA, the name
of a region in central Japan, and an era in Japanese history around
600 AD during which Asuka was the capital.  It is also an ancient
Japanese word meaning ``flying bird''.} shortly after launch in
February 1993 {\it ASCA} had two major innovations which
revolutionized the study of many astrophysical X-ray sources.  It was
the first observatory to possess (four) focusing X-ray telescopes that
worked up to 10\,keV (previous focusing X-ray telescopes only operated
in the soft X-ray band).  But, most importantly, {\it ASCA} had X-ray
sensitive CCDs placed at the focal plane of two of these telescopes,
giving X-ray spectra with an energy resolution $E/\Delta E \approx
50$. This allowed, for the first time, X-ray line widths to be
resolved, as well as details of broader features to be discerned.
Many of the results of AGN studies discussed below were obtained via
{\it ASCA} observations.  NASA and ESA also had a significant contribution
to the {\it ASCA} project.

The two premier X-ray observatories currently operating are the {\it
Chandra X-ray Observatory} (NASA) and {\it XMM-Newton} (ESA).  Before
discussing future planned missions, we briefly describe these two
observatories in a little more detail.

\subsubsection{The Chandra X-ray Observatory}  The Advanced X-ray 
Astrophysics Facility (AXAF) was launched in July 1999 as one of
NASA's ``Great Observatories'', and was subsequently renamed {\it Chandra},
in honor of Subrahmanyan Chandrasekhar.  Its prime characteristic is
its extremely high spatial resolution.  Its X-ray mirrors (consisting
of nested shells of grazing incidence optics) are able to resolve
angular scales of only 0.2\,arcsecs.  Even with all of the
practicalities of observing, resolutions of 0.5\,arcsec can be
achieved.  The primary detector array (ACIS) comprises of an array of
CCDs at the focal plane of the telescope.  When operated in
``imaging'' mode, the X-rays strike these CCDs directly, producing a
high-quality image in which the energy of each photon is tagged with
moderate energy resolution ($E/\Delta E\sim 50$).  In ``spectroscopy''
mode, one of a set of (transmission) diffraction gratings is placed in
the X-ray beam, allowing high resolution spectra (up to $E/\Delta
E\sim 1000$ at soft X-rays; $E/\Delta E\sim 200$ @ 6\,keV) to be
obtained at the expense of one spatial dimension.

\subsubsection{XMM-Newton} The X-ray Multi-Mirror Mission
(now called {\it XMM-Newton}) was launched by the European Space Agency in
December 1999.  It possesses three grazing incidence X-ray telescopes
with CCD detectors (the European Photon Imaging Camera; EPIC) in each
focal plane.  {\it XMM-Newton} is complementary to {\it Chandra} in the sense
that, while not possessing the high spatial resolution of {\it Chandra} (it
can achieve spatial resolutions of $\sim 5$\,arcsec), it has a
significantly larger collecting area thereby allowing high-throughput
spectroscopy.  Since many studies of X-ray reflection in accreting
black holes are signal-to-noise limited, {\it XMM-Newton} has become an
extremely important observatory for these purposes.  Two of the
telescopes also have (reflection) diffraction gratings that
permanently intercept half of the X-ray beam in two of the telescopes,
providing simultaneous high-resolution (up to $E/\Delta E\sim 500$)
X-ray spectra in the soft X-ray band (below 2\,keV).

\subsection{Future missions}  

\subsubsection{Astro-E II} 
Astro E was supposed to be the fifth in the series of Japanese X-ray
satellites.  Its overall soft X-ray effective area and mirror design
were similar to those of {\it ASCA}, except that its mirrors had a factor of
two improvement in spatial resolution. In addition to soft X-ray CCD
detectors, Astro E had a hard X-ray detector that was sensitive to
energies as high as 700\,keV.  Most importantly, however, Astro E was
to utilize the first space-based X-ray micro-calorimeters.  Rather
than achieve high spectral resolution via dispersion gratings, X-ray
photons would deposit heat into calorimeter elements with eV energy
resolution, thus achieving $E/\Delta E \approx 1000$ or more at
6\,keV.  The calorimeters required cryogenic cooling, limiting their
useful lifetime to approximately two years.  Thereafter, however, the
remaining Astro E detectors would have had comparable energy
resolution to that of the {\it ASCA} CCDs, and broader energy coverage
(0.3-700\,keV) than either {\it RXTE} or BeppoSAX.  The first Astro E
satellite was launched in early 2000.  Unfortunately, attitude control
of the launch vehicle was lost, and Astro E was placed in a shallow,
elliptic orbit, from which it fell to Earth a short while later.
Following much the original design plan, however, the Astro E2 mission
is currently being built.  It will have all of the originally intended
capabilities of Astro E, and a successful launch is expected in 2005.
NASA is a major collaborator in this endeavor.
 
\subsubsection{Constellation-X} 

NASA's {\it Constellation-X} is an observatory that is in the early stages
of design and planning, with a proposed for launch date of c.2009.
Similar to Astro E, the prime detector instruments are expected to be
X-ray micro-calorimeters, albeit with total effective area greater
than any current or previous X-ray satellite ($A\sim 3\,{\rm m}^2$).
The planned energy range will be $0.2-40$\,keV, with a spatial
resolution of 6 arcsec. The large effective area is to be achieved by
``mass-producing'' and launching four separate satellite systems with
their own independent mirror and detector assemblies.  The combination
of high spectral resolution with large effective area has raised the
prospects that {\it Constellation-X} can be used for detailed iron line
variability studies, including ``reverberation mapping'' between
continuum and line features, as we further discuss below.
 
\subsubsection{XEUS} 

ESA's {\it XEUS}, the X-ray Evolving Universe Mission, is a potential
successor to {\it XMM-Newton}.  The current plan envisions two phases for
the mission.  The mirrors and detectors are planned to be housed in
two separate spacecraft assemblies, with both being in close proximity
to the International Space Station.  The detectors will cover the
0.05--30\,keV range, with a collecting area of 6\,m$^2$ (i.e.,
approximately 10 times the peak effective area of {\it RXTE}) at energies of
1\,keV in the first phase of the mission.  This is planned to be
expanded to 30 m$^2$ at 1\,keV and 6\,m$^2$ at 8\,keV in the second
phase of the mission.  The instruments will have energy resolution of
1 to 10 eV (i.e., $\Delta E/E \approx 1000$ at 6\,keV) and spatial
resolution of 2 arcsec.  Like {\it Constellation-X}, this would be a very
powerful combination of large effective area and high spectral
resolution, making {\it XEUS} ideal for reverberation mapping studies, as we
further describe below. 
\subsubsection{MAXIM} 
\label{sec:maxim}
The proposed Micro-arcsecond X-ray Interferometry Mission (MAXIM)
aims to obtain sub-microarcsecond resolution X-ray images of nearby
galactic nuclei (especially our Galactic Center, M87, and nearby
Seyfert galaxies).  At this resolution, one can resolve X-ray emitting
structures on the scale of the event horizon of the central massive
black hole.  Furthermore, one may be able to map, in both spatial and
velocity coordinates, the iron line fluorescence across the surface of
the accretion disk.  This would provide unprecedented and direct
constraints on the dynamics of the disk.  The technological
developments required to construct this ambitious observatory are
challenging, requiring one to perform X-ray interferometry in deep
space across baselines of hundreds of meters.  However, these
difficulties are not insurmountable, and it is reasonable to suppose
that such an observatory could be constructed within 20 years, with a
less able ``pathfinder'' mission in 10 years or so.

\section{Iron lines from active galactic nuclei}
\label{sec:agn}

The cleanest examples of observed relativistic iron lines have been
found in AGN.  There are three principal reasons why AGN tend to
present cleaner disk reflection signatures than GBHCs.  Firstly, the
primary X-ray continuum is often found to be very featureless, being
well modeled by a simple power-law form.  This makes it
straightforward to subtract the continuum in order to study the
underlying disk reflection signatures.  The physical reason underlying
the simplicity of the continuum form in AGN is the wide separation of
temperatures between the accretion disk surface (with $kT\sim 10\eV$)
and the disk corona (with $kT\sim 100\keV$) --- thermal Comptonization
of the optical/UV disk emission by the corona produces a featureless
power-law spectrum across much of the X-ray band.  The accretion disks
in GBHCs, on the other hand, can get so hot ($kT\sim 1\keV$) that the
Wien tail of the thermal disk emission can overlap with the red-wing
of any iron line, making continuum subtraction much more challenging.
Secondly, we expect there to be a wide range of parameter space over
which AGN accretion disks are not strongly ionized, thereby
simplifying the study of the X-ray reflection signatures.  On the
other hand, GBHCs may well be generically ionized due to the high
temperature of the disk.  Finally, since GBHCs typically lie in the
plane of the Galaxy, they are usually much more heavily absorbed than
type-1 AGN, further complicating the modeling of the continuum
radiation.

\subsection{A case study of MCG--6-30-15}
\label{sec:mcg6}

There are a small number of AGN that have become test-beds for the use
of X-ray observations to probe black hole and accretion disk physics.
Here, we shall describe studies of one such object --- the Seyfert 1
galaxy MCG--6-30-15.  This is a somewhat unremarkable S0-type galaxy
in the constellation of Centaurus, with a redshift of $z=0.008$,
placing it at a distance of $D=37\Mpc$.  As we will describe, X-ray
studies of this object have allowed us access to some of the most
exotic black hole physics yet observed.

Unfortunately, the mass of the supermassive black hole in this AGN
remains rather poorly determined --- some general arguments place it
in the range $10^6-2\times 10^7\Msun$\cite{reynolds:00a}.  Most of the
astrophysical conclusions drawn below are robust to this uncertainty
since, to first order, the mass of the black hole merely scales the
size of the system rather than changing any of the underlying physics.
However, constraints on the mass are important if we are to physically
interpret observed temporal variability, since all of the fundamental
time scales of the accretion disk scale linearly with black hole mass.

\subsubsection{The first detection of a relativistic accretion disk}

MCG--6-30-15 was the first AGN where X-ray reflection from fairly
neutral material was clearly detected using the {\it EXOSAT}
\cite{nandra:89a} and {\it Ginga} \cite{pounds:90a} observatories.
This demonstrated that there was cold and optically-thick material in
the vicinity of the SMBH that was being irradiated by the power-law
X-ray continuum.  It was suspected by many that this material was
indeed the accretion disk.  However, direct confirmation of this had
to await the superior spectral capabilities of {\it ASCA}.

As has already been described, {\it ASCA} was the first observatory
capable of obtaining medium resolution 0.5--10\,keV X-ray spectra of
cosmic sources.  Thus, it must be appreciated that the demands for
observing time on {\it ASCA} were intense, with many different
sub-communities of the astrophysical world wishing to utilize this new
resource to explore the universe (indeed, {\it ASCA} has made major
contributions in fields as diverse as stellar coronae, the diffuse
interstellar and intracluster medium, and gamma-ray bursts).
Furthermore, at the time when {\it ASCA} started operating, there was
a certain level of skepticism within the astrophysical community
concerning the possibility of detecting strong relativistic effects
via spectroscopic studies of AGN --- thus, since the majority of
observing time on {\it ASCA} was allocated in open peer-reviewed
competitions, it was not a foregone conclusion that {\it ASCA} would
spend significant time taking high-quality spectra of AGN.  The fact
such spectra {\it were} obtained, thereby allowing the direct
detection of relativistic effects around supermassive black holes, is
largely due to the perseverance of Yasuo Tanaka (the Project
Scientist) and Andy Fabian, for which they were awarded the Rossi
Prize of the American Astronomical Society in 2001.

Returning to the history of MCG--6-30-15, a rather short (half day)
{\it ASCA} observation of MCG--6-30-15 showed the iron line to be
significantly broadened in energy space, being well modeled as a
Gaussian profile with centroid energy $E_0=6.2\keV$ and a
standard-deviation of $\sigma=0.7\keV$\cite{fabian:94b,reynolds:95a}.
The improved signal-to-noise from a much longer (4.5
days\footnote{During the early stages of the ASCA mission, this was an
  observation of unprecedented length.  The typical ASCA observation
  rarely exceeded one day.}) {\it ASCA} observation was required to
determine that the iron line did, indeed, possess the profile expected
from the surface of a relativistic accretion disk\cite{tanaka:95b}.
As shown in Fig.~1, this line profile is well fit with line emission
from a disk around a Schwarzschild black hole with a disk inclination
of $\theta\approx 27^\circ$ (measured away from the normal to the disk
plane), with line flux distributed across the disk proportional to
$r^{-3}$, extending down to $r=6M$.

\begin{figure}
\centerline{
\hspace{0.5cm}
\psfig{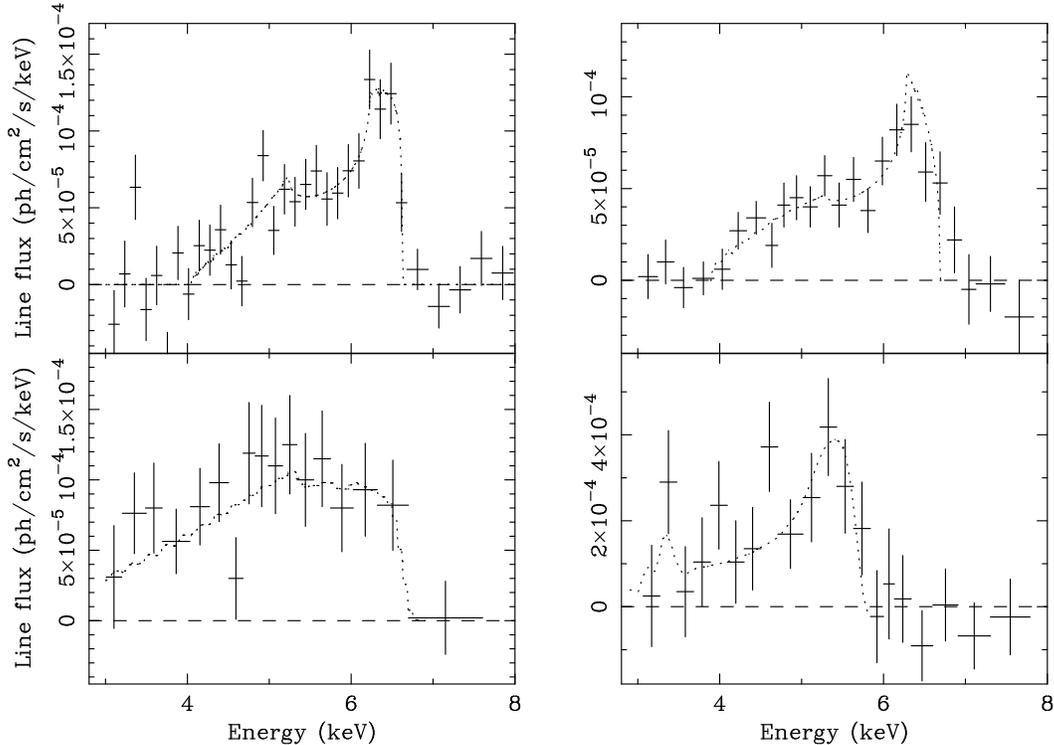}
}
\caption{Time-averaged (upper panels) and peculiar line profiles
(lower panels) of the iron K emission from MCG--6-30-15 seen in the
two long {\it ASCA} observations in 1994 (left) and 1997 (right). In the
1994 observation, a very broad profile with a pronounced red-wing is
seen during a period of Deep Minimum of the light curve (lower left),
compared to the time-averaged line profile shown in the upper
panel. In contrast, during a sharp flare in the 1997 observation,
whole line emission is shifted to energies below 6 keV and there is no
significant emission at the rest line-energy of 6.4 keV (lower right).
Both peculiar line shapes can be explained by large gravitational
redshift in small radii on the accretion disk.}
\label{fig:iwasawa}
\end{figure}

In fact, these {\it ASCA} data were a great deal richer than just
described.  As is common with type-1 AGN, the X-ray flux of this AGN
varied rapidly as a function of time.  Via a detailed analysis of this
and subsequent {\it ASCA} observations, Kazushi Iwasawa found that the
iron line flux and profile also undergoes dramatic changes with
time\cite{iwasawa:96a,iwasawa:99a}. The detailed iron line changes are
complex and defy simple characterization (see Fig.~\ref{fig:iwasawa}).
In the 1994-July {\it ASCA} observation, it was found that, during a
large flaring event, the iron line became quite narrow with an energy
close to the intrinsic value.  This suggests that the observed
fluorescence was dominated by fairly large radii in the disk during
the flare.  On the other hand, during another flare observed in a 1997
{\it ASCA} observation, the line emission redshifted entirely to
energies below 6\,keV suggesting dominance by small radii in which
gravitational redshifts are large.

Of particular interest was the identification of the so-called ``deep
minimum'' (DM) state.  During the DM state, the continuum flux falls
below its average value by a factor of 2--3, and the fluorescent iron
line appears to become much broader and stronger.  In fact, the
redshifts needed to explain the observed line profile are so great as
to require line emission from within $r=6M$.  As we shall see, we
believe that the DM state gives us a window into the more exotic
astrophysics of rapidly rotating black holes. 

Of course, the claim that iron line studies are probing the region
within a few gravitational radii of the black hole is a bold one, and
should be tested against other models at every opportunity.  Given the
quality of data, the July-1994 MCG$-$6-30-15 line profile has become a
traditional test bed for such comparisons.  The first issue to assess
is the correctness of the continuum subtraction that underlies the
determination of the iron line profile.  Andy Fabian and collaborators
systematically examined the effect of unmodeled spectral components
(e.g., absorption edges) and found that the line profile was robust to
all physically reasonable possibilities \cite{fabian:95a}.  Given the
reality of the spectral features, one must then critically assess its
interpretation as an emission line arising from a relativistic
accretion disk.  The most serious challenger is a model in which a
source of intrinsically narrow iron line emission is surrounded by a
rather optically-thick ($\tau\sim 5$), cool ($kT\sim 0.2\keV$) and yet
highly-ionized plasma.  Repeated Compton recoil suffered when the
photons scatter off the electrons can then broaden and redshift the
observed line\cite{misra:99a}.  However, such a scenario
predicts that the continuum spectrum passes through the same Compton
cloud, filtering out high-frequency temporal variability and imposing
a continuum spectral break at $E\sim m_{\rm e}c^2/\tau^2$.  The
presence of high-frequency temporal variability and the absence of a
continuum spectral break strongly argues against such a scenario
\cite{reynolds:00b,ruszkowski:00a}.

In another alternative model, it has been has proposed that energetic
protons transform iron in the surface of the disk into chromium and
lower $Z$ metals via spallation which then enhances their fluorescent
emission \cite{skibo:97a}.  With limited spectral resolution, such a
line blend might appear as a broad skewed iron line.  This model
suffers both theoretical and observational difficulties.  On the
theoretical side, high-energy protons have to be produced and slam
into the inner accretion disk with a very high efficiency (one
requires 10\% of the binding energy of the accreted material to be
channeled into this process alone).  On the observational side, it
should be noted that the broad line in MCG--6-30-15 \cite{tanaka:95a}
is well resolved by the {\it ASCA} SIS (the instrumental resolution is
about 150~eV at these energies) and it would be obvious if it were due
to several separate and well-spaced lines spread over 2~keV. There can
of course be Doppler-blurring of all the lines, but it will still be
considerable and require that the redward tails be at least 1 keV
long.

\subsubsection{Is the black hole rapidly spinning?}
\label{sec:is_bh_spinning}
The very broad line seen in the {\it ASCA} DM state of MCG--6-30-15
leads us to consider rotating black holes.  If, for now, we make the
assumption that there is no fluorescent emission line from within the
radius of marginal stability, the DM spectrum requires a rapidly
rotating black hole in order to have $r_{\rm ms}$ appreciably less
than $6M$ \cite{iwasawa:96a}. In fact, if one assumes that the iron
line flux is distributed across the disk according to
eqn.~\ref{eq:pt}, the formal limit on the dimensionless spin parameter
of the black hole is $a>0.94$ \cite{dabrowski:97a}.

There has been a great temptation within the X-ray astrophysics
community to interpret very broad iron lines, with $r_{\rm in}<6M$, as
direct evidence for black hole rotation.  However, when one attempts
to make these arguments rigorous, a loophole is rapidly discovered.
By its basic nature, one can see that it is possible to produce
arbitrarily redshifted spectral features from around any black hole if
{\it any} region of the disk (even those within the plunging region)
is permitted to produce observable line emission \cite{reynolds:97d}.
As the emitting region tends to the event horizon, the redshifts grow
arbitrarily large.  Thus, an attempt to determine the space-time
geometry (and, in particular, the spin parameter of the black hole)
from the time-averaged iron line profile necessarily involves
considerations of the {\it astrophysics} of the inner accretion disk
and, in particular, the extent to which emission/reflection from the
plunging region of the disk might be important.

Within the standard model for black hole accretion (i.e., that of
Page, Novikov and Thorne \cite{page:74a,novikov:74a}), there is no
angular momentum transport and no dissipation of energy within the
plunging region.  Hence, this region of the disk cannot support an
X-ray active corona.  Furthermore, if the corona of the dissipative
part of the disk has a small scale-height, the surface of the plunging
region will not be subjected to X-ray irradiation for purely geometric
reasons, and will not produce X-ray reflection signatures.  These
considerations would validate the assumption of no X-ray reflection
signatures from within the plunging region thereby allowing the spin
of the black hole to be determined in the manner discussed above.
However, there are two scenarios in which appreciable X-ray reflection
might occur from part of the plunging region.  Firstly, it is possible
that the outer disk corona might be geometrically-thick, thereby
allowing it to irradiate a dissipationless plunging region.  Given
this scenario, the {\it ASCA} data for the DM-state of MCG--6-30-15
data could be consistent with even a non-rotating (Schwarzschild)
black hole \cite{reynolds:97d}.  Secondly, as discussed in
\S\ref{sec:innerdisk_mhd}, the accretion flow within the plunging
region may well be able to dissipate a significant fraction of its
binding energy.  One could easily imagine the formation of a powerful
X-ray emitting corona sandwiching the plunging region of the disk.
Although this scenario has not been worked out in detail, such an
inner corona might be responsible for locally produced X-ray
reflection within the plunging region.  It should be noted that the
plunging region is likely to be highly ionized, and so any detailed
study of X-ray reflection from this region must use self-consistent
ionized reflection models as were discussed in \S\ref{sec:reflection}
\cite{young:98a}.

\begin{figure}
\centerline{
\psfig{figure=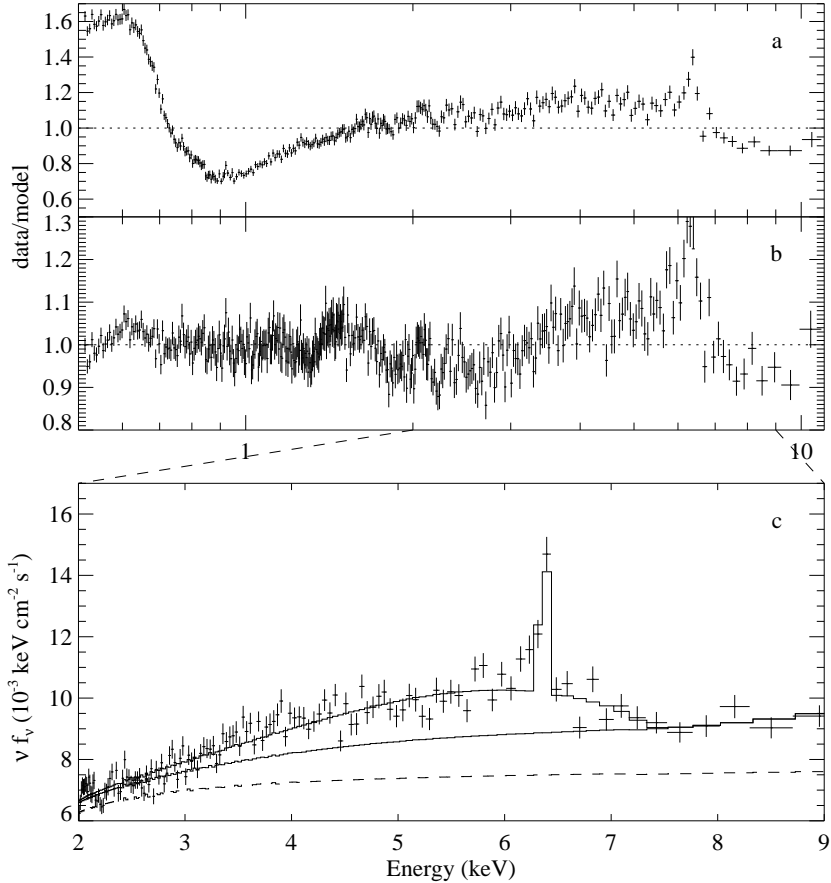,width=0.8\textwidth}
}
\caption{{\it XMM-Newton} data from the first observation of
MCG--6-30-15.  Panel (a) shows the ratio between the data and a simple
model consisting of a power-law fitted to the 0.5--11\,keV data. Panel
(b) shows the ratio of data to a more detailed model which also
accounts for line-of-sight absorption towards the accreting black
hole.  The residuals in this plot are attributed to X-ray reflection
from the relativistic disk.  (c) Deconvolved spectrum of the iron K
band portion of the spectrum, showing a very broadened and redshifted
iron line.  Figure from \cite{wilms:01b}}
\label{fig:mcg6xmm}
\end{figure}

The above history made MCG--6-30-15 a prime target for study with the
high-throughput EPIC instruments on board the {\it XMM-Newton}
observatory.  The first {\it XMM-Newton} observation of this source
happened to catch it in a prolonged DM state.  The superior
sensitivity of these instruments allowed a much more detailed study of
this enigmatic state to be conducted \cite{wilms:01b}.

In agreement with the {\it ASCA}-based conclusions of Iwasawa, the
X-ray reflection features were found to be very broad and very strong
(see Fig.~\ref{fig:mcg6xmm}).  The detailed modeling of the {\it
XMM-Newton/EPIC} spectrum required the modeling of the full reflection
spectrum, not just the fluorescent iron line, with relativistic
shifting/smearing applied to both the fluorescent line and the
reflection continuum.  Again, the reflection features were found to be
so broad and redshifted as to require reflection from within $r=6M$.
Assuming a background spacetime corresponding to a near-extremal Kerr
black hole (with $\as=0.998$, giving $r_{\rm ms}=1.23M$), the data
constrain the inner edge of the reflecting region to lie closer than
$r_{\rm in}<2M$, with a steep emissivity profile outside of that
radius of $4.5<\beta<6.0$.  This line profile {\it can} be
successfully modeled assuming a Schwarzschild (non-rotating black
hole) background (see discussion in \S~\ref{sec:is_bh_spinning}), but
very extreme parameters are demanded --- the inner edge of the
reflection region is constrained to be $r_{\rm in}<3M$, very deep
within the plunging region, with an extremely large emissivity index
$\beta>10$.  Thus, in the Schwarzschild scenario, essentially {\it
all} of the fluorescence has to originate from an annulus in the
innermost parts of the plunging region, $2M-3M$.

So, the possibility of X-ray reflection signatures from within the
radius of marginal stability remains a wild-card affecting the
detailed interpretation of these data.  While there are no models that
treat such emission in detail, the extreme parameters required to
explain the {\it XMM-Newton} data within the context of a
Schwarzschild black hole ($r_{\rm in}<3M$ and $\beta>10$) seem
entirely unreasonable.  At the very least, one would expect this
material to be in an extremely high ionization state and incapable of
producing iron fluorescence and other X-ray reflection features.
Thus, the case for a rapidly spinning black hole in this object is now
very strong, even accounting for the possibility that there may be
some emission from the plunging region.

\subsubsection{Is the spinning black hole energizing the disk?}

To explore the implications of the MCG--6-30-15 {\it XMM-Newton} data
further, we employ the standard assumption that there is little or no
X-ray reflection from within the plunging region.  To make a
connection with theoretical accretion disk models, we must relate the
distribution of observed X-ray reflection to the distribution of
primary energy dissipation in the accretion disk (since it is the
latter that is predicted by the models).  The simplest assumption is
that the local intensity of X-ray reflection from the disk surface is
proportional to the local dissipation in the underlying disk.  This
would be the case if the ionization of the disk surface was reasonably
uniform with radius, and a fixed fraction of the locally dissipated
energy was transported into a geometrically-thin X-ray emitting
corona.  In reasonable disk models, any appreciable ionization changes
would be in the sense of the innermost regions being more highly
ionized, therefore producing weaker X-ray reflection signatures and
requiring an even steeper dissipation profile than inferred from the
X-ray reflection studies.  We also note that, in MCG--6-30-15, a large
fraction (25--50\%) of the electromagnetic luminosity is emitted in
the X-ray band by the Comptonizing corona.  Thus, it cannot be grossly
wrong to assume that the X-ray activity traces the underlying
energetics of the disk.

\begin{figure}
\centerline{
\psfig{figure=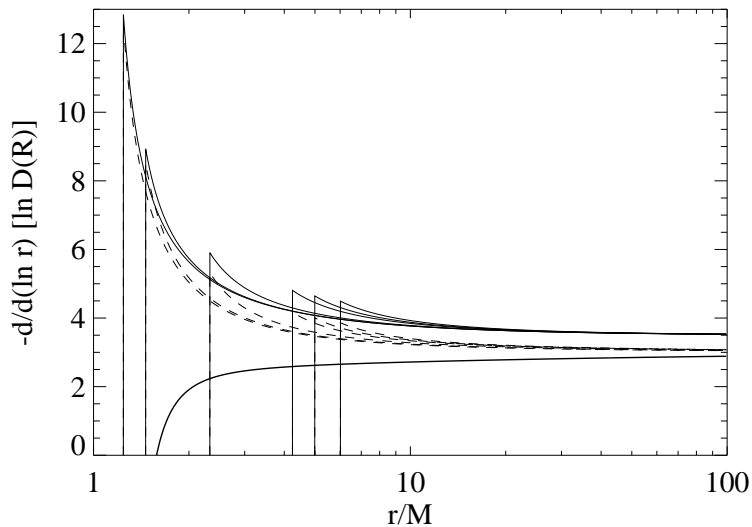,width=0.8\textwidth}
}
\caption{Logarithmic derivatives (i.e. local emissivity indices) for
  the dissipation profiles shown in Fig.~\ref{fig:pt_agol}.  As in
  Fig.~\ref{fig:pt_agol}, the thick line shows the dissipation profile
  for a standard non-torqued disk (i.e., one in which the ZTBC
  applies) around a near-extremal Kerr black hole with spin parameter
  $a=0.998$.  The thin solid lines show the torque-induced dissipation
  component $D_{\rm tor}$ for spin parameters of (from left to right)
  $a=0.998,0.99,0.9,0.5,0.3$, and $0$.  The dashed lines show the
  torque-induced component of the emitted flux, including the effects
  of returning radiation.  To explain the MCG--6-30-15 result, one
  requires a torqued accretion disk around a rapidly rotating black
  hole.  }
\label{fig:pt_agol_beta}
\end{figure}

Taking the X-ray reflection as a proxy for the underlying disk
dissipation, we see that these data are clearly discrepant with the
standard models for black hole accretion disks for any spin parameter.
As discussed in \S\ref{sec:rel_disks}, the flux emitted per unit
proper area of a standard disk around a near-extremal Kerr black hole,
$D_{\rm PT}(r)$, is zero at $r=r_{\rm ms}$ by virtue of the ZTBC,
peaks at $r\sim 1.6r_{\rm g}$ and then gradually steepens to approach
$D(r)\propto r^{-3}$.  At no point does $D_{\rm PT}(r)$ become as
steep as $r^{-4.5}$, as required by the {\it XMM-Newton} observations.
This is true for any assumed BH spin.  However, as discussed in
\S\ref{sec:innerdisk_mhd}, magnetically-induced torques at the radius
of marginal stability can significantly enhance the amount of
dissipation in the innermost regions of the accretion disk.  The
`best' possible case (in terms of producing steep dissipation
profiles) is where the magnetic torque is applied entirely at the
radius of marginal stability.  In Fig.~\ref{fig:pt_agol_beta}, we plot the
logarithmic derivative of the dissipation profiles shown in
Fig.~\ref{fig:pt_agol}.  One finds that it is still impossible to
achieve the required steep dissipation profiles in the case of
non-rotating or slowly rotating black holes.  Only in the case of a
rapidly-rotating black hole does the dissipation profile become
sufficiently steep.  In fact, one finds that sufficiently steep
dissipation profiles can only be obtained in the regions of parameter
space where an appreciable fraction of the extra energy is derived
directly from the spin of the black hole.  As discussed in
\S\ref{sec:innerdisk_mhd}, the physical origin of this torque is
likely a magnetic connection between the accretion disk and either the
plunging region or the rotating event horizon itself.

Relativistic effects may also reveal themselves through the strength
of the reflection features.  These {\it XMM-Newton} data show that,
during the DM state, the relative strength of the reflection is about
twice that expected from a plane-parallel coronal geometry.  While
there may be geometrical explanations, such an enhancement is in fact
a natural consequence of returning radiation
\cite{cunningham:75a,wilms:01b}.

It must be reiterated that, while the case for a rapidly-rotating hole
in MCG--6-30-15 is solid, the argument for a torqued accretion disk is
vulnerable to the possibility of powerful fluorescence/X-ray
reflection from within the plunging region.  More theoretical work on
the plunging region is required to assess the impact it might have on
these arguments.  There is also another way of alleviating the need
for a torqued accretion disk.  Martocchia \& Matt have recently
suggested that the DM spectrum of MCG--6-30-15 could be explained via
the gravitational focusing of X-rays emitted by a source that is
$r\sim 3M$ from the black hole at high latitudes (i.e. close to the
spin axis of the black hole) \cite{martocchia:02a}.  One problem
suffered by this scenario is that it {\it overpredicts} the relative
strength of the X-ray reflection by a factor of two.  Thus, for this
picture to be correct, only half of the disk surface area can be in a
state capable of producing X-ray reflection signatures.  Even if this
picture is correct, the presence of such a powerful X-ray source on
the spin axis of the black hole (presumably corresponding to the base
of an MHD jet) is itself a strong circumstantial indicator of black
hole spin-energy extraction.

\begin{figure}
\centerline{
\psfig{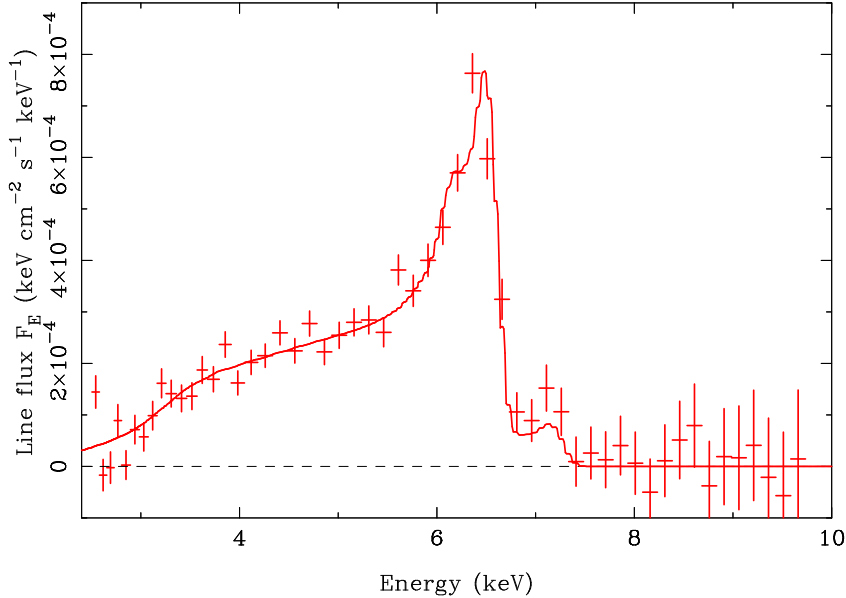}
}
\caption{Continuum subtracted iron line profile from the second 
  {\it XMM-Newton} observation of MCG--6-30-15 \cite{fabian:02a}.
  This is the highest signal to noise relativistic line profile yet
  measured.}
\label{fig:mcg6_fab02}
\end{figure}

The first results from a later and very long (400\,ks) {\it
  XMM-Newton} observation of MCG--6-30-15 were reported recently by
Andy Fabian and collaborators \cite{fabian:02a}.  This observation has
produced the best iron line profile yet obtained from any object
(Fig.~\ref{fig:mcg6_fab02}).  There is strong evidence for line
profile variability between the two {\it XMM-Newton} observations,
with the latter observation showing a prominent blue horn to the line
profile.  This horn indicates the presence of emission from larger
radii in the disk.  However, the extreme red-wing to the line profile,
requiring a very steep central emissivity profile, was also present in
these data.  We await further analysis of this tremendously important
dataset.

\subsubsection{Relativistically-broadened soft X-ray emission lines}

Before leaving MCG--6-30-15, we will briefly discuss a fascinating
debate that is currently raging about the soft X-ray spectrum of this
object.  The soft X-ray spectrum of this, and many other type-1 AGN,
is rather complex.  Prior to the launch of {\it XMM-Newton}, it was
widely accepted that this complexity was due to absorption by
photoionized material along the line of sight to the central engine
(the so-called warm absorber; \cite{nandra:92a,reynolds:97a}), with
the principal observables being the photoelectric absorption edges of
OVII and OVIII.  In many objects, this hypothesis has been confirmed
by high-resolution grating spectroscopy with {\it Chandra} and {\it
  XMM-Newton}; such spectra clearly display the resonant absorption
lines of oxygen that are expected to accompany the photoelectric edges
(e.g., see the fabulous X-ray spectrum of NGC~3783 obtained by Kaspi
and collaborators; \cite{kaspi:02a}).  However, in MCG--6-30-15, an
{\it XMM-Newton} grating spectrum showed that the simple warm absorber
model fails --- there are anomalous ``edges'' and an apparant lack of
strong oxygen resonance absorption lines \cite{branduardi:01a}.  On
the basis of this, Branduardi-Raymont and collaborators suggested that
the soft X-ray spectral complexity is due to strong, relativistically
broadened oxygen, nitrogen and carbon recombination lines rather than
a warm absorber \cite{branduardi:01a}.  The team led by Julia Lee
analysing the non-simultaneous {\it Chandra} grating data can,
however, explain the soft X-ray spectrum in terms of a warm absorber
which contains embedded iron-rich dust \cite{lee:01b}.  It remains
unclear whether the dusty warm absorber model can explain the {\it
  XMM-Newton} grating data.

Given the rapid and on-going development of these arguments (both for
and against the idea of soft X-ray recombination lines that possess
relativistic disk profiles), we refrain from drawing conclusions at
this time.  The interested reader is pointed to the astrophysical
literature on X-ray spectroscopy of MCG--6-30-15.

\subsection{Other ``normal'' Seyfert nuclei}

While MCG--6-30-15 has been subjected to particularly detailed study,
there have also been intensive studies of other Seyfert galaxies.
Given the limited collecting area of {\it ASCA}, multi-day
observations of bright AGN were required in order to obtain reasonable
signal-to-noise in the relativistic iron line.  Prior to the existence
of many such long datasets, Paul Nandra and co-workers showed that the
relativistic iron line is present in the ``average'' {\it ASCA}
spectrum of a sample of two dozen Seyfert-1 galaxies \cite{nandra:97a}.
However, in the past 5 years, deep observations of Seyfert-1 galaxies
with {\it ASCA} and, more recently, {\it Chandra} and {\it XMM-Newton}
have been performed.  Here we summarize the results of these studies.

In addition to MCG--6-30-15, relativistically broad iron lines with
very high signal-to-noise have been found by {\it ASCA} in the Seyfert
galaxies NGC~3516 and NGC~4151.  In NGC~3516, the red-wing of the iron
line appears to track changes in the continuum flux, as expected by
the simple reprocessing model.  The blue-wing of the iron line,
however, displays higher amplitude variability that is uncorrelated
with the continuum flux, suggesting more complex changes in the
pattern of fluorescence across the disk \cite{nandra:97d,nandra:99a}.
More interestingly, the {\it ASCA} data suggested the presence of an
absorption line at $\sim 5.9\keV$ (absorbing flux from the broad iron
emission line).  If real, this feature may arise from redshifted
resonance absorption by ionized iron in tenuous plasma above the disk,
with the energy shift due to either gravitational redshift
\cite{ruszkowski:00b} or infall of the material \cite{nandra:99a}.  If
the latter interpretation is true, this may be a rare detection of
material actually in the process of accreting into the black hole.
Recent simultaneous {\it Chandra} grating and {\it XMM-Newton}
observations of NGC~3516 have also revealed fascinating substructure
within the broad line profile in the form of narrow emission spikes at
5.6\,keV and 6.2\,keV \cite{turner:02a}.  These may be the first
indications of the expected transient non-axisymmetric illumination.

The high quality {\it ASCA} data on NGC~4151 also reveals interesting
variability and complexity within the line profile
\cite{yaqoob:95a,wang:99a,wang:01a}.  During the 1995-May {\it ASCA}
observation, the iron line profile displayed significant variability,
especially in the strength of the red-wing, despite relatively small
changes in the observed continuum \cite{wang:01a}.  On the other hand,
during the 2000-May observation, the opposite conclusion appears to
hold, with the continuum undergoing significantly larger changes than
the iron line \cite{takahashi:02a}.  Thus, it appears that the iron
line in a given AGN can change its variability behavior, although the
non-uniformity of the data analysis methods employed in these
(independent) studies prevents one from making a straightforward
conclusion.

In comparison with the very broad accretion disks iron lines in
MCG--6-30-15, NGC~3516 and NGC~4151, the well-known Seyfert galaxy
NGC~5548 has developed a reputation for displaying a rather narrow
iron line.  Fitting accretion disk models to {\it ASCA} data for this
object suggested that the line emitting region of the disk was
truncated at an inner radius of about $r=10M$
\cite{mushotzky:95a,chiang:00a}.  However, we now know that this
source displays a composite iron line \cite{yaqoob:01a}.  The {\it
  Chandra-HETG} has revealed a ``narrow'' core to the line (with a
velocity dispersion of $4500^{+3500}_{-2600}\kmps$) originating via
fluorescence of material a substantial distance from the black hole.
The large equivalent width of this narrow core ($130^{+60}_{-50}\eV$)
suggests that this material subtends a large part of the sky as seen
by the primary X-ray source.  Once this component is accounted for in
medium-resolution spectra, the {\it Chandra} data are entirely
consistent with a relativistic accretion disk extending down to the
radius of marginal stability \cite{yaqoob:01a}.  However, the last
word has yet to be spoken about NGC~5548.  Very recently (in fact,
after the initial submission of this review), there was a convincing
non-detection of the broad line in NGC~5548 by {\it XMM-Newton}
\cite{pounds:02a}.  In order to make sense of results from these
different observatories, we are forced to conclude that broad iron
lines, in at least some objects, are transitory.  

\begin{figure}
\centerline{
\psfig{figure=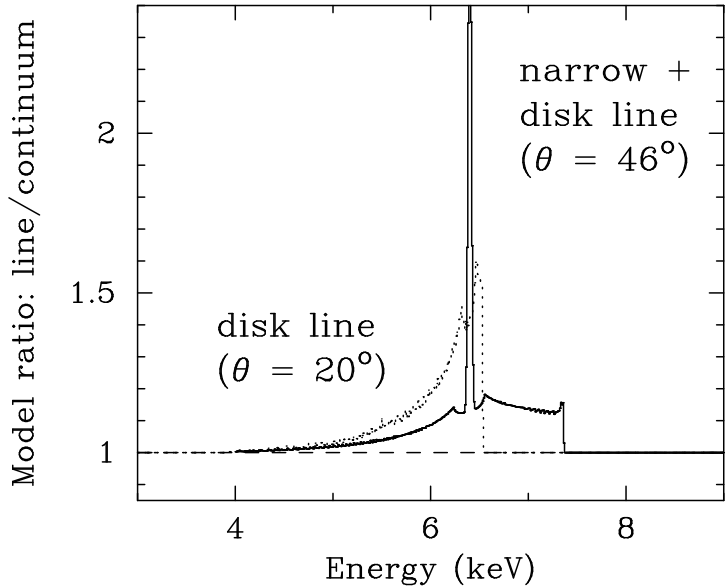,width=1.0\textwidth,angle=270}
}
\caption{Model line profiles for a narrow iron line superposed on an
iron line from an accretion disk viewed at $46^\circ$ (solid line),
compared with a line from a disk viewed at an inclination of $20^\circ$.
These two cases would be difficult to distinguish in data of limited
resolution and signal-to-noise.  Figure from \cite{weaver:98c}.}
\label{fig:nelg_lines}
\end{figure}

However, our experiences with NGC~5548 does illustrates a point of
some general practical importance --- at medium spectral resolution,
narrow (i.e., low-velocity) iron line components can be blended
together with the relativistic line profile and, unless account for,
produce significant systematic errors in measurements of, for example,
the disk inclination, emissivity profile and inner emitting radius.
For example, it has been claimed on the basis of {\it ASCA} measured
iron line profiles that the rather absorbed ``intermediate'' Seyfert
galaxies are viewed face-on \cite{turner:98a}, a result that runs
counter to our current understanding of AGN geometry (see
Fig.~\ref{fig:nelg_lines}).  However, this result is likely to be an
artifact of blending between relativistic iron line profiles and
narrow lines --- including narrow lines with reasonable strengths in
the spectral models shows that the data for these systems are
consistent with fairly edge-on disks \cite{weaver:98c}.  {\it Chandra}
has made a major contribution in this respect --- high resolution
grating spectra with the {\it Chandra}/HETG can unambiguously identify
the narrow iron line component, thereby allowing it to be subtracted
from the medium resolution spectra of the relativistic line profiles.
This is another sense in which MCG--6-30-15 is a particularly good
system for studying relativistic line profiles --- {\it Chandra}/HETG
spectra show that it has, at most, a very weak narrow line component
\cite{lee:02a} (which, interestingly, suggests that the larger-scale
environment of this SMBH is very ``clean'' with little evidence for an
obscuring torus).

\begin{figure}
\centerline{
\psfig{figure=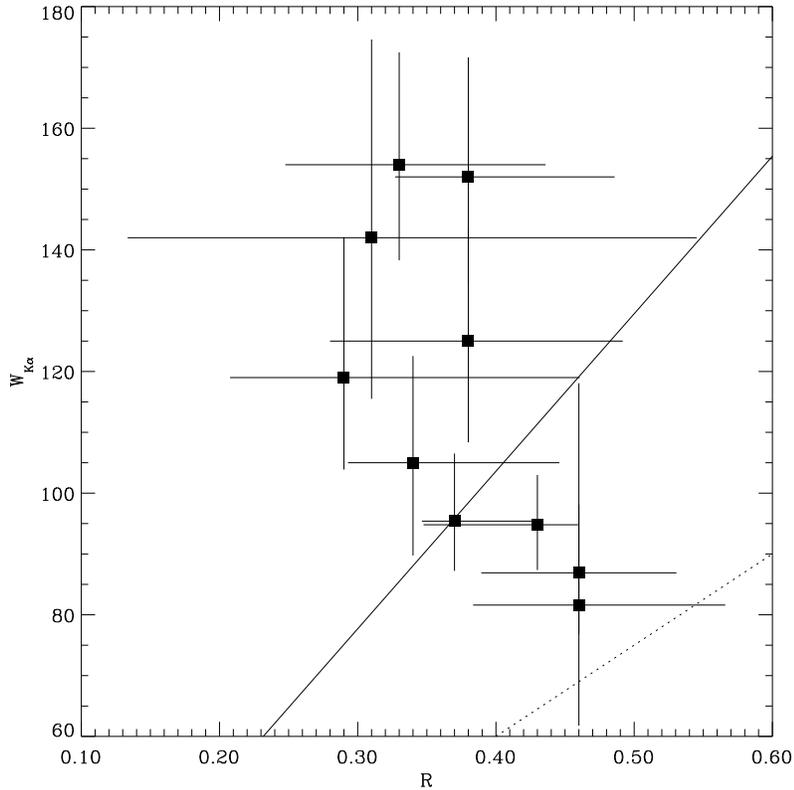,width=0.8\textwidth}
}
\caption{Iron line equivalent width $W_{\rm K\alpha}$ against the relative 
  reflection normalization ${\cal R}$ for the joint {\it ASCA}/{\it
    RXTE} campaign on NGC~5548 \cite{chiang:00a}.  The dotted line is
  the proportionality relationship expected, assuming reflection from
  a planar slab of cold gas with cosmic abundances.  The solid line is
  the best fit proportionality relationship.  The data show an obvious
  anti-correlation, possibly arguing for flux-correlated changes in
  the ionization state of the disk.}
\label{fig:ngc5548_anticorr}
\end{figure}

NGC~5548 was also the first source for which a particular mystery
associated with spectral variability was discovered.  Jim Chiang and
collaborators performed contemporaneous observations of NGC~5548 with
{\it ASCA} and {\it RXTE}, thereby allowing details of the iron line
and the reflection continuum to be constrained simultaneously
\cite{chiang:00a}.  As the continuum source varied, it was found that
the iron line flux was fairly constant or, stated another way, the
equivalent width was inversely proportional to the continuum flux.
This is partially explained by the narrow component to the iron
emission line noted above, which occurs in material light weeks or
more from the black hole.  The light travel time eliminates any line
variability over the course of the month-long {\it ASCA}/{\it RXTE}
monitoring campaign.  However, at the same time, it was found that the
continuum reflection fraction as measured by {\it RXTE} was positively
correlated with the continuum flux.  In other words, the relative
strength of the fluorescent iron line appeared to be anti-correlated
with the relative strength of the reflection continuum (see
Fig.\ref{fig:ngc5548_anticorr}).  Since the iron line and reflection
continuum are different facets of the same phenomenon (i.e. X-ray
reprocessing in the disk atmosphere), this result is most puzzling.
In principle, unmodeled flux correlated changes in the ionization
structure of the accretion disk atmosphere might decouple the iron
line and apparent reflection continuum strength in this way
\cite{reynolds:00a}, although the success of detailed ionized disk
models in explaining this observation has yet to be demonstrated.
Very similar behavior is also found in {\it RXTE} studies of
MCG--6-30-15 by Julia Lee and collaborators \cite{lee:00a}.

To summarize the above discussion, iron lines from relativistic
accretion disks appear to be generic features in the X-ray spectra of
normal Seyfert-1 galaxies, although NGC~5548 and the intermediate
Seyferts demonstrate that some care might be needed to extract the
relativistic line profile from the data.  In one source, MCG--6-30-15,
the strength of the relativistic effects argues strongly for a
rapidly-rotating black hole.  Furthermore, the data suggest that the
central regions of the accretion disk may be extracting the black hole
spin energy.  However, substantial uncertainties remain.  In
particular, we do not possess even a phenomenological characterization
of iron line variability that can predict how the line changes as a
function of continuum flux or other source properties.  Line
variability is of particular interest since it undoubtedly is the key
to substantial understanding of accretion disk physics.

The lack of a simple correlation between the iron line strength and
the observed X-ray continuum has lead some researchers to throw out
the whole picture of a relaticistically-broadened fluorescence iron
line, leading them to favour rather complex and seemingly contrived
X-ray continua as an explanation for the observed X-ray spectrum
(e.g., \cite{takahashi:02a}).  However, in the opinion of one of us
(CSR), this is an extreme over-reaction to the failure of a
particularly simple-minded realization of the fluorescent iron line
model.  A realistic MHD turbulent disk with have a ``surface'' that,
at a local level, may be highly complex both in terms of geometry and
ionization state.  For example, it is easy to image that the
(turbulent) surface is comprised of filaments and/or sheets which have
a variety of ionization states.  The resulting response of the iron
line flux from such a complex surface to a change in the irradiating
coronal flux (even assuming no direct thermal/dynamical coupling
between the corona and the disk surface) is very unlikely to be
linear.

The on-going operation of {\it XMM-Newton} promises to make
significant progress in the near future.  In addition to MCG--6-30-15
and NGC~3516, there are a small number of other normal Seyferts for
which {\it XMM-Newton} data has already been reported
\cite{gondoin:01a,gondoin:01b,gondoin:02a,pounds:02a}.  On the basis
of these early studies, it has been suggested that the {\it ASCA}
studies may have over-estimated the generality of relativistic iron
lines in the spectra of Seyfert galaxies.  However, robust conclusions
must await a more systematic analysis of current and future {\it
  XMM-Newton} datasets, part of which would include a reconciliation
with the older yet still valid data from {\it ASCA} and {\it RXTE}.

\subsection{Very-high and very-low luminosity AGN}

>From the observations discussed above, we can deduce that the
accretion disks in most Seyfert-1 nuclei are only modestly ionized and
remain optically-thick (and capable of producing iron line features)
down to the radius of marginal stability.  Seyfert-1 nuclei provide a
useful reference against which we can compare and contrast the
accretion disks of other types of AGN.  By making such comparisons, we
can hope to learn how the properties and physics of accretion disks
change as a function of the gross observables of the system.  Here, we
examine constraints on the dependence of the accretion disk properties
on the overall radiative luminosity of the AGN.

We begin by discussing low-luminosity AGN (LLAGN), roughly defined as
sources with a total radiative luminosity of $L<10^{42}\ergps$.  Being
low-luminosity, even the nearest examples are comparatively faint
X-ray sources.  Furthermore, stellar processes in the galaxy (such as
X-ray binaries and vigorous star-formation) can produce a comparable
X-ray flux.  This galactic emission can be spatially blended with the
LLAGN emission and somewhat hamper studies.

One of the best studied LLAGN is the nearby galaxy NGC~4258
(M~106)\footnote{More precisely, this source is classified as a
  low-ionization nuclear emission region (LINER)} --- this is the same
object for which masers give an excellent constraint on the black hole
mass.  The first good constraints on the iron line in this source were
obtained with a deep {\it ASCA} observation \cite{reynolds:00c}
(although previous shorter observations had hinted at such a line
\cite{makishima:94a}).  It was found that the iron line was narrow,
implying that the bulk of the fluorescent emission originated from
$r>100M$.  Given the fairly large equivalent width of the line ($\sim
100\eV$) and the lack of any evidence (from any waveband) for cold,
optically-thick material away from the plane of the accretion disk,
this emission line very likely originates from the accretion disk
itself.  Thus, to produce such a narrow line, the X-ray emitting
corona must be extended, producing appreciable X-ray illumination of
the disk across a range of radii exceeding $r>100M$.  This is in stark
contrast to the higher-luminosity Seyfert galaxies whose broad lines
require a very compact X-ray emitting corona ($r<10M$).  This result
is consistent with a sphere+disk model for this LLAGN with a
transition radius of $r_{\rm tr}>100M$.  However, due to the limited
signal-to-noise, the presence of a weaker relativistic iron line in
addition to the narrow component could not be ruled out.  Recently it
has been reported that a rather short {\it XMM-Newton} observation of
NGC~4258 failed to detect any iron line, setting an upper limit on the
line flux to be lower than that detected by {\it ASCA}
\cite{pietsch:02a}.  Such variability suggests that the iron line does
indeed originate from the accretion disk (rather than more distant
material) and, furthermore, implies that the disk/corona system
undergoes significant changes in structure and/or ionization state on
year time scales.

X-ray studies of the general population of LLAGN have been hampered by
their comparative faintness.  The best such study to date, based upon
{\it ASCA} data, detects significant iron line emission in several
LLAGN \cite{terashima:00a,terashima:02a}.  In no object, however, is
the line emission found to be significantly broadened.  Unfortunately,
the signal-to-noise of these datasets is insufficient to draw robust
astrophysical conclusions --- relativistic iron lines such as those
seen in higher-luminosity Seyfert nuclei could readily be hiding in
the noise of these data.  However, there are excellent prospects for
characterizing any broad line component with current and future {\it
  XMM-Newton} and {\it Chandra} observations.  Such observations will
allow us to compare the properties of accretion disks (especially
ionization state and truncation radii) in LLAGN with those in normal
Seyfert nuclei.

We now turn to a brief discussion of high-luminosity AGN (HLAGN).
Again, due to the maturity of the data, much of our understanding is
still largely based on {\it ASCA} observations.  In an important
study, Nandra and co-workers studied the average {\it ASCA} spectrum
of a large sample of AGN as a function of source luminosity ranging
from normal Seyfert nuclei to powerful quasars \cite{nandra:97b}.
They found that broad iron line becomes appreciably weaker once one
considers sources with an X-ray luminosity greater than $L_X\sim
10^{44}-10^{45}\ergps$.  The most reasonable interpretation for this
trend is that the HLAGN possess more highly ionized accretion disks
--- this is expected if the luminosity of an AGN is determined
primarily by the Eddington fraction rather than the black hole mass.
Support for this picture is growing with subsequent {\it XMM-Newton}
observations.  {\it XMM-Newton} studies of the high-luminosity Seyfert
galaxies Mrk~205 and Mrk~509 (which possess an X-ray band luminosity
of $\sim 10^{45}\ergps$) find significant broadened K$\alpha$ iron
lines, with profiles well described with accretion disks models, and
energies corresponding to helium- or hydrogen-like iron
\cite{reeves:01a,pounds:01a}.  These observations likely represent
direct detections of ionized accretion disks.

Direct evidence for ionized disks have also been found in the X-ray
spectra of the so-called narrow-line Seyfert-1 galaxies (NLS1s,
\cite{vaughan:99a,vaughan:02a}).  NLS1s are a peculiar sub-class of
Seyfert-1 galaxies which are defined by their rather narrow optical
emission lines, together with very soft and highly variable X-ray
emissions \cite{boller:96a}.  While they do not possess a high luminosity
in absolute terms, it is thought that NLS1s may have rather small
black hole masses and luminosities rather close to their Eddington
limit.  The detection of highly ionized disks in these objects nicely
fits in with this scenario.

\subsection{Iron lines from radio-loud AGN : a clue to the dichotomy?}

One of the greatest mysteries in the field of AGN research is the
physical mechanism underlying the radio-quiet/radio-loud dichotomy.
Why do some black hole systems choose to launch, accelerate and
collimate powerful relativistic jets, whereas other systems do not?
The natural environment in which to form an energetic relativistic jet
is the relativistic region close to an accreting black hole.  A first
step in addressing this problem from an observational standpoint is to
compare and contrast the central engine structures of radio-quiet and
radio-loud AGN.  Since they are direct probes of the inner accretion
disk, broad iron lines provide a powerful way of facilitating such a
comparison.

Radio-loud AGN are rarer, and hence typically fainter, than their
radio-quiet counterparts.  Furthermore, many of the best candidates
for study are found in clusters of galaxies and it can be difficult to
observationally distinguish AGN emission from thermal X-rays emitted
by $10^7-10^8\K$ gas trapped in the cluster's gravitational potential
well. For these reasons, the quality of the observational constraints
are rather poorer than for Seyfert-1 galaxies.  There {\it does},
however, appear to be a difference between the iron line properties of
radio-loud nuclei and radio-quiet nuclei.  Broad iron lines, and the
associated Compton reflection continua, are generally weak or absent
in the radio-loud counterparts
\cite{eracleous:96a,allen:97a,wozniak:97a,reynolds:97f,sambruna:99a,grandi:99a,eracleous:00a}.
This effect might be due to the swamping of a normal `Seyfert-like'
X-ray spectrum by a beamed jet component (similar to the swamping of
optical emission lines in a blazar spectrum).  Alternatively, the
inner disk might be in the form of a very hot and optically-thin
radiatively-inefficient accretion flow.  Finally, the inner disk might
be radiatively-efficient and optically-thick, but not produce
prominent X-ray reflection features due to a very high ionization
\cite{ballantyne:02a}.  Again, future observations with {\it
  XMM--Newton} should be able to distinguish these possibilities by
searching for very weak broad components to the iron line and iron
edges with high signal-to-noise.

\section{Iron lines from Galactic Black Hole Candidates}

A crucial difference between GBHCs and AGN are their characteristic
time scales.  Except for the least massive AGN, the viscous time
scales characterizing the inner disk can be many tens of years.  Since
the viscous time is the timescale on which the mass accretion rate can
change appreciably, we see that it is impossible to sample a range of
mass accretion rates within a particular AGN on a human timescale.  On
the other hand, the viscous time scales in GBHCs can range from days
to weeks, so individual observations can, and likely do, represent
different intrinsic accretion rates.  The same is true for dynamical
and thermal time scales.  A typical X-ray observation lasting several
tens of thousands of seconds represents only a few thermal and
dynamical time scales in an AGN, but many thousands of such time
scales in GBHCs.  In fact, a reasonable estimate of the thermal time
scale in a stellar mass black hole corresponds to a few seconds, which
often also corresponds to the time scale of the peak variability power
in the X-ray lightcurve of such sources.

In many ways, because of these differences in characteristic time
scales, the study of stellar mass black hole candidates and AGN are
complementary. One figure of merit is the square root (since one is
typically dealing with Poisson statistics) of the flux received over a
characteristic time scale, i.e., $[(L/D^2)(GM/c^3)]^{1/2}$, where $L$
is the source luminosity and $D$ is its distance.  For sources with
the same fractional Eddington luminosity, the luminosity scales as
mass, so this figure of merit scales as $(M/D)$.  Thus, although AGN
are one thousand or more times further away than stellar mass black
holes, they are also up to one hundred million times more luminous and
therefore are much more effectively observed over their characteristic
dynamical time scales.  Stellar mass black holes, however, are much
more readily observed over multiple viscous time scales.  Ten thousand
seconds in the life of a GBHC can be equivalent to as much as a
thousand years in the life of an AGN.

\subsection{A case study of Cygnus X-1}

The GBHC Cyg~X-1 has been extensively studied by every major X-ray
satellite of the past 30 years, and it has been the focus of extensive
theoretical modeling.  It spends a large fraction of its time in a
spectrally hard, highly variable X-ray state; however, it occasionally
transits to a softer, less variable state
\cite{dolan:79a,ogawara:82a,ling:83a}.  Several such transitions to a
soft state have been observed with \textsl{RXTE}
\cite{zhang:97b,cui:96a,cui:02a}.  The geometry within each of these
accretion states, as well as the causes of the transitions among the
different states, has been the focus of much theoretical speculation
and observational study.  It is precisely these issues that studies of
X-ray lines and reflection features hope to illuminate.

Before discussing the specifics of the line studies, it is worthwhile
to review some of the suggested models for the Cyg~X-1 system, and to
briefly consider their implications for predicted line properties.  As
Cyg~X-1 spends most of its time in the hard state, models have focused
on explaining this state.  Soon after the discovery of hard state
black hole candidates, it was suggested that these spectra were due to
Comptonization of soft photons in a hot, low optical depth plasma
\cite{thorne:75a}, and such spectra were soon successfully applied to
the hard state of Cyg~X-1 \cite{sunyaev:79a,liang:79a}.
Comptonization models have invoked, variously, slab geometries (and
their variations; see Fig.~\ref{fig:coronal_geom}), sphere+disk geometries,
and have posited soft seed photons coming from either the optically
thick, geometrically thin disk or from synchrotron photons generated
within the corona itself.

For the slab or sandwich geometries, it is expected that the generated
reflection fraction will be of order unity (half the coronal flux is
intercepted by the disk), and thus iron line equivalent widths should
also be large ($\sim 200\eV$).  If the corona extends inwards toward
small disk radii, then relativistic effects will also be prominent.
The `pill box' geometry preserves the reflection and line features of
this geometry, but reduces the amount of Compton cooling of the
corona.  `Sphere+disk' geometries were in part specifically considered
since, like the pill box, these coronae are less Compton cooled than
slab models.  In contrast to the pill box models, they also have low
reflection fractions ($f\approx0.3$) and produce weak iron line
features.  The degree to which relativistic effects are prominent in
such line features is dependent upon the transition radius between
corona and disk.  ADAF models essentially adopt the `sphere+disk'
geometry; however, their hard spectra tend to be more centrally
concentrated, and they often posit large transition radii between
inner corona and outer disk.  Thus, although not a necessary feature
of ADAF models in general, many specific realizations of ADAF models
produce very small reflection fractions, as well as very weak and
narrow fluorescent iron line features.

The observational nature of the reflection and line features observed
in Cyg~X-1 has been controversial.  In order to illustrate how
technology, theoretical prejudice and sociology influence our
understanding of such a complex system, we will present a detailed
historical account of the iron line in Cyg~X-1. The first reported
detection of a broad Fe line feature in Cyg X-1 was made by Paul Barr
and collaborators using data from the {\it EXOSAT} Observatory
\cite{barr:85a}.  The line was seen to be of moderate strength (an
equivalent width of 120\,eV), have a slightly redshifted peak from
that of neutral iron (6.2\,keV, as opposed to 6.4\,keV), and be fairly
broad (Gaussian width of $\sigma = 1.2$\,keV).  This line, however,
was \emph{not} interpreted as indicating a relativistic profile.
Instead, it was interpreted as due to Compton scattering in a
moderately optically thick ($\tau_{\rm es} \sim 5$), low temperature
($\sim 5$\,keV) corona.  Scattering in such a corona would indeed both
broaden and slightly redshift an intrinsically narrow line.  It was
further hypothesized that the line was predominantly due to
recombination radiation in a highly ionized corona, as opposed to
being due to fluorescence radiation.

Several years later, however, these exact same observations were
reinterpreted as relativistic broadening of a fluorescent line by Andy
Fabian and collaborators \cite{fabian:89a}.  In fact, this was the
first attempt to claim such a line in any X-ray source.  It was noted
that, due to the poor intrinsic resolution of {\it EXOSAT} in the iron line
region ($\Delta E/E> 10\%$), the characteristic ``double horned''
features of relativistic lines would be smeared out, regardless of the
inclination, and yield the broad, Gaussian feature previously fit
\cite{barr:85a}. This interpretation was challenged, however, by prior
observations performed with the Tenma X-ray satellite
\cite{kitamoto:90a}.  These observations, carried out in 1983 and
spanning several phases of the Cyg X-1 binary orbit, revealed a weaker
line (equivalent width 60-80\,eV) that furthermore was consistent with
being somewhat narrower ($\sigma < 1$\,keV) and being dependent upon
orbital phase.  This suggested, instead, fluorescence from the surface
of the companion (donor) star, not the black hole accretion disk.  A
broader component emanating from the inner disk regions, however,
could not be ruled out by the data.

As we have discussed extensively, a fluorescent iron line is produced
along with other spectral signatures --- one also expects to observe
the iron K edge, as well as Compton recoiled photons at higher energy.
When these features were added to the models of {\it EXOSAT} data (as well
as to models of earlier HEAO 1-A2 data), Christine Done and
collaborators obtained fits that yielded reflection fractions of
$\Omega/2\pi \sim 0.5$ and \emph{narrow} Gaussian lines with
equivalent widths $\sim 40$\,eV \cite{done:92a}. These results were
confirmed by observations performed by BBXRT, which was an X-ray
telescope with a resolution $E/\Delta E\sim 30$ that was flown aboard
the space shuttle in 1990.  Specifically, it was found that the
inclusion of a reflection component from material with twice the solar
abundance of iron obviated the need for anything other than a weak,
narrow iron line \cite{marshall:93a}.  This model was then also
applied to {\it ASCA} data by Ken Ebisawa and collaborators, and, again,
when including an iron edge due to reflection, a narrow, weak iron
line was found \cite[although see the further discussion
below]{ebisawa:96b}.

Unambiguously constraining spectral models that include line features
(at 6\,keV), reflection features (at 10--30\,keV), and thermal
Comptonization features (especially the high energy cut-off at
$\approx 100$\,keV), requires broad band data.  Such a data set for Cyg
X-1 was obtained in 1991 with a simultaneous Ginga
($\approx2$--30\,keV) and OSSE ($\approx 50$--1000\,keV) observation.
{\it Ginga}, similar to {\it EXOSAT}, had poor spectral resolution ($E/\Delta E
\approx 10$). Marek Gierlinski and collaborators described these
observations \cite{gierlinski:97a} with models comprised of low
reflection fractions ($\Omega/2\pi =0.2-0.5$), and narrow iron lines
with moderate equivalent widths (EW$=90$--140\,eV).  As described in
\S\ref{sec:coronae}, there was a growing realization that in order for
coronal models to yield very hard spectra with thermal Comptonization
cutoffs at very high energy, photon starved geometries were required
\cite{dove:97a,dove:97b,poutanen:97b}.  A sphere+disk geometry, for
example, could describe the Cyg X-1 data, and furthermore would imply
a narrow iron line if the transition between the coronal sphere and the
outer thin disk occurred at sufficiently large radius
\cite{gierlinski:97a,dove:97b,poutanen:97b}.

It has been the application of more sophisticated coronal and
reflection models (with the latter including such effects as
ionization of the disk atmosphere and relativistic smearing of the
reflection features), coupled with the advent of the new generation of
X-ray telescopes ({\it RXTE}, BeppoSAX, {\it Chandra}, {\it XMM-Newton}), that has
led researchers to reconsider the possibility of a relativistically
broadened iron line in the spectrum of Cyg X-1.  An early observation of
Cyg X-1 with {\it RXTE} by James Dove and collaborators gave ambiguous
results in this regard \cite{dove:98a}.  The spectrum above $\approx
10$\,keV was extremely well fit by a pure power law with an
exponential cutoff, without any signs of spectral curvature associated
with Compton recoil photons from a reflection model.  Extrapolating
the exponentially cutoff power law to energies below 10\,keV, however,
yielded a strong, power law-like excess. This excess could have been
produced by three effects: a mismatch in the calibration of spectral
slopes between the two instruments that comprise {\it RXTE} (PCA and HEXTE;
PCA typically yields softer fits than HEXTE to the power-law spectrum
of the Crab nebula and pulsar \cite{wilms:99aa}), reflection of a
continuum spectrum more complicated than the presumed exponentially
cut-off power law, or a combination of an unmodeled soft excess and a
strong, broad line.  In retrospect, it was likely that all three
effects were playing a role in these initial results.  The same data,
modeled with a sphere+disk coronal model (which included reflection of
the Comptonized spectrum with an effective covering fraction of
$\Omega/2\pi \approx 0.3$), fit the PCA and HEXTE spectra
simultaneously; however, such models yielded strong, broad residuals
in the iron line region \cite{dove:98a}.  At that time, these
line-like residuals could not be attributed definitively to a
relativistically broadened feature owing to the then poorly known {\it RXTE}
calibration near energies of $\approx 6$\,keV \cite{dove:98a}.

With further improvements to the calibration of the {\it RXTE} response in
the 6\,keV region of the spectrum, numerous fits to {\it RXTE} observations
of Cyg~X-1 have continued to indicate strong, broad line features.
For one set of (predominantly hard state) {\it RXTE} observations of
Cyg~X-1, Marat Gilfanov and collaborators used a crude Gaussian
convolution of a reflection model to simulate relativistic effects
\cite{gilfanov:99a}.  It was found that both the line (equivalent
widths $\approx 70$--160\,eV) and reflection features ($\Omega/2\pi
\approx 0.3$--0.7) were well-described by a Gaussian smearing width
$\sigma \approx 0.3$--0.9\,keV.  Such large smearing widths would be
consistent with relativistic motions and gravitational redshifts in
the inner disk.

The {\it EXOSAT}, {\it Ginga}, and {\it ASCA} data were later re-examined by Christine
Done and Piotr Zycki with a more complex set of reflection models
\cite{done:99a}.  Similar to prior analyses \cite{done:92a}, these
models included reflector ionization using the constant density
assumption; however, they also included the effects of relativistic
distortions.  Furthermore, the models considered a non-relativistic
(narrow) set of line/reflection features in conjunction with the
relativistically smeared line and reflection features.  It was found
that all the datasets could be well-fit by models wherein the disk
ionization was not very high, and the relativistically smeared line
and edge features dominated ($\Omega/2\pi \approx 0.1$--0.2) over the
non-relativistic features ($\Omega/2\pi< 0.05$) \cite{done:99a}. Andy
Young and collaborators reanalyzed these datasets with an independent
model and affirmed the description with a large degree of relativistic
smearing of the line and reflection features; however, this latter
model postulated a larger reflection fraction ($\Omega/2\pi \approx
1$), albeit with a higher degree of ionization \cite{young:01a}.

Analyses of BeppoSAX observations of the Cyg X-1 hard state also were
well described by broad line features \cite{disalvo:01a,frontera:01a}.
BeppoSAX had extremely broad energy coverage from $\approx
0.5-200$\,keV; therefore, it was capable of simultaneously
constraining models of the distribution of seed photons for
Comptonization (e.g., the disk with temperatures
$kT\approx0.1$--2\,keV), the line/reflection region (6--30\,keV), and
the hard, Comptonized tail (10--200\,keV). As for the re-analyses of
{\it EXOSAT}, {\it Ginga}, and {\it ASCA} data \cite{done:99a},
analyses of BeppoSAX by Frontera and collaborators revealed a
relativistically smeared reflected power law ($\Omega/2\pi \approx
0.1$--0.3), with the degree of smearing being indicative of the
reflection being dominated by the innermost regions of the accretion
disk \cite{disalvo:01a}.  Similarly, sophisticated Comptonization
models, including the effects of (non-relativistic) reflection
($\Omega/2\pi \approx 0.25$), also required the presence of a strong
(equivalent width $\approx 350$\,eV), relativistically broadened line
\cite{frontera:01a}.

One can see that there has been something of a split, with the
prevailing wisdom prior to 1997 being dominated by the ``narrow''
features point of view, while the prevailing wisdom post-1997 seems to
be dominated by the ``broad'', i.e., relativistic, features point of
view (with some researchers having been on both sides of the issue).
Has this shift been one of sociology, improved theoretical modeling,
improved observational data, or some other effects?  Improved models
have certainly played an important role.  For example, in order to
explain the {\it ASCA} data at the lowest energies, Ken Ebisawa and
collaborators invoked a broken power law (with a softer, i.e. steeper,
low energy slope and a break energy at $\approx 4$\,keV) in addition
to the disk component with peak temperature of $kT \approx 150$\,keV
\cite{ebisawa:96b}.  The presence of such a low energy ``soft excess''
(above the requirements of the disk component) in the spectrum of Cyg
X-1, as well as other hard state black hole spectra, continues to be
debated \cite{done:99a,disalvo:01a,frontera:01a}.  It has been argued,
however, that the apparent hardening at energies above 4\,keV in fact
might be due to the red tail of the relativistically broadened iron
line \cite{nowak:02a}.  The combination of simultaneous broad band
data with models that attempt to (at least somewhat self consistently)
model this entire range leaves less leeway to introduce
phenomenological model components (e.g., a spectral break at 4\,keV)
that might remove evidence of a relativistically broadened line.  It
is worthwhile noting that Comptonization models
\cite{dove:98a,frontera:01a} seem to require the broad line, even when
including the other effects of reflection.

The above discussion highlights the complexity inherent in stellar
mass black hole spectra.  Unlike the case of AGN, it is clear that a
simple, exponentially cutoff power law is insufficient to adequately
describe the underlying X-ray continuum.  This is especially true at
soft energies, as the hottest GBHC accretion disks ($kT \approx
2$\,keV) can have significant spectral contributions into the red tail
region of any broadened iron line.  A further complication arises from
the fact that stellar mass black hole disks likely represent a wide
range of disk inclinations.  It is typically thought that Seyfert 1
disks are viewed reasonably close to face-on (see
Fig.~\ref{fig:agn_geom}); therefore, the gravitational redshift and
transverse Doppler shifts typically cause the blue wing of a broadened
K$\alpha$ line to lie at energies below its associated ionization
edge.  This is not the case with potentially more highly inclined
stellar mass black hole systems.  There, especially if one
misidentifies the continuum power law slope, it is much easier to
mis-model the blue wing of such a line with a reflection edge.

\begin{figure}
\centerline{
\psfig{figure=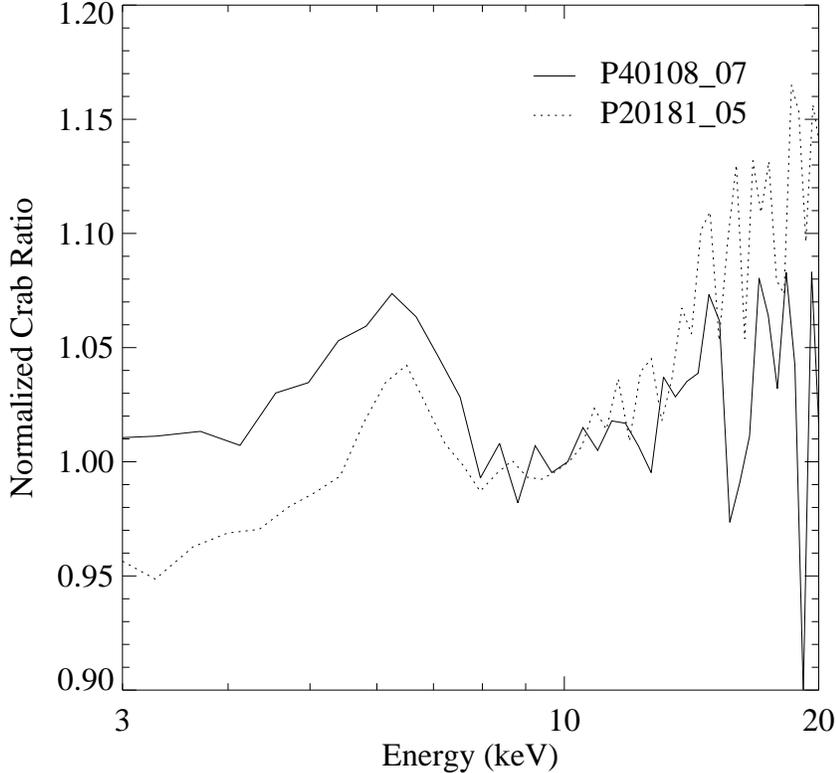,width=0.8\textwidth}
}
\caption{Ratio of {\it RXTE} observations of GX339$-$4 to {\it RXTE}
  observations of the Crab nebula and pulsar, normalized to unity at
  10\,keV.  The spectrum of the Crab nebula is widely believed to be a
  featureless power-law continuum.  Showing the observations this way
  gives an indication of the intrinsic resolution of the detectors, as
  well as minimizes the worry that the line features are artifact of a
  poorly determined detector response.  Note that despite the fact
  that these two observations have nearly identical fluxes and average
  spectral slopes, their line profiles are different.}
\label{fig:gx339_rxte_crab}
\end{figure}

Most importantly, due to their extremely short viscous and thermal
time scales (compared to AGN), the environment of the accretion flow
in a stellar mass black hole system can be very different from
observation to observation.  The existence of ``state changes''
clearly points toward this; however, even within a given state large
changes in the spectral properties are observed \cite[for
example]{zdziarski:99a,gilfanov:99a,pottschmidt:02a}.  Thus, some of
the seemingly different results quoted above are undoubtedly due to
changes in the structure of the accretion flow between observations
(e.g., changes in the ionization state, the disk inclination to our
line of site, the disk/corona geometry, etc.).  These changes, in
fact, can be sometimes associated solely with the line profile.  As an
example, in Fig.~\ref{fig:gx339_rxte_crab} we show two {\it RXTE}-PCA
observations of the stellar mass black hole candidate GX339$-$4 in its
hard state \cite{nowak:02a}.  To minimize features due to
uncertainties in the calibration and response of {\it RXTE}, the data are
presented as a ratio to {\it RXTE} observations of the Crab nebula and
pulsar (which are thought to possess a synchrotron hard X-ray spectrum
well approximated by a featureless power-law).  Even though these two
observations represent nearly the same flux, power law slope, and
(low) fitted reflection fraction, one exhibits a broad line residual,
while the other one exhibits a line residual consistent with being
narrow.

\begin{figure}
\centerline{
\psfig{figure=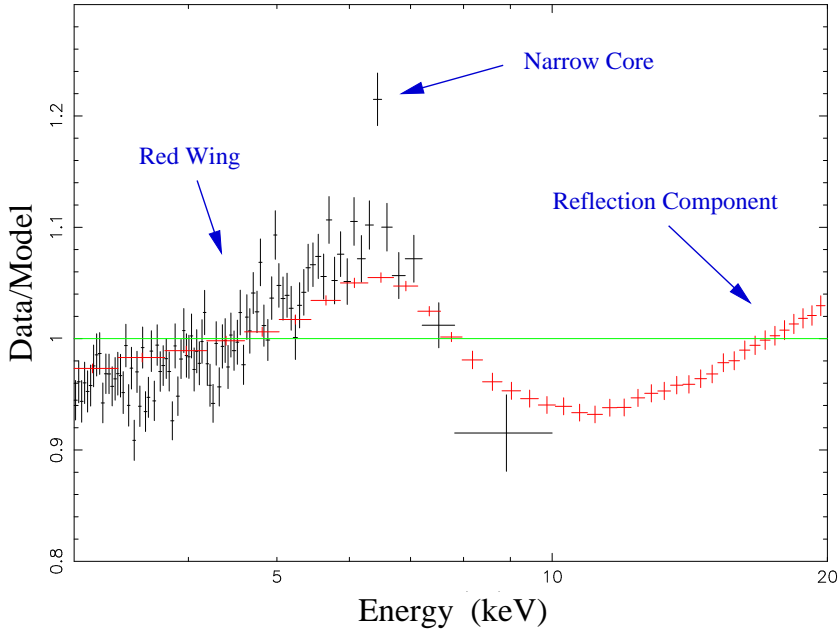,width=0.8\textwidth}
}
\caption{Simultaneous {\it Chandra} and {\it RXTE} observation of Cygnus X-1, fit
with a simple power law model (J. Miller, priv. comm.; see also
\protect\cite{miller:02a}).  The {\it Chandra} data have been rebinned to
show both the narrow core at 6.4\,keV, as well as the broad component.
The broad component in {\it Chandra} agrees well with that seen in {\it RXTE}.
{\it RXTE} data further show an upturn at high energy likely due to
reflection.}
\label{fig:cygx1_cxo_rxte}
\end{figure}

A final important example of spectral variability in stellar mass
black hole systems is that due to orbital variation.  A binary orbital
dependence for the strength of the iron line has been previously
reported for Cyg~X-1, which was taken as evidence that the line was
due to fluorescence by the secondary, not by the relativistic regions
of the inner disk \cite{kitamoto:90a}.  Recent {\it Chandra} observations,
however, have dramatically demonstrated that the line region in fact
is comprised of both narrow and broad components
\cite{miller:02a,schulz:02a}.  For the first time, instruments exist
with high enough spectral resolution to definitively resolve the
narrow component of the line; however, as shown by Jon Miller and
collaborators, the {\it Chandra} sensitivity and energy coverage also
convincingly reveal the broad component of the line. In
Fig.~\ref{fig:cygx1_cxo_rxte} we show the profile, comprised of both
narrow and broad features, obtained with the {\it Chandra} observations
performed by Miller et al.  Further {\it Chandra} observations reveal that
the the narrow line component is not always present in the data
(H. Marshall 2001, priv. comm.; see also \cite{marshall:01a}).  The
narrow line has an equivalent width of only 80\,eV, while the
parameters of the broad line component (equivalent width $\approx
140$\,eV) agree well with those previously found with, e.g., {\it RXTE}
observations \cite{miller:02a}.  Thus, whereas the presence of a
(variable) narrow line component complicates analyses performed with
broader resolution instruments, it apparently does not obviate the
need for a broad line to be present.

All of the above observations refer to spectrally hard states (often
called ``low'' to ``intermediate'' states); however, Cyg~X-1
occasionally transits into a much softer state (often called a ``high
state'').  Although rarer, observations of this soft state have been
performed with {\it ASCA}, {\it RXTE}, BeppoSAX
\cite{cui:96a,dotani:97a,gierlinski:99a,gilfanov:99a,frontera:01a},
and more recently {\it Chandra}.  All of the above caveats about determining
line profiles in the hard state apply to the soft state as well. We
further note that due to the increased strength and temperature of the
soft ``disk'' component (with maximum temperature $kT \approx
300$\,eV), continuum modeling near the red wing of the line region is
even more difficult.  Given these caveats, however, several
researchers have claimed increased reflection fractions ($\Omega/2\pi
\approx 0.6$--1.5), and increased line widths ($\sigma \approx
1$--2\,keV), for moderately strong lines (equivalent widths $\approx
80$--190\,eV) \cite{gierlinski:99a,frontera:01a}.  There have even been
claims for evidence that the line energy increases to $\approx
6.7$\,keV, i.e., consistent with heavily ionized Fe, in the soft state
\cite{cui:96a}.  

\subsection{Other GBHC and Spectral Correlations}\label{sec:other}

\begin{figure}
\centerline{
\psfig{figure=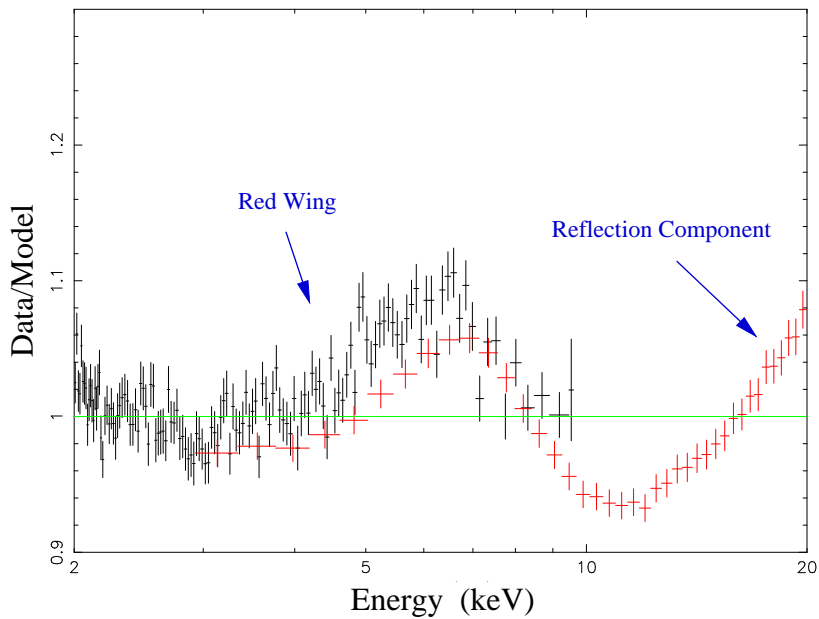,width=0.8\textwidth}
}
\caption{Simultaneous {\it Chandra} (rebinned to show broad features) and
{\it RXTE} observation of GRS~1650$-$500 (courtesy J. Miller; see also
\cite{miller:02b}) fit with a power-law and a soft disk component.
The residuals of this fit show a broad component in the Fe line
region, and an upturn at high energy, possibly due to reflection.}
\label{fig:grs1650_line}
\end{figure}

Evidence continues to accumulate that many GBHC systems exhibit broad
Fe lines, both in their hard and soft spectral states.  As discussed
above, and as shown in Fig.~\ref{fig:gx339_rxte_crab}, the hard state
of GX~339$-$4 can show a broadened line with a substantial red wing.
Additionally, during a transition from its soft state into its hard
state, GX~339$-$4 exhibited a broad line that was as skewed redward as
the line observed in MCG$-$6-30-15 during its DM state
\cite{nowak:02a}.  Similarly strong, broad, very redward-skewed lines
have been detected by J\"orn Wilms and collaborators in {\it RXTE}
observations of the soft states of LMC~X-1 and LMC~X-3; however, as
these objects are somewhat faint (and given some of the systematic
uncertainties described above) these authors could not unambiguously
describe the line properties \cite{wilms:01a,nowak:01a}.  Observing
the somewhat brighter soft state of GRS~1650$-$500 with {\it XMM-Newton}
(which has lower background and greater spectral resolution than
{\it RXTE}), Jon Miller and collaborators found very good evidence for an iron
line qualitatively and quantitatively (in terms of physical width,
equivalent width, emissivity index, etc.) similar to the DM line of
MCG$-$6-30-15 \cite{miller:02b}.  This line was also taken as evidence
for spin-energy extraction from a rotating black hole.  In
Fig.~\ref{fig:grs1650_line}, we show further observations by Miller
and collaborators, performed simultaneously with {\it Chandra} and {\it RXTE}
(J. Miller, priv. comm.), which yield independent evidence for such a
strong, broad line.  If, as discussed previously, soft state spectra
represent disks extending down to, or even within, their marginally
stable orbit, barring the optical-thick material being completely ionized, one
naturally expects to find the most broadened lines within these
states.

Broadened iron lines have been detected with BeppoSAX observations of
{V4641~Sgr} \cite{miller:02c} and {GRS~1915+105}
\cite{martocchia:02a}.  Both of these sources have exhibited strong
radio ``jet ejection events'' with apparent super-luminal motion and
inferred velocities of ~$0.9\,c$ \cite{hjellming:00a,mirabel:94a}.  As
radio and X-ray studies continue to explore the physics of jets in
these sources, and as X-ray spectroscopy continues to probe the
innermost regions of the accretion flow near the event horizon, there
is the hope that these observations can be combined to reveal answers
to such fundamental physical issues such as the connection between
black hole spin and jet production.  In this regard, GBHCs provide a
unique advantage over AGN systems.  In a number of these systems,
X-ray variability reveals high-frequency, stable quasi-periodic
oscillations (QPO), such as the 67\,Hz oscillation in GRS~1915+105
which is stable in frequency over a factor of a few in source flux
\cite{morgan:97a}, or the 450\,Hz QPO observed in GRO~J1655$-$40
\cite{strohmayer:01b}.  Although there is great debate as to the
correct model for these oscillations, there is a general consensus
that the high, stable frequencies point towards a direct relationship
to the gravity of the black hole in the ``strong field'' regime of
GR. The oscillations therefore may act as independent probes of black
hole mass and spin
\cite{nowak:97a,stella:98a,stella:99a,abramowicz:00a,wagoner:01a}.  It
is hoped that models of variability can be combined with models of
broadened lines to yield even tighter constraints on black hole mass
and spin.

To what extent are GBHC and SMBH line and reflection profiles the
same? Can observations of one class lead to inferences about the
other? Given that the two systems complement each other in terms of
spectrally resolvable time scales (i.e., the dynamical time scales of
AGN are readily observed, while multiple viscous time scales can be
observed in GBHC), commonalities between the two could prove very
powerful in constraining the underlying physical mechanisms. Recently,
Andrzej Zdziarski and collaborators have claimed such a correspondence
between the two classes.  Specifically, they describe a correlation
between the spectral slope of a fitted power law with the amplitude of
reflection, with softer spectra yielding greater reflection fractions
\cite{zdziarski:99a}. The observations that made up this correlation
consisted of both AGN and GBHC, mostly observed in relatively hard
spectral states. (Ueda and collaborators had previously described such
a correlation with four {\it Ginga} observations of GX~339$-$4
\cite{ueda:94a}.) Considering solely a sample of Seyfert galaxies,
Piotr Lubinski and Zdziarski claimed that the correlation could also
be carried over to fits of the line region as well, with Seyfert lines
being composed of a fixed equivalent width, narrow core line, plus a
broad, variable line whose equivalent width increased with spectral
softening of the continuum \cite{lubinski:01a}. As discussed
previously for the case of NGC~5548, the correlations between line
strength and spectra can be quite complex, with that particular case
showing a decreasing line equivalent width (i.e., a constant line
flux) with a softening and brightening of the spectrum.

Further questions about this possible correlation arose over two
issues.  First, possible systematic correlations exist between the
fitted power laws and reflection fraction. An increased reflection
fraction phenomenologically hardens a soft power law, therefore, error
bars for the two parameters naturally show such a correlation for a
fixed spectrum. Second, the effect had been shown convincingly only by
combining a large number of observations of different objects. Both of
these objections have been alleviated to some extent by considering
multiple observations of \emph{single} objects that represent a wide
range of flux and spectral slopes.  Furthermore, these observations
have been fit with a variety of models (that bear in mind possible
systematic errors), all of which yield the basic correlation
\cite{gilfanov:99a,revnivtsev:01a,nowak:02a}.  The ability to observe
GBHC over a very wide variety of accretion rate states has proven
crucial for such studies.

One suggestion for how the correlation is produced was described by
Andzrej Zdziarski and collaborators (elaborating upon earlier
suggestions of Juri Poutanen and collaborators, \cite{poutanen:97a}),
and it bears upon the coronal models described in \S\ref{sec:coronae}.
It was suggested that there is a certain amount of overlap between the
quasi-spherical corona and the outer, thin accretion disk.  As this
disk moves inward, the corona becomes more effectively Compton cooled
and produces softer spectra, while the increase of overlap between
disk and corona yields stronger lines and reflection features
\cite{zdziarski:99a}.  As for the case of NGC~5548, however, the
details of the overall correlation are more complicated than any
simple model.  Analyzing {\it RXTE} observations of Cyg~X-1 and GX~339$-$4,
Marat Gilfanov, Michael Revnivtsev, and Eugene Churazov used a crude
Gaussian convolution of a reflection model (to mimic relativistic
effects) to deduce a very strong correlation between spectral slope
and amplitude of reflection \cite{gilfanov:99a,revnivtsev:01a}.
Furthermore, they found that the width of the Gaussian smearing (or
equivalently, the amplitude of the relativistic distortions) increased
as the spectrum softened and the reflection amplitude increased.  This
would be consistent with the corona-disk overlap model described
above.

\begin{figure}
\centerline{
\psfig{figure=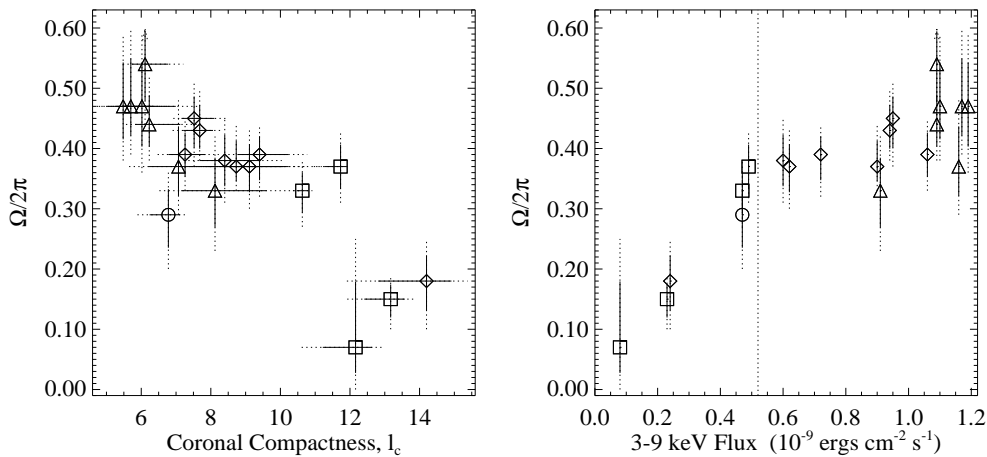,width=1.0\textwidth}
}
\caption{Fits to multiple {\it RXTE} observations of the galactic black hole
candidate, GX~339$-$4.  Left: Reflection fraction, $\Omega/2\pi$
vs. ``coronal compactness'' (i.e., spectral hardness, with higher
compactness indicating harder spectra).  Right: Reflection fraction
vs. 3-9\,keV flux, with the vertical line being the lower flux
boundary of the regime of hysteresis in the spectral properties (see
text) \cite{nowak:02a}.}
\label{fig:gx339_rxte}
\end{figure}

Other analyses of the same observations of GX~339$-$4, however,
yielded different results \cite{nowak:02a}.  Instead of fitting a
power law, a more realistic Comptonized spectrum was employed
(hardness was instead described by a ``coronal compactness
parameter'', $\ell_c$, with larger compactnesses corresponding to
harder spectra).  As shown in Fig.~\ref{fig:gx339_rxte}, although the
overall correlation was found, it was not as strong as described by
the models of Revnivtsev et al. \cite{revnivtsev:01a}.  Furthermore, a
slightly better correlation was found between amplitude of reflection
and soft X-ray flux \cite[and
Fig.~\ref{fig:gx339_rxte_crab}]{nowak:02a}. The reflection parameters
were also found to be fairly ``flat'' in a regime of soft X-ray flux
corresponding to hysteresis in the source.  This regime corresponds to
fluxes where the source appears spectrally hard if it had already been
spectrally hard, and the source appears soft if it had already been
spectrally soft \cite{miyamoto:95a}.  Additionally, although a weak
correlation between the width of the line and amplitude of reflection
was found, this correlation was weakened further if ionized reflection
models were instead considered \cite{nowak:02a}.

Clearly, the exact correlations between broad continuum spectrum,
reflection, and line properties, are extremely complicated, both
physically and observationally. The existence of such correlations,
however, are now more widely accepted as being both real and very
important.  Our understanding of these correlations are likely to grow
as both our theoretical models improve (especially as regards our
ability to describe highly ionized atmospheres) and as the
observations continue to be refined with ever more sensitive
instruments.

\section{Future prospects for iron line studies}

This is a technology-driven field whose advances will parallel those
in astrophysical X-ray instrumentation.  While we will avoid the
temptation for lengthy speculation of what we will find, it is
important to say a few words about some of the science issues that
will be opened up by further observations.

For the next few years, most of the progress is likely to be made by
the continuing operation of {\it XMM-Newton} and {\it Chandra} (with
these instruments likely often being assisted by simultaneous {\it
RXTE} observations to constrain the high energy spectra).  The
accumulated body of data from these observatories will allow a
detailed categorization of the X-ray reflection properties of all
classes of accreting black hole in the nearby universe.  By applying
detailed considerations similar to those described in
\S\ref{sec:mcg6}, such data will allow the first observational
investigation of black hole spin in a sample of both GBHCs and AGN.
Investigation of variability in the iron line profile or strength can
also be studied in more detail than ever possible before.  In the case
of AGN, line variability can be studied on time scales down to $\sim
10$ orbital periods (at which point the spectra become too
photon-starved to provide useful constraints) --- line variability
seen on these time scales most likely corresponds to spatial changes
in the X-ray emission from the disk corona.  These observatories will
also provide the first constraints on the astrophysical environment of
accreting SMBHs at very high redshift.

After launch in 2005, {\it Astro-E2} will complement {\it XMM-Newton}
and {\it Chandra}.  As well as being able to study broad iron lines at
high signal-to-noise even when observing alone, its ability to provide
very high resolution spectra in the iron line band (thereby completely
characterizing any low-velocity components to the iron line) will be a
powerful complement to any {\it Chandra} and {\it XMM-Newton}
observation.  Even with high-quality data in the iron line band, good
constraints on ionized disks greatly benefit from having simultaneous
hard X-ray ($>10$\,keV) observations of the Compton reflection
continuum.  At present, only the Proportional Counter Array (PCA) on
{\it RXTE} can obtain such data.  In the near future, our ability to
constrain the continuum shape above 15\,keV will be greatly enhanced
by the recent successful launch of ESA's {\it Integral} (October 2002)
and the future launch of NASA's {\it Swift} (September 2003).

At the end of this decade or the beginning of the next, it is hoped
that the {\it Constellation-X} and {\it XEUS} observatories will come
on-line.  With their very large collecting areas, they will represent
a major leap in sensitivity.  These observatories will make at least
two major advances in AGN research.  Firstly, we will be able to
produce good X-ray spectra of the highest redshift AGN currently
known.  This opens the real prospect of being able to constrain the
cosmic history of SMBH properties as well as the chemical history of
the matter in the vicinity of the SMBH.  Secondly, we will be able to
study spectral variability on time scales of the order of the light
crossing time of the black hole.  By searching for detailed changes in
the iron line profile, we will be able to search for the
``reverberation'' of X-ray flares, i.e., the X-ray light echo that
propagates across the accretion disk due to the finite speed of light
\cite{stella:90a}.  These reverberations signatures encode detailed
information about the space-time geometry, and might allow for a
quantitative test of General Relativity in the very strong field limit
\cite{reynolds:99b,young:99a}.  The proposed timing capabilities of
{\it XEUS} could also make a major impact in our understanding of GBHCs.  In
particular, we will be able to study the temporal power-spectrum of
the X-ray variability in exquisite detail.  The ultimate goal is to
use the form of these power spectra and, in particular, the families
of quasi-periodic oscillations that are known to exist
\cite{dimatteo:99a,strohmayer:01b,belloni:02a}, in order to study
accretion disk physics and the space-time geometry.

On a longer time scale (maybe two decades), we look forward to
ultra-high resolution imaging of black hole systems.  The MAXIM
concept (\S\ref{sec:maxim}) will allow us to image nearby accreting
SMBHs (e.g., the SMBHs in our Galactic Center, M87 or nearby Seyfert
galaxies) on spatial scales comparable to, or smaller than, that of
the event horizon.  In some versions of this concept, one will also be
able to map out the spectrum of the system (including disk reflection
features) across the image.  This will clearly lead to major advances
--- the geometry and physical state of the disk, corona and jet can be
inferred directly from such an image.  Combining the spectral
information, the dynamics of the relativistic disk can be measured and
compared with that expected from material orbiting in a Kerr metric.
In particularly clean systems (e.g., Seyfert galaxies) these
measurements could be used as a test of the Kerr metric itself --- such
a study would be complementary to the Kerr-testing experiments that
will be possible using gravitational wave signatures with the Laser
Interferometry Space Antenna (LISA).

\section{Conclusions}

The past few years have seen a shift in the mind-set of many
observationally-oriented black hole researchers.  A small number of
important observations (the mass of the compact object in the binary
star system Cyg~X-1, stellar motions in our Galactic Center, the gas
disk in M87, and the maser disk in NGC~4258) have led most astronomers
to the conclusions that black holes do indeed exist beyond any
reasonable doubt.  With this established, the interest has shifted to
the demographics of black holes in the Universe, and the detailed
astrophysics of black hole systems.

As we have described, X-ray spectroscopy currently provides the best
understood method of exploring the astrophysical environment in the
immediate vicinity of an accreting black hole.  The accretion disk is
the engine that drives all observable phenomena from accreting black
hole systems.  In many systems, the surface layers of the accretion
disk are expected to produce X-ray spectral features (so-called X-ray
reflection signatures) in response to external hard X-ray illumination
from a disk-corona --- the most prominent spectral feature is often
the fluorescent K$\alpha$ emission line of iron.  This line, which has
an intrinsically small energy width, will be dramatically broadened
and skewed by the rapid orbital motions of the accreting material and
strong gravity of the black hole.  These relativistic spectral
features can be identified {\it unambiguously} in many of the AGN for
which X-ray data of sufficient quality exist.  Similar spectral
features can also be found in many AGN and GBHCs for which poorer
quality data exist --- while it may not be possible to rule out other
forms of spectral complexity in these objects, relativistic disk
features are often the most compelling and physical of the
possibilities.  Given how generic these features are, they provide a
powerful way to study the near environment of accreting black holes.

Moderate luminosity radio-quiet AGN, the Seyfert galaxies, present the
cleanest examples of these spectral features.  Analysis of X-ray data
from {\it ASCA} and, more recently, {\it XMM-Newton} finds evidence
for a rather cold accretion disk extending all of the way down to the
radius of marginal stability around the black hole.  In one case
(MCG--6-30-15), the data argue strongly that the black hole is
rapidly-rotating.  Furthermore, there may be suggestions that the
accretion disk is torqued by processes associated with the spinning
black hole and, in particular, may be tapping into the rotational
energy of the spinning black hole. The case of MCG--6-30-15, which we
discussed in some detail, illustrates the exotic black hole physics
that can be addressed via these techniques.
 
The difference in iron line properties between Seyfert galaxies and
other types of AGN allows us to probe the physical state of the
accretion flow as a function of, in particular, the mass accretion
rate.  The fact that higher-luminosity AGN display weaker spectral
signatures is very likely due to increasing ionization of the disk
surface as the accretion rate (relative to that rate which would
produce the Eddington luminosity) is increased.  The weakness of the
X-ray reflection displayed by radio-loud AGN and low-luminosity AGN is
much less certain (primarily due to the paucity of high quality
datasets).  Possibilities include ionization of the disk surface,
dilution of the disk spectrum by X-ray emission from a relativistic
jet, and/or the transition of the disk into a radiatively-inefficient
state (which would be extremely hot and optically-thin thereby
producing no X-ray reflection features).

Although it is somewhat counter-intuitive given that they can be
three orders of magnitude brighter than AGN, it has been much more
difficult to study relativistic disk signatures in GBHCs.  In addition
to the fact that the X-ray spectra of GBHC are inherently more complex
(since the X-ray band contains the thermal disk emission and the disk
is almost always strongly ionized), the high X-ray fluxes from these
sources can readily overwhelm sensitive photon counting spectrometers.
However, with {\it Chandra} and {\it XMM-Newton} we can now study
relativistic X-ray reflection in GBHCs free of these instrumental
constraints, at medium-to-high spectral resolution, and high
signal-to-noise.  Consequently, relativistic iron lines have been
found in several GBHCs with a variety of ``spectral states''.  With
further study of these features over the coming few years, we should
be able to constrain the gross geometry of the accretion flow as a
function of spectral state.

The future is bright.  Until just a few years ago, the physics of
relativistic accretion disks and the astrophysics of black hole spin
was firmly in the realm of pure theory.  Modern X-ray spectroscopy
has given us a powerful tool with which we can peer into this exotic
world.  There is still much to be learnt --- much of the physics
determining the properties of accretion disks, the demographics of
black hole spin (which is closely related to black hole formation and
history), and the prevalence of spin-energy extraction (which is of
fundamental interest) are just a few of the open questions.  Continued
investigation with increasingly capable instruments promise to answer
these, and many more, questions.

\section*{Acknowledgments}

The authors thank Philip Armitage, Mitchell Begelman, Andy Fabian,
David Garofalo, Kazushi Iwasawa, Julia Lee, Herman Marshall, Cole
Miller, Jon Miller, Norbert Schulz, Joern Wilms and Andrew Young for
their careful reading, and insightful comments, of various drafts of
this article.  We are particularly grateful to the referee, Eric Agol,
whose report led to a substantial improvement of the manuscript.  In
addition, we also express our appreciation to Tomaso Belloni, Omer
Blaes, Jim Chiang, Paolo Coppi, Stephane Corbel, James Dove, Rob
Fender, Julian Krolik, Tom Maccarone, Phil Maloney, and Katja
Pottschmidt for stimulating discussions and help in preparing some of
the figures for this article.  We acknowledge and thank NASA for
financial support under grants NAG5-10083 (CSR) and SV-61010 (MAN).

\end{document}